\documentclass[sigconf,authorversion,nonacm,table]{acmart}
\AtBeginDocument{%
  }

\usepackage{soul}

\colorlet{usercolorname}{yellow!0}
\sethlcolor{usercolorname}

\usepackage{svg}

\usepackage{multicol,multirow}

\newcommand{\textdi}[1]{\textbf{\textsc{#1}}}

\newcounter{numquote}

\copyrightyear{2026}
\acmYear{2026}
\setcopyright{cc}
\setcctype{by-nc-nd}
\acmConference[CHI '26]{Proceedings of the 2026 CHI Conference on Human Factors in Computing Systems}{April 13--17, 2026}{Barcelona, Spain}
\acmBooktitle{Proceedings of the 2026 CHI Conference on Human Factors in Computing Systems (CHI '26), April 13--17, 2026, Barcelona, Spain}
\acmPrice{}
\acmDOI{10.1145/3772318.3790659}
\acmISBN{979-8-4007-2278-3/2026/04}

\begin{document}

\title[Radical Gender Neutrality: Agender Euphoria in Gaming and Play Experiences]{Radical Gender Neutrality: \texorpdfstring{\\}{} Agender Euphoria in Gaming and Play Experiences}

\author{Katie Seaborn}
\orcid{0000-0002-7812-9096}
\affiliation{%
  \institution{Institute of Science Tokyo}
  \city{Tokyo}
  \country{Japan}
}
\affiliation{%
  \institution{University of Cambridge}
  \city{Cambridge}
  \country{UK}
}
\email{katie.seaborn@cst.cam.ac.uk}

\author{Shano Liang}
\orcid{0000-0003-1976-9680}
\affiliation{%
  \institution{Worcester Polytechnic Institute}
  \city{Worcester}
  \country{United States}}
\email{sliang1@wpi.edu}

\author{Rua M. Williams}
\orcid{0000-0002-6182-8923}
\affiliation{%
  \institution{Purdue University}
  \city{West Lafayette}
  \state{Indiana}
  \country{United States}}
\email{rmwilliams@purdue.edu}

\author{Phoebe O. Toups Dugas}
\orcid{0000-0002-6192-2004}
\affiliation{%
    \institution{Exertion Games Lab}
  \institution{Monash University}
  \city{Melbourne}
  \state{Victoria}
  \country{Australia}}
\email{Phoebe.ToupsDugas@Monash.edu}

\renewcommand{\shortauthors}{Seaborn et al.}


\begin{abstract}
  \textit{Agender euphoria} is a new term representing the powerful feelings of happiness, joy, and contentment derived from experiences in gender-free embodiments, spaces, and activities. 
  People with \emph{and} without agender and adjacent identities (e.g., genderless, gender-free, non-binary, gender-apathetic) may have such experiences under the right circumstances. 
  Video games can offer gender minorities a safe haven for gender euphoric experiences. 
  However, the possibility of agender euphoric experiences was unexplored. 
  We considered this overlooked frame of self-actualization with 142 people who identified as having or desiring agender euphoric experiences. Using the critical incident technique (CIT), we uncovered how games and play experiences create (and inhibit) agender euphoria. We surface this experiential phenomenon and provide empirically-grounded criteria for the design of games to elicit agender euphoric experiences for everyone, but especially agender and agender adjacent players. This work adds to the growing critical literatures on marginalized experiences in games research and human-computer interaction.
\end{abstract}

\ccsdesc[500]{Human-centred computing~Empirical studies in HCI}
\ccsdesc[500]{Human-centred computing~User studies}
\ccsdesc[500]{Software and its engineering~Interactive games}
\ccsdesc[500]{Social and professional topics~Gender}

\keywords{Agender Euphoria, Gender, Video Games, Player Experience, Human-Computer Interaction}

\begin{teaserfigure}
  \includegraphics[width=\textwidth]{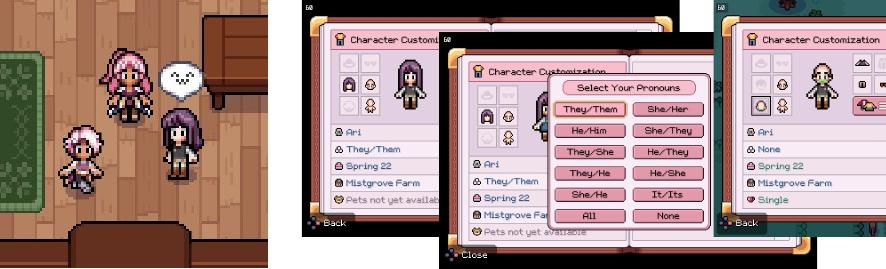}
  \caption{Screenshots from \textit{Fields of Mistria}, a game that elicited agender euphoria experiences. Left: \textbf{P18} noted that the pixilated, ``chibi'' art style enables flexible gendering of characters. 
  Right: Three stacked screenshots showing the character creation interface,
  which is always available and offers a range of pronouns changeable at-will. 
  The default pronouns are ``They/Them.'' 
  Any combination of trait and gender options are available: pronoun choice, name, and/or character design. 
  \textit{Fields of Mistria} represents an exemplar for designing agender euphoria experiences grounded in our findings: \textbf{one-size-fits-none}, \textbf{gender-free play}, \textbf{gender on/off switch}, and \textbf{sandbox-style freedom}.
  Screenshots taken by the authors.}
  \Description{Left: A screenshot showing three pixilated anime-style characters in a house. Right: Three screenshots: A long-haired anime character wearing a shirt and slacks (named ``Ari'' with pronouns ``They/Them''); a modal dialogue with the title ``Select Your Pronouns'' and options for ``They/Them,'' ``She/Her,'' ``He/Him,'' ``She/They,'' ``They/She,'' ``He/They,'' ``They/He,'' ``He/She,'' ``She/He,'' ``It/Its,'' ``All,'' ``None;'' and a bald anime character with a large green moustache wearing a shirt and slacks (named Ari wit pronouns set to ``none'').}
  \label{fig:teaser}
\end{teaserfigure}


\maketitle


\section{Introduction}

Video games are designed to give rise to positive, pleasurable, and often meaningful experiences. 
A diversity of players enjoy games~\cite{shaw2015gaming}. Growing recognition of just how diverse gamer populations are~\cite{Williams_2009rep,shaw2015gaming} has spurred critical work on representation in games~\cite{Liang2023misrepresentation,_isler_2008} and player identity~\cite{shaw2015gaming,Shaw_2011identity,Christoph_2009self}. Games can be uncomfortable spaces when identity is portrayed in stereotyped and limited ways~\cite{Williams_2009rep,shaw2015gaming} or even harmful to certain identity groups~\cite{Brock_2011racism,_isler_2008,liang-2025-three-steps}, to say nothing of identity-based toxicity in multiplayer games~\cite{Beres_2021toxic}. Still, games can also include marginalized identities and elicit experiences of delight and even euphoria connected to those identities~\cite{reyes2022theimpacts,Whitehouse_2023agender,liang-2025-euphoria}.

Identities and modes of experience 
described as \emph{\textbf{agender}} are among the most marginalized and understudied~\cite{Whitehouse_2023agender}.
People who adopt the agender label may reject gender entirely, reject the binary of masculine and feminine specifically, or feel apathetic towards gender~\cite{ketola2022identity,Whitehouse_2023agender}.
Agender and adjacent \emph{identities} (hereafter \textbf{\textit{agender+}}) include those who identify as genderless, gendervoid, gender-free, gender-apathetic, non-binary, and more. 
The agender identity reveals that gender and/or sex\footnote{We distinguish gender as social and sex as morphological, considering both as social constructs that often intersect~\cite{Hyde_2019sexgender}; refer to \autoref{sec:bg}.} %
is profoundly infused into personal and social experiences, even though they need not be. 
People of all gender identities may have agender \emph{experiences} that are typically characterized by:

\begin{enumerate}
    \item a desire to experience the world without the confines of gender; 
    \item a rejection of gender norms (especially the binary model);
    \item a lack of interest in gender as identity; and/or 
    \item apathy or indifference about gender~\cite{Whitehouse_2023agender,galupo2017like,ketola2022identity,hcigender,Spencer_Hall_2021terms}.
\end{enumerate}

Agender identities are often omitted or categorized within the non-binary\footnote{``Non-binary'' refers to people who define their identity as beyond the gender binary; refer to \autoref{sec:bg}. Notably, non-binary people may not entirely reject binary genders, while agender people may reject gender entirely.} or transgender\footnote{Those who identify as ``transgender'' have a different gender than the one assigned at birth; refer to \autoref{sec:bg}.
We do not conflate the two, but note that these identities are often grouped together.} umbrellas. 
Examples include eleven agender participants in the RPG-centred work of \citet{VanWert2024}, three agender+ participants the non-binary-oriented study by 
\citet{Smith2025binbarriers}, and an unknown (but seemingly small) number of agender(+) participants in \citet{maletska2024queer}.
This resonates with gender multiplicity (i.e., agender+) but also obscures the agender identity and mode of experience. 
Only one study (\citet{Whitehouse_2023agender}) has explored the intersection of games and agender identities by distinguishing and then focusing on the reports of six agender-identifying players. These agender players reported strategies on using gender absence or ambiguity in avatar presentations or performances to achieve gender-aligning experiences. This set the stage for our work, where we scoped up to a \emph{larger sample} and reorientated the focus more inclusively on agender euphoric \emph{experiences} that all people can have.

\textbf{Gender euphoria} 
is the intense feeling of joy and affirmation resulting from experiencing one's gender identity
physically and/or socially~\cite{austin2022gender,beischel2022gender,Blacklock-gender-euphoria-scale-2025}. 
This can happen in games, where people can connect meaningfully with the sense of self and inhabit various avatars, characters, relationships, and perspectives~\cite{banks2015object,george2023,Livingston2014how,reyes2022theimpacts,collecting-toups-dugas}. 
Most work has focused on transgender people~\cite{ashley2018favor,bradford2021hair,beischel2022little,skelton2024itjust,toups-dugas-pole-2024}, even though anyone can experience gender-based euphoria, including cisgender people\footnote{``Cisgender'' refers to people whose gender assigned at birth matches their gender identity.}. 
The preliminary work of \citet{Whitehouse_2023agender} and \citet{austin2022gender} suggest that \textbf{\underline{a}gender euphoria} can arise from \textit{not} experiencing gender, feeling unconstrained by gender, or being in a gender-neutral world. 
As such, agender euphoria may be a markedly different phenomenon from other gendered experiences (e.g., gender euphoria) and have different design requirements, but this has yet to be explored fully, in games or elsewhere.

We addressed this gap by exploring \textbf{agender euphoria experiences}---intense feelings of happiness, joy, and contentment borne from gender-free embodiments, spaces, and activities. 
We believe that everyone, but especially marginalized individuals, deserve to experience gender euphoria as a hallmark of well-being~\cite{Grant_2024euph}. 
However, at present, we do not know to what extent \underline{\textbf{a}}gender euphoria features in games.
We also have no empirical basis for designing games that ensure, if not elicit, \underline{\textbf{a}}gender euphoria.
To this end, we aimed to discover the nature of agender euphoric experiences in games. We carried out a critical user experience (CRUX) survey using the critical incident technique 
(CIT)~\cite{woolsey_critical_1986}, a method for capturing UX in general~\cite{hassenzahl_experience_2010,muller_facets_2015,mekler_momentary_2016} and games UX specifically~\cite{Seaborn_2023maldai,Seaborn_2024maldai}. Our overarching research question (RQ) was: 
\emph{\textbf{Can games enable agender euphoric experiences?}} 
Our findings from \textbf{142} agender+ players recounting agender euphoric experiences offer key contributions on games and identity:

\begin{itemize}
    \item \textbf{Knowledge}: Thick descriptions of agender euphoria contextualized to a range of games and play experiences for a diversity of players who identify within and beyond the agender continuum.
    \item \textbf{Design}: Guidelines for game designers to enable these experiences, notably how to design the radically gender-neutral player experiences many desire (and how to avoid common pitfalls that prevent euphoria or relate to dysphoria).
    \item \textbf{Social}: Recognition and empirical validation of agender euphoric play experiences as distinct and overlapping 
    with other varieties of gender euphoria.
\end{itemize}

Our work is the first of its kind and an important addition to the growing literature on games and identity~\cite{Shaw_2011identity,Gabriela_T_Richard_2018,Kuss_2022femgamer,Kivij_rvi_2021,George_2021,Trammell_2023,liang-2025-euphoria}. Our findings can inform game design practice---from the indie world to the AAA level---in pursuit of ensuring an inclusive, entertaining, and even uplifting experience for a diversity of players and modes of experience~\cite{Storr_2021,Freeman_2022}. We build on the work identifying barriers and limitations, e.g., \cite{Jaroszewski_2018,Drenten_2022,Freeman_2022,Liang2023misrepresentation,Trammell_2023}, with our focus on how to elicit euphoric experiences, a nascent topic~\cite{George_2021}, and specifically identity-transcendent agender euphoric experiences. 
Our work is significant in elevating the underexplored dimension of agender euphoria. 
Moreover, 
our findings offer ways in which to achieve the often sought-after ``gender neutrality'' in HCI and related spaces~\cite{Vorvoreanu_2019neut,Keyes_2018,Seaborn_2022neut,Sutton_2020ambig}.


\section{Background}
\label{sec:bg}

We begin with a primer on gender and agender identities, 
games, and gender euphoria, which heavily informed our approach and analysis. We also cover the emergent literature on gender expression and player experience. 

\subsection{Gender: A Primer}

\textit{Gender identity} is one's understanding of being in the world personally and socially. While the gender binary of man and woman is prevalent~\cite{Hyde_2019sexgender,rode-2025-reframing-gender}, there are infinite, concurrent possibilities, like agender and gender-apathetic, non-binary and genderfluid, and trans woman~\cite{dypshoriabible,transhistory,gendermap,third-gender-werft}.
We distinguish gender identity from gender/ed \emph{norms} that arise when societies use sex and assigned gender to segregate, control, and oppress people via appearance, behaviour, and/or modes of dress~\cite{Hyde_2019sexgender,Tannenbaum_2019}. 
\textit{Sex} is a named physical configuration of the body based on external appearance---genitals and post-puberty characteristics---that result from internal processes---exposure to sex hormones---categorized into intersex, female, or male~\cite{dypshoriabible,blackless2000}. 
People are often assigned a sex and binary gender at birth, despite the existence of 
other options~\cite{Rahilly_2022},
based on the infant's externally observed sex
~\cite{blackless2000,fausto2000sexing,Turner1999,carpenter-intersex-2024}. 
People whose gender identity differs from this assignment are \textit{transgender}, while those who accept the assignment are \emph{cisgender}. 

\subsection{Agender Identities and Experiences: Terms, Definitions, and Modes of Expression}

Scholarship in agender and non-binary studies~\cite{ketola2022identity,richards2016non,galupo2017like,hcigender,Spencer_Hall_2021terms} defines an agender \textit{individual} as a person who does not find the 
notion of gender relevant to their self-identity. 
The term's origin has been traced to the \textit{alt.messianic}\footnote{This web page is now offline.} chatroom on \textit{UseNet}, where it was used to denote the gender-amorphous nature of ``God''~\cite{them2018inqueery,ketola2022identity}. Now, ``agender'' as a term and concept transcends time and degree of gender-relevance.
Agender \emph{identities} can be defined as: a non-binary gender identity, i.e., with reference to the gender binary; without a gender identity, i.e., being genderless, gender-free, or gender-neutral~\cite{papisova2016wahtitmeans,ketola2022identity}; or categorically free from gender, i.e., gender irrelevance~\cite{ketola2022identity}.
\citet[p.~4]{Whitehouse_2023agender} discovered agender variants that touch on \emph{experiences} of gender.
\textit{Gender-distanced} describes a feeling of being far away or disconnected from gender. \textit{Gender-absence} describes feeling devoid or absent of gender~\cite{moistmogai2023agenderandadjacent}. \textit{Gender-apathetic} refers to apathy about gender~\cite{bachert2023apathetic,grindr2024apathetic}. \textit{Genderfluid} refers to a person's identity shifting over time or situation~\cite{galupo2017like,diamond2020gender}.
Hence, ``agender identity'' is a continuum \emph{and/or} a singular identity, distinguished from but connected to ``agender experience'' as mode of being or quality of existence. The ambiguity of the Greek prefix ``a-'' (without) is reflected in the evasiveness of a firm definition or singular locus in identity or embodiment.

Agender identities and experiences are often ``hidden'' within other gender categories~\cite{ketola2022identity}.
Notably, the agender label is often placed under the non-binary umbrella~\cite{richards2016non,hcigender}. Yet, ``non-binary'' covers other genders (gender \emph{presence}), while ``agender'' rejects this premise (gender \emph{absence}). 
Many agender people also have other gender identities~\cite{ketola2022identity,Whitehouse_2023agender}. 
In \citet{galupo2017like}, people 
described ``not comfortably fitting into male or female,'' having ``a blend of male and female characteristics,'' and being ``somewhere beyond rather than between''~\cite[pp.~170--173]{galupo2017like}. 
This pluralism leads people to use multiple descriptions to articulate their identity and experiences. 
Indeed, agender self-identification is unique and challenging: defined by an \emph{absence} of identity or experience~\cite{ketola2022identity} and averse to \emph{conventional} gender categories or modes of being, if not gender itself~\cite{ketola2022identity}. 
This complexity guided our choice of an inclusive framing, 
whereby we composed our prompts to be about agender euphoric \emph{experiences} rather than identities (\autoref{sec:procedure}).
We also use ``\textit{agender+}'' to represent the plurality and fluidity in those who experience agender euphoria, ensuring representation of people with similar gender perceptions and experiences in this study.

\subsection{Terms, Concepts, and Theory for Identity in Games}

The novelty of agender euphoric experiences in general and in games presents a challenge: this is virtually untrodden ground. 
We recognize that this literature approaches identity within game contexts as matters of representation (\emph{in} games) and construction (\emph{through} games). 
\textit{Game mechanics}---designed choices players can invoke during play~\cite{rules-of-play,Adams:2012:GMA:2385822}---and \textit{player representation}---how a player inhabits a game~\cite{Livingston2014how,gameworld-jorgensen,tyler-mmorpg,McArthur-avatar-affordances-2015}---are the means by which players shape their virtual identity.  
Game rules, environments, and player choice architectures 
intimately tie these core elements of games to player self-expression, identity recognition, and the immersive experience of gameplay~\cite{banks2015object,reyes2022theimpacts,Liang2023misrepresentation,tyler-mmorpg,McArthur-avatar-affordances-2015}. 
As a baseline, we first 
establish a shared understanding of relevant game terms, concepts, and theory.

\subsubsection{Game Mechanics and Player Representation}
\label{sec:gmplayer}

Games consist of rules that constrain choice and play that enables it \cite{rules-of-play}. 
Game \textit{rules} address what a player can and cannot do, reified through a combination of physical, digital, and/or social constraints, e.g., avatars prescribed by the avatar-creation UI~\cite{McArthur-avatar-affordances-2015,tyler-mmorpg}, characters unable to move through the ground, or 
pushing a control stick moves an avatar in the direction pushed. 
Game \textit{play} addresses the player's freedom to act within the rules' constraints, e.g., the player's freedom to choose an avatar (subject to the rules), their choice of movement direction and speed in open space or water, or how much and what direction to push a control stick to move. 
\textit{Game mechanics} are designed moments of choice in games: repeated action-outcome cycles consisting of the player observing the game state (as defined by the rules), playing by choosing action(s), and observing the new state~\cite{rules-of-play,Adams:2012:GMA:2385822}. 

Gameplay occurs within special space and time, 
implying that regular life and play are separated. 
This is commonly called the \textit{magic circle}\footnote{
\citet{huizinga_homo_2000} listed a number of other sites, e.g., the court of law or a stage for plays, within which ``special rules obtain.'' 
Yet, the term is technically no different than any other special place and time on the list, and may have been chosen due to its evocativeness and peculiarity.}~\cite{tekinbas2003rules,huizinga_homo_2000}.
The magic circle may interweave with regular life, especially in the case of mixed reality and mobile games. 
The concept usefully captures how players freely engage in behaviours and experimentation that is otherwise not possible, opening the door for explorations of agender euphoria. 


Crucial is how a player controls, influences, and/or interacts with a game through some entity, e.g., avatar, character, embodiment, locus of control~\cite{Bayliss2010Videogames-Inte,DAloia:2009,reyes2022theimpacts}. 
We use \textit{player representation} to describe any scenario in which an entity or entities are controlled by and/or meant to embody the player. 
A player may or may not identify with the player representation; sometimes this is by designer intent. 
Examples 
include via the camera, such as a character being inhabited through the first-person perspective; an avatar crafted at the start of a game; a rotating party of pre-designed characters; a spaceship; a cursor; and so on. 
We use \textit{non-player character (NPC)} to refer to entities that the player may, through their player representation, 
interact with (e.g., speak to, heal, fight)~\cite{dnd2ephb}. 



\subsubsection{Queer Game Studies}

Game mechanics and player representations in games allow players to experience themselves reflected in play~\cite{Whitehouse_2023agender,liang-2025-euphoria,collecting-toups-dugas,Livingston2014how}, create scenarios in which players are marginalized~\cite{Liang2023misrepresentation,liang-2025-three-steps}, or set norms for 
the real world~\cite{Liang2023misrepresentation,tyler-mmorpg,gardner-tanenbaum-census-2018}. 
Much of the discourse on the intersection of identities and games originates in \textit{queer game studies}, which arose from the convergence of game studies, gender studies, and early queer movements. 
Queer game studies focuses on identity and gender representation in games, as well as how insights from this intersection challenge the normalizing tendencies of game studies and ways of comprehending games~\cite{ruberg2019videogameshavealwaysbeenqueer,ruberg2017queergamestudies}.
Much work has critically
examined how LGBTQI+ identity representations in games can perpetuate marginalization~\cite{phillips2017welcome,shaw2016whereisthequeerness,braganca2016twinegame,utsch2017queer,lawrence2018if}.
Others have explored the presence of accessible digital spaces within game realms that facilitate the inclusion of marginalized identities and community expression~\cite{jenson2010gender,halberstam2000tellingtalesbrandonteena,halberstam2017queergaminggaming,nakamura2012queer,pulos2013confronting}. 
We take a cue from \citet{ruberg2020thequeergamesavant}, who argues that the lens of queer theories could imbue game design and analysis of games, which should apply to agender euphoria. 

\subsection{(A)gender Euphoria: Joyous Alignments of Self and Experience}

As the positive homologue of \textit{gender dysphoria}~\cite{dypshoriabible,transhistory,Liang2023misrepresentation,lev2013gender,Blacklock-gender-euphoria-scale-2025}, \textbf{gender euphoria} is 
defined as a range of positive feelings 
arising from the congruence between one's internal identity and the gendered features of their experience~\cite{beischel2022little,austin2022gender,skelton2024itjust,ashley2018favor,reisner2023exploring,blacklock2024gender}. 
Gender euphoria has a significant positive impact on mental health for transgender and non-binary people (e.g., \cite{lambrou2020learning,cosgrove2021service,austin2022gender}), even while it is idiosyncratic, having a dynamic, complex, and multi-dimensional nature~\cite{skelton2024itjust,tacit2020joyful}.
Some work has attempted to untangle this complexity and identify specific features of gender euphoria. 
For instance, \citet{beischel2022little}, using survey data, identified three sources of gender euphoria: (i) \textsc{external} (euphoria in relation to physical aspects of a person's gender); (ii) \textsc{internal} (euphoria from self-affirmation or thinking of oneself in certain gendered way); and (iii) \textsc{social} (euphoria from interactions with others).
Similarly, \citet{skelton2024itjust}, who focused on transgender euphoria experiences, echo this understanding while contributing more specific sources for euphoria, from e.g., self-reflection, healthcare, close relationships, the broader community. 
\citet{Blacklock-gender-euphoria-scale-2025} published a protocol for a 
``Gender Euphoria Scale,'' as yet unvalidated. 
Gender euphoria can be a pillar of well-being for trans* and gender-diverse folks~\cite{Grant_2024euph}, which guided our inclusion of well-being measures.

Very few have explored \textbf{agender euphoria} or 
expressions of agender+ identities. 
In the context of virtual worlds, \citet[p. 429]{ketola2022identity} discovered three ways that agender individuals express identity: (i) ``gender neutralization,'' or carrying out a hyper-performance of the ``opposite'' gender to counter their assigned one; (ii) ``gender sterilization,'' or removing all gender-related elements from expressions; and (iii) ``agender performance,'' or acts of agender-specific expressions and activities. \citet{ketola2022identity} also captured three features of digital worlds that promote agender identity expressions: (i) sense of control over identity presentation; (ii) segregation and aggregation of identity performances; and (iii) identity performance shortcuts. 
To the best of our knowledge, \citet{Whitehouse_2023agender} offers the only research on agender play to date. This foundational work involved interviews with 40 trans* and gender diverse players, of which six identified as agender, focusing on their general experiences with identity alignment and expression in games and virtual worlds. 
\citet{Whitehouse_2023agender} elevated the notion of agender euphoria while not precisely characterizing it as such and limiting the framing as one result of transgender and gender diverse people's feelings of gender alignment with avatars. 
We expanded on this work by seeking to specifically capture agender euphoric experiences, regardless of identity. 
The dearth of work so far within and outside of games and play leaves much to be explored and understood.  
As a baseline, we sought to understand the \emph{nature} of agender euphoric experiences, asking: 
\begin{quote}
\textbf{RQ1}: \emph{What kinds of agender euphoric experiences do games enable?} 
\end{quote}

We also recognized that games can offer experiences of \emph{dysphoria}~\cite{Liang2023misrepresentation,kai2022euphor}, where forced gendering and misgendering embedded in the design of games or meted out in a play context by others (NPCs and real players) can lead to discomfort, wrongness, distress, and even trauma. Considering the potential for agender \emph{dysphoric} experiences, we asked: 
\begin{quote}
\textbf{RQ2}: \emph{What barriers are there to agender euphoric play?} 
\end{quote}

In \citet{Whitehouse_2023agender}, \emph{avatars}, a form of player representation (\autoref{sec:gmplayer}), emerged as a crucial factor in euphoric experiences.
Indeed, avatars can be powerful engines for gender euphoria~\cite{liang-2025-euphoria,kai2022euphor} (and \emph{dysphoria}~\cite{Liang2023misrepresentation,kai2022euphor}). 
We therefore created prompts specifically on avatar creation and avatar experiences. 
Given the nascency of this work, we also included prompts on what other features of games as designed artefacts elicit (or created barriers to) agender euphoria.
With the aim of developing guidelines and best practices for game designers, we asked: 
\begin{quote}
\textbf{RQ3}: \emph{What designs support agender euphoric play?} 
\end{quote}


\section{Methods}
\label{sec:methods}

We conducted an online survey
, replicating and expanding on the procedures for eudaimonic UX~\cite{mekler_momentary_2016} and maldaimonic game UX~\cite{Seaborn_2023maldai,Seaborn_2024maldai}. We pre-registered our protocol on January 18\textsuperscript{th}, 2025 via OSF\footnote{Protocol: \url{https://osf.io/tv8jp}}. The protocol was approved by the institutional ethics board on December 27, 2024 (\#2024262).

The main method in the survey was the CIT~\cite{woolsey_critical_1986}, a qualitative approach to systematically gathering self-reports on \emph{incidents}---events and experiences---that had a significant---whether positive or negative---impact on an individual. Prompts are used to elicit and guide reporting of a single incident that the person deems critical within the context of the larger topic, here agender euphoria in games. The elicited experiences are critical in the sense that the outcomes for the person were substantial, typically affecting their attitudes and behaviours in the content domain. This method has long been employed within HCI~\cite{hassenzahl_experience_2010,mekler_momentary_2016,muller_facets_2015} and specifically for understanding the player experience~\cite{Seaborn_2023maldai,Seaborn_2024maldai}.

We also supplemented this qualitative approach with quantitative measures related to well-being orientations, affect, and player experience, in line with the literature~\cite{muller_facets_2015,mekler_momentary_2016,Seaborn_2023maldai,Seaborn_2024maldai} and the link between euphoria and well-being~\cite{Grant_2024euph}.

\subsection{Participants}
\label{sec:participants}

Participants were recruited from Prolific\footnote{\url{https://www.prolific.com}}, an online panel service that vets individuals and collects detailed demographics for fine-tuned screening that cannot be quickly or easily changed, preventing identity manipulation for a given study\footnote{\url{https://researcher-help.prolific.com/en/article/412c0a}}. Recruitment had two phases: (i) an all-encompassing agender+ screener via ``Gender identity non-binary only''\footnote{Operationalized as: Bigender, Non-binary, Demiwoman, Demiman, Gender fluid, Gender queer, Agender, Genderless, Questioning/ unsure, Option not included here, Rather not say.} and ``Gender
Non-binary'' ($N=91$); and (ii) a targeted screener via ``Gender identity non-binary only
Agender, Genderless'' ($N=59$). Both phases also screened for English as the ``Primary language'' and ``Playing games,'' including:  Computer games, Console games, Handheld console games, Free-to-play mobile games, Premium mobile games (pay to download), Esports games, Virtual reality games, Online casino games, Collectible card games, Board games. We excluded participants who provided insufficient or nonsensical responses ($n=8$).

\begin{figure*}[!ht]
  \centering
  \includegraphics[width=.7\linewidth]{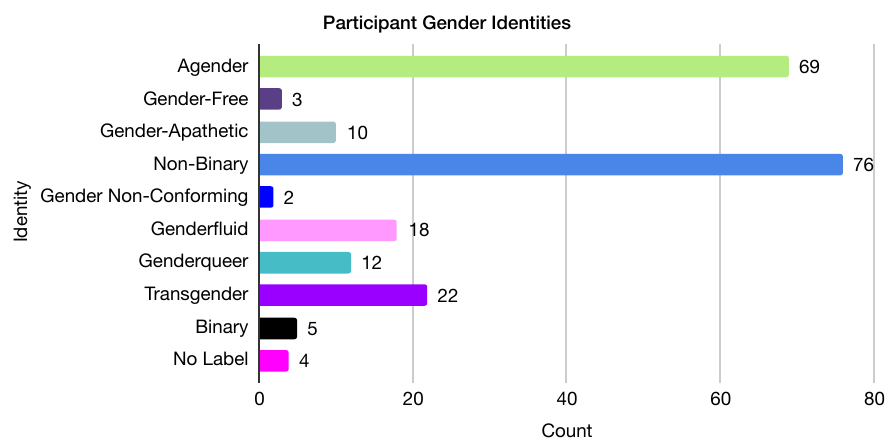}
  \caption{
  Gender identities self-reported by participants. 
  Note: Many participants reported multiple identities.}
  \Description{A graph indicating a range of agender+ identities, with agender and non-binary showing the highest representation.}
  \label{fig:identities}
\end{figure*}

Participant gender identities are represented in \autoref{fig:identities}. 
Anyone who had an agender experience was invited. 
Most identified as non-binary ($n=76$) and/or agender ($n=69$). The most common multiple identities were: ``Agender or a related identity (e.g., genderless, gender-free, gender-apathetic, gendervoid, etc.)'' and ``Non-binary'' ($n=30,	20\%$); ``Agender or a related identity (e.g., genderless, gender-free, gender-apathetic, gendervoid, etc.),'' ``Non-binary,'' and ``Transgender'' ($n=10,	6.7\%$); and ``Agender or a related identity (e.g., genderless, gender-free, gender-apathetic, gendervoid, etc.)'' and ``Transgender'' ($n=5,	3.3\%$). The average age was 29.2 ($SD=7.8, MD=27, IQR=9, min=18, max=68$; one did not report their age). This indicates that agender euphoric experiences cross generations. 

\subsection{Procedure and Qualitative Instrument (RQ1--3)}
\label{sec:procedure}

Participants were given a link to a Google Form. The first page included the information sheet, the consent form, and an attention check. Those who passed the attention check were given the following definition of agender euphoria:

\begin{quote}
    ``Agender euphoria'' refers to the powerful feelings of happiness, joy, and contentment that people with agender and adjacent identities (genderless, gender-free,  gender-apathetic, etc.) experience when they perceive themselves to be in gender-free embodiments and spaces. We are asking you to share an example from your personal experience, if you have any.
\end{quote}
\noindent
This acted as an intentional prime to focus recall on a specific, relevant experience~\cite{woolsey_critical_1986}.

Participants were then asked to describe their gender identity in an open field: 
\begin{quote}
``First, how would you characterize your identity in terms of gender?'' 
\end{quote}

Next, they were asked (``Please select the option that applies to you:'') whether they could provide an account of such an experience, and if not, whether providing a non-euphoric experience or ideas for agender euphoria was possible. 

Those who could provide an account were given this prompt, adapted from \citet{Seaborn_2024maldai}:

\begin{quote}
    Can you bring to mind a single ``agender euphoric'' experience that you have had in a game? Think of ``agender euphoria'' in whatever way that makes sense to you. You can choose any type of game on any platform, including games played on a smartphone, a computer, a console, etc. You can share a solo or multiplayer experience. It can be one where you chose to act in an agender euphoric way or someone else did, or it was required by the gameplay.
\end{quote}

Those who had a non-euphoric experience were given the same prompt but with that term in place of ``agender euphoria.''
As in \citet{Seaborn_2024maldai}, participants were asked to provide the name of the game, the platform, and whether they were playing with other people. They were then given the two main open-ended items:

\begin{quote}
    Now, please describe the ``agender non-euphoric'' experience you had in the game. Please be as detailed as possible.
\begin{quote}
\vspace*{-2.5mm}
\end{quote}    
    More specifically, what aspects of the game were non-euphoric to you? You can provide a list of features, actions, play styles, skill trees, strategies, party configuration, or other elements, if that makes sense.
\end{quote}

In recognition of the affordances games can offer as magic circles~\cite{tekinbas2003rules} and creative spaces~\cite{ruberg2025queer,collecting-toups-dugas,Livingston2014how}, especially for queer people~\cite{ruberg2025queer}, we also asked:

\begin{quote}
    Were there any elements of the game that prevented you from exploring things you cannot do in real life, in relation to agender euphoria? This could involve being unable to question or challenge the authority of the game in terms of the gameplay or the story (or something else).
\end{quote}
\noindent
After these, participants were asked to report how they felt during the experience. 

Then, participants were asked an open-ended item about creating a player representation, distinguished from the general experience due to its saliency~\cite{Whitehouse_2023agender}:

\begin{quote}
    Now, we would like ask about your avatar or character creation experience in relation to agender euphoria, if applicable. Please describe in as much detail as possible.
\end{quote}

Those who did not have an experience to share were given the following prompt:

\begin{quote}
    Even though you have not had an experience that you can share, can you imagine an agender euphoric experience that you would like to have in a game? Please describe. 
\end{quote}
\begin{quote}
    If you do not have an idea, please explain what makes imagining an experience difficult for you.
\end{quote}

After answering the qualitative items, participants were then presented with the quantitative instruments (\autoref{sec:measures}) and asked to provide demographics (their gender identities, including and beyond agender, alongside age). They received a Prolific compensation code after submission.

\subsection{Quantitative Measures (RQ1)}
\label{sec:measures}

We supplemented the qualitative insights with quantitative measures, borrowing from \citet{mekler_momentary_2016} and \citet{Seaborn_2024maldai}. 
We adapted all instructions to refer to the reported experience (whether agender euphoric or not).

\subsubsection{Positive and Negative Affect (I-PANAS-SF)}
The very notion of euphoria rests on positive affective experiences. Moreover, positive affect is correlated with subjective well-being and notably hedonia and eudaimonia~\cite{Deci_2006}. As such, we used the International Positive and Negative Affect Schedule Short-Form (I-PANAS-SF) instrument~\cite{Karim_2011panas}, a cross-culturally validated version of the 10-item PANAS-SF~\cite{Thompson_2007panas}, which measures mood in halves of positive and negative items. We chose the international version to ensure future replicability across cultural cohorts. Items were presented in random order, with response items in the form of 5-point Likert scales (0~=~``strongly agree,'' 4~=~``strongly disagree'').

\subsubsection{Hedonic and Eudaimonic Orientations (HEMA-R)}
Subjective well-being and well-being orientations are linked to euphoric feelings for gender-diverse folk~\cite{Grant_2024euph}. Orientations direct our actions and activities in the world~\cite{ahmed2020orientations}, including game worlds~\cite{ruberg2025queer}.
We used the trait-level version of the 10+1-item Hedonic and Eudaimonic Motivations for Activities-Revised (HEMA-R) instrument with the optional item about comfort~\cite{Huta_2013,Huta_2016}. These operationalizations of well-being represent pleasure and pain avoidance drivers (hedonia) and self-actualization and meaningful experience drivers (eudaimonia).
The instrument captures general well-being orientations in a 1~=~``not at all'' to 7~=~``very much'' response scale. Example items include ``seeking comfort'' and ``seeking to do what you believe in.'' We expect correlations with positive affect and between hedonia and eudaimonia~\cite{Huta_2013}, as well as correlations with player experience via the instrument we used (described next), based on previous research~\cite{Seaborn_2024maldai}.

\subsubsection{Player Experience (PXI)}
We used the validated Player Experience Inventory (PXI)~\cite{abeele_development_2020}, aligning with previous work~\cite{Seaborn_2024maldai}. 
We used the Psychosocial Consequences and Functional Consequences subscales, plus the optional Enjoyment scale. Psychosocial Consequences may reflect eudaimonia orientations~\cite{Seaborn_2024maldai}.  Functional Consequences and Enjoyment are expected to reflect euphoric and positive experiences. We used a seven-point Likert scale, but due to limitations with Google Forms, we could not use the recommended 0±3 format~\cite{abeele_development_2020}.

\subsubsection{Long-Term Meaning Potential (LTMP)}
This factor is linked to eudaimonia and critical incidents~\cite{kim_pleasure_2014,mekler_momentary_2016,Seaborn_2024maldai}, establishing the weight and meaning of the account. We used the one-item scale from this previous work on  a 7-point importance scale (1~=~``not important at all,'' 7~=~``very important''). The item was:

\begin{quote}
If you consider your life one year from now, how important will you find this experience?
\end{quote}

\subsection{Data Analysis}
\label{sec:analysis}

We carried out a convergent parallel mixed methods analysis, which involves collecting and analyzing qualitative and quantitative data separately before interpreting and discussing the results of each together with respect to the RQs~\cite{creswell2007mixed}. This allowed us to capture a rich set of data on multiple concurrent but different facets of the experiential accounts.

\subsubsection{Qualitative}
\label{sec:analysis_qual}

We used reflexive thematic analysis (RTA) to evaluate the qualitative data~\cite{Braun_2019,Braun_2022}. The unit of analysis was all responses to the items related to each prompt: on the experience, this was three items at $\sim$67, $\sim$31, and $\sim$27 words (in total: $\sim$125 words); on the avatar, there was one item at $\sim$40 words; and on the ideas, there was one item at $\sim$95 words.

For the accounts, the first and third authors took a collaborative hybrid inductive and deductive approach~\cite{Proudfoot_2022}. We employed Braun and Clarke's methodological frame, with codes as the building blocks of themes representing patterns of meaning across accounts~\cite{Braun_2019,Braun_2022}. We also treated our subjective interpretations of the data as valid knowledge, leaning into our identities as agender+ gamers (\autoref{sec:positionality}). We separately read through the data and developed codes, working from opposite ends of the sheet. We also referred to the stated games, social context, and feelings data. We then met to discuss the codes in real-time to explain our interpretations. Thereby, we identified unique concepts from each other's analysis, and found that these missing links folded in well to each other's developing thematic frame. We then separated to finalize coding and provide commentary on the data, building towards a final cohesive frame. The first researcher then read through the codes and commentary, supplying comments that the other later checked. They then developed the codes into themes based on a shared understanding of a simple matrix between agender and gameplay dimensions. At this stage, they drew from the literature to ensure representation of existing themes deductively, notably from \citet{ketola2022identity} and \citet{Whitehouse_2023agender}. The third researcher signed off on the resulting first draft of the ``prism,'' which was then discussed and signed off in a live session with all authors.

For the avatar and ideas data sets, the first researcher developed codes and themes alone, but continuously referred to the account themes, given substantial overlap, despite participants being instructed not to discuss avatars until later. The third researcher checked and agreed on the final thematic framework.

The identity statements were coded by the first and second authors using hybrid analysis~\cite{Proudfoot_2022}. Gender identity codes were derived deductively from the literature on agender+ identity labels~\cite{Whitehouse_2023agender,ketola2022identity,Spencer_Hall_2021terms,richards2016non} and inductively from participant self-descriptions (reported in \autoref{sec:participants}).
The first author coded one-quarter and the second coded the rest. There were no disagreements.

All researchers collaboratively checked the analyses multiple times, even if they did not participate in coding or thematic development at any stage. No major disagreements were had, and minor disagreements were resolved by integrating multiple perspectives (e.g., discussing and then agreeing that ``social'' includes NPCs) or collaboratively finding consensus on descriptors (e.g., rather than ``illegibility'' or ``confusion'' deciding on ``evasion'').


\subsubsection{Quantitative}
\label{sec:analysis_quant}

We generated descriptive statistics and carried out inferential analyses where applicable. We used Google Sheets and \citet{statskingdom}, which provides an online interface for R statistics packages\footnote{\url{https://www.r-project.org/}}. We generated counts and percentages for the demographics. We generated means ($M$), standard deviations ($SD$), medians ($MD$), and interquartile ranges ($IQR$) for the main measures (\autoref{sec:measures}): the I-PANAS-SF, HEMA-R, PXI, and LTMP.
We used the Shapiro-Wilk test to test normal distribution of the data. 
In line with previous work, and for future meta-syntheses, we used Pearson correlations, a common test,  to explore the expected relationships between measures (\autoref{sec:measures}).

\subsection{Positionality Statement}
\label{sec:positionality}

The experiences and positionality of each researcher inherently influences how data is interpreted and analysis is conducted~\cite{Braun_2019,Braun_2022,Bardzell_2010,Haraway_1988,Dalton_2013}. We briefly describe our experiences of gender in the context of our research.
The lead researcher is a gender-apathetic gamer who often feels gendervoid. 
They are ambivalent about gender, having negative feelings about societal norms while also fascinated by and fully in support of diverse gender identities. 
The second author is a trans binary femme who studies gender euphoria and marginalized group culture in games. 
Her gender was mis-assigned at birth and she spent decades in gender-incongruous environments and groups. 
The third author is a non-binary person who has written about experiencing ``gendervoid'' in online communities. 
They worked in video games as a technical artist and designer for 10 years. As a child, when asked about their gender, they once responded, ``I am a gamer.''
The fourth author is a queer trans woman. She is a senior academic with expertise in HCI and game design on trans inclusivity and serious games, notably gender euphoria. 
Our authors have diverse experiences of gender, making us competent at evaluating and analyzing the agender+ experiences of our participants. Each member of the research team was educated in a Western context, and our survey was limited to English. Therefore, our analysis may not generalize to gender expansive experiences in other cultural contexts.


\section{Findings}
\label{sec:findings}

We begin with the qualitative findings on the critical incident accounts and supplementary insights. 
From 150 initial submissions, we included \textbf{142} (removing eight or 5.3\%). Of these, 109	(72.7\%) reported on agender euphoria, 
four (2.7\%) covered non-euphoria, and
29	(19.3\%) offered ideas or reasons why it was difficult to report.

\subsection{Thematic Frameworks (RQ1)}
\label{sec:themes}

Our core framework is presented in \autoref{tab:prism_i} and \autoref{tab:prism_ii}. We used a minimalist approach for ease of reference, relying on the codes developed to capture the nature of the reported experiences. We arranged our themes as a prism, aligning an \emph{agender dimension} with a \emph{game dimension} and placing types of experiences as \emph{angles} at their intersections. This allowed us to maintain the important connection between agender experiences and game play.

\begin{table*}[!ht]
\caption{Agender euphoria in game experiences prism: agender and game dimensions and their angles (Part I).}
\label{tab:prism_i}
\begin{tabular}{p{3cm}p{5.5cm}p{3.9cm}p{4cm}}
\toprule
 & \textdi{Character} & \textdi{Mechanics} & \textdi{Narrative} \\
 & \textsc{Avatar, Camera} & \textsc{Actions, Abilities} & \textsc{Plot, Lore}\\
\midrule

\multicolumn{4}{l}{\textdi{Agender Baseline}} \\
\textsc{Gender-Free} & No mention of gender, no gender sliders, no pronouns & No relation to gender & No stigma even from evil characters  \\
\midrule

\multicolumn{4}{l}{\textdi{Agender by Design} } \\ \textsc{Representation\textsuperscript{1}} & Playable agender character & N/A & Agender backgrounds \\

\textsc{Validation} & Agender by default, not binary by default, ``a choice, not a given,'' ``seeing myself'' & No relation to gender & No relation to gender \\
 
\textsc{Irrelevance} & No character, no self-representation, no need to choose gender, ``appearance, not traits'' & No relation to gender, no gender adaptations, no characters (e.g., rhythm game) & No relation to gender, or not a major plot point, no gender consequences  \\
 
\textsc{Choice\textsuperscript{2}} & Option for agender+, non-humans, add-ons & N/A & N/A \\
\midrule

\multicolumn{4}{l}{\textdi{Agender Play\textsuperscript{3}}} \\ 
\textsc{Projection} & Via first-person POV, gender-free despite apparent gender, parasocial engagement & Via no ascribed gendering & Via no ascribed gendering, gender-free despite apparent gender  \\

\textsc{Evasion\textsuperscript{4}} & Masks that disguise face, burden-free via first-person POV, random assignment of characters & Via no ascribed gendering, secondary control (cat owner) & Via no ascribed gendering \\
 
\textsc{Exploration\textsuperscript{5}} & On-demand customization, random assignment of characters, create cast of characters & Multiple playable characters & Multiple playable storylines  \\
 
\textsc{Subversion\textsuperscript{6}} & Mix 'n matching, choice of non-birth gender, crossdressing/inversion of prescription & Choosing unexpected gender-marked classes & Reinterpreting gendered plotlines  \\
 
\textsc{Reflection} & Agender envy, realizing agender identity/questioning & Gendering not enforced & Evil but not gender evil, relief when gender-free \\

\bottomrule

\addlinespace
\multicolumn{4}{@{}p{\dimexpr\linewidth}@{}}{\footnotesize 
\textsuperscript{1} Deduced from ``representation'' via \citet{Whitehouse_2023agender}. 
\textsuperscript{2} Deduced from ``sense of control'' via \citet{ketola2022identity}. 
\textsuperscript{3} Deduced from ``gender-aligning performances'' via \citet{Whitehouse_2023agender}. 
\textsuperscript{4} Deduced from ``gender sterilization'' via \citet{ketola2022identity}.
\textsuperscript{5} Deduced from ``agender performance'' via \citet{ketola2022identity}. 
\textsuperscript{6} Deduced from ``gender neutralization'' via \citet{ketola2022identity}.
}
    
\end{tabular}
\end{table*}

\begin{table*}[!ht]
\centering
\caption{Agender euphoria in game experiences prism: agender and game dimensions and their angles (Part II).}
\label{tab:prism_ii}
\begin{tabular}{p{3cm}p{6.5cm}p{7.2cm}}
\toprule
  & \textdi{World} & \textdi{Social} \\
  & \textsc{Place, Aesthetics} & \textsc{Other Players, NPCs} \\
\midrule

\multicolumn{3}{l}{\textdi{Agender Baseline}} \\
\textsc{Gender-Free} & Non-humans, everyone wearing the same clothes & Gender not a characteristic or profile option, dialogue does not use gender referents \\
\midrule

\multicolumn{3}{l}{\textdi{Agender by Design} } \\ \textsc{Representation\textsuperscript{1}} & Agender identities exist or ``seeing others'' & Agender characters, agender handles, normalization \\

\textsc{Validation} & Agender identities are the default or at least exist & Referents, romance options, sense of belonging, affirmation or ``being seen'' \\
 
\textsc{Irrelevance} & No relation to gender, no gender adaptations & Gender-free interactions, gender-free romance and pregnancy, queer families \\
 
\textsc{Choice\textsuperscript{2}} & N/A & Handle disclosure, option for agender+ \\
\midrule

\multicolumn{3}{l}{\textdi{Agender Play\textsuperscript{3}}} \\ 
\textsc{Projection} & Via text-only, pixels & Via no ascribed gendering or forced disclosures \\

\textsc{Evasion\textsuperscript{4}} & Via text-only, pixels & Via no ascribed gendering or forced disclosures, empty handles, unselected gender options \\
 
\textsc{Exploration\textsuperscript{5}} & Ability to enter different gendered domains & Ability to change settings at will, try identities out with others \\
 
\textsc{Subversion\textsuperscript{6}} & Traversing any gender-marked zone & Cause gender illegibility or confusion \\
 
\textsc{Reflection} & Desires never to leave the world, safe and fear-free magic circle, misogyny but not me & Identity exploration with family members, ambivalence about self-representation \\

\bottomrule

\addlinespace
\multicolumn{3}{@{}p{\dimexpr\linewidth}@{}}{\footnotesize 
\textsuperscript{1} Deduced from ``representation'' via \citet{Whitehouse_2023agender}. 
\textsuperscript{2} Deduced from ``sense of control'' via \citet{ketola2022identity}. 
\textsuperscript{3} Deduced from ``gender-aligning performances'' via \citet{Whitehouse_2023agender}. 
\textsuperscript{4} Deduced from ``gender sterilization'' via \citet{ketola2022identity}.
\textsuperscript{5} Deduced from ``agender performance'' via \citet{ketola2022identity}. 
\textsuperscript{6} Deduced from ``gender neutralization'' via \citet{ketola2022identity}.
}
    
\end{tabular}
\end{table*}

\subsubsection{Theme 1: \emph{\textdi{Agender Baseline}}}

\textsc{Gender-free} experiences were among the most euphoric.
Games with no playable \textdi{character} were ideal. Explains \textbf{P143}: ``There are no characters (...) therefore I do not have to choose a gender. [The puzzle \textdi{mechanic}] doesn't require me to be male or female or even have a gender to be able to use my mind to solve the problem.''
Games with a first-person POV or occlusion of the avatar allowed some to experience agenderedness via immersion~\cite{denisova2015first} and the inability to perceive any gendered aspects of the player \textdi{character}. This effect was magnified if the game \textdi{narrative} avoided binary gendering, which would break the illusion.
Euphoria was also dependent upon the game's overall attitude to gender. 
Players found joy when games did not assert gendered roles or expectations onto the player or other \textdi{characters} (\textdi{narrative}, \textdi{mechanics}). 
In fact, games that featured gender ambiguous aesthetics for all characters (such as non-human characters or characters wearing space suits) allowed players to avoid reading gender into the \textdi{world}. 
This 
echoes the discourse around gender-neutral designs in queer game studies, such as not explicitly defining character sexualities~\cite{shaw2016whereisthequeerness}, allowing non-normative gender presentations, including non-gender but intersecting customizations like fatness~\cite{maletska2024queer}, offering non-human or pseudo-human avatars like the fantasy race Miqo'te in \emph{Final Fantasy XIV Online}~\cite{Smith2025binbarriers}, providing androgynous traits for avatar customization to enable queer gender presentation~\cite{kosciesza2025doing,blanco2023video}. Here, the avoidance of gendered readings specifically contributed to agender euphoric experiences.
This occurred with the ``chibi'' aesthetic experienced by \textbf{P18} in \emph{Fields of Mistria} (\autoref{fig:teaser}), where there were no ``specific body details (...) so I don't have to focus on gendered physical characteristics.'' A lack of gender bigotry within the \textdi{narrative}, ``not even,'' wrote \textbf{P120}, from villains (who could still be evil without gratuitously misgendering people), was a primary aspect enabling joy. Games that avoided \textdi{character} gendering also created a ``safe space'' in the \textdi{social} environment~\cite{purdie2008social,gesler2003healing,tucker2010mental,liang-2025-euphoria}, where the default was to not assume 
gender. Identity affirmation was influenced by societal and communal judgments and available typologies~\cite{tatum2017all,cooley1983human,erikson1968identity}, or ``who the world says I am''~\cite[p.~5]{tatum2017all}. Hence, gender ambiguous game \textdi{worlds} and \textdi{social} contexts enhanced the ability of agender players to self-observe, reflect, and self-affirm, which are primary antecedents of euphoria~\cite{skelton2024itjust}.

\subsubsection{Theme 2: \emph{\textdi{Agender by Design}}}

\textsc{Representation} mattered to participants, who wished for playable agender \textdi{characters}, agender NPCs, and even agender relationships (\textdi{social}): ``The ability to select my gender as non-binary. Having a non-binary companion who you can bond with and relate to'' (\textbf{P22}). This was considered a baseline for inclusion. \textdi{Narrative} arcs that featured agender storylines led \textbf{P51} to ``never stop playing'' \emph{Episodes}.

\textsc{Validation} was multifaceted. An ``agender by default'' or a non-binary \textdi{character} was ``relieving'' (\textbf{P34}), if not euphoric: ``I absolutely love being able to see myself'' (\textbf{P41}). 
The \textdi{mechanics} and \textdi{narrative} had no relation to gender for a truly agender experience. A \textdi{world} without or beyond gender, like that of \emph{The Talos Principle}, was a euphoric validation: as a sentient robot, ``[t]he aspect of what it means to be human without any genders attached to them'' (\textbf{P45}) was special. The \textdi{social} angle was critical, as for \textbf{P50}: ``Gender not being a big deal, especially in romance when it can be very intimate, makes me comfortable and relaxed.
'' When real players were afoot, identity affirmation was key: \textbf{P65} ``was happy to be around people that understood me and allowed me to be gender free.''
Agender identities are often considered deviant and are \textit{othered} by the dominant cis/heteronormative social group~\cite{jensen2011othering}. Hence, \textsc{validation} of \textdi{agender identities by design}, i.e., validation by the creators, who hold the greatest power, supports identity affirmation in a process that Chimakonam calls ``\textit{de-othering}''~\cite{chimakonam2020othering}. Games that integrate agender identities in a way that shapes digital public spaces effectively counter marginalization.

\textsc{Irrelevance} was encapsulated by the feeling of gender being ``no big deal.'' \textbf{P8} found joy in \emph{Elden Ring} not having gender-locked equipment and clothing (\textdi{character}). \textbf{P35} described the base \textdi{character} models in \emph{Splatoon 3} as ``negligible and their unique style made me feel extremely happy.''
Agender euphoria was reliant on gender choice not being made a major plot point (\textdi{narrative}).
\textsc{Irrelevance} was also tied to the \textsc{social} context, whether NPCs or players. As \textbf{P84} described, the NPCs ``treated my character neutrally, focusing on my actions and personality instead of my appearance.'' Players also took pleasure in queer reproduction, like in \emph{The Sims}: ``asexual reproduction is also available in the game via a `science baby'\thinspace'' (\textbf{P73}).
Such \textsc{irrelevance} echoes \citet{skelton2024itjust}: euphoria emerges in the presence of decentring or the absence of resistance to a given identity by the social group, who ``think nothing of it'' (p.~8).

\textsc{Choice} lays bare the power involved in the ``player's degree of freedom''~\cite[Location No. 2196]{ruberg2025queer}, surfacing the spatial and political navigability~\cite{wolf2011theorizing,ruberg2025queer} and manoeuvring \cite{murray2021video,ruberg2025queer} set by the creators of the game. \textsc{Choice} in games like \emph{Baldur's Gate 3}, \emph{CyberPunk 2077}, and \emph{Animal Crossing: New Horizons} was specifically noted for the flexibility and fluidity in \textdi{character} creation. Games that did not begin with an immediate gender binary logic gate and did not restrict appearance, \textdi{narrative} arcs, abilities (\textdi{mechanics}), or handles (\textdi{social}) were euphoric.

\subsubsection{Theme 3: \emph{\textdi{Agender Play}}}

Players recounted the many creative and crafty ways of making do within the confines of the game. 
Players are, after all, \emph{interactors,} or active participants in construction player experiences~\cite{fernandez2009play}. 
This constructive process acknowledges but also tempers the magic circle~\cite{huizinga_homo_2000,tekinbas2003rules}, betraying the flux and disassembly of the game as a boundary~\cite{ruberg2025queer,munoz2019cruising}. Such retaking of power provides an agender form of ``queer comfort''~\cite{ahmed2013cultural} and consequently produces agender euphoric experiences by challenging gender normative institutions~\cite{beischel2022gender}.

\textsc{Projection} and \textsc{evasion} emerged as entwined strategies. Some players relied on cloaks and masks to 
satisfy their desire for an indeterminable gendered experience (\textdi{character}).
Minimalist game \textsc{aesthetics} (\textdi{world}), like ``old-school pixel / low-poly art style[s]'' (\textbf{P5})
, allowed players to \textsc{project} or \textsc{evade} identity, a strategy of being illegible~\cite{pozo2018queer}.
The 3D pixel \textdi{world} in \emph{Minecraft} allowed \textbf{P13} to feel ``that the characters were androgynous, it made me feel happy (...) it represented me.'' Similarly, for \textbf{P30}, the game was ``not detailed enough for specifics of gender to matter (...) playing without those things being rubbed in your face.'' Some, like \textbf{P31}, expressed para\textdi{social} engagement with \textdi{characters} and \textdi{narrative}, explaining ``I make them as one would create characters in a story.''
This promotes self-review~\cite{horton1956mass,derrick2008parasocial}, manifesting in favourable, positive self-assessments and identity-based enjoyment and satisfaction~\cite{nakashima2012group,smith2016understanding,weitekamp2013more,seaborn2023link}.
When there were few or no customization options, a sense of gender play and euphoria could be achieved by choosing a gender ``opposite'' to the one assigned at birth and performing a sort of mental interpolation and \textsc{projection} for the androgeneity they craved~\cite{ketola2022identity}. Otherwise, players \textsc{evaded} gender through \textdi{character} garb and appearance, as with \textbf{P121}: ``I play as a thief, they're covered up in a way that isn't overly masculine or feminine.'' \textdi{Worlds} and \textdi{mechanics} that were completely free from gendering allowed \textsc{evasion}, too: ``I felt liberated by the lack of any imposed gender role or expectations, which allowed me to focus entirely on strategy, decision-making, and expressing my creativity in the game'' (\textbf{P43}). \textdi{Social} context was important for testing out ideas or performing the self in a ``networked'' fashion~\cite{papacharissi2012without}. \textbf{P147} highlights how ``an agender profile customization in-game (...) made me feel happiness to show people I’m agender, including to my sister and friend's, I felt pride in my identity.''

Participants embraced \textsc{exploration},
describing workarounds for games with limited \textdi{character} customization options. ``Mixing'' (\textbf{P9, P20, P26, P27, P32, P33, P60, P61, P65, P84, P107, P145}) gendered body aspects with ``opposite'' gender attributes or clothing was important for making binary bodies seem more agender or genderqueer~\cite{krobova2015dressing,maletska2024queer,sunden2012queer}. 
Games that randomly assigned player \textdi{character} bodies made reading gender unreliable and presented an opportunity for gender exploration. \textbf{P72} left it up to the game, while \textbf{P60} ``would select gender randomly because there was just man and woman, and then mix styles and stereotypes.'' Time was also a factor, with many finding joy in games that allowed ``on-demand'' customization (\textdi{mechanics}): players could represent their gender as fluid with their moods and tastes, a capability deemed a euphoric source for agender individuals~\cite{beischel2022little}. 
\textbf{P48} and \textbf{P15} sum it up: ``Transitioning on demand'' (\textdi{mechanics}): ``The ability to drastically change my appearance at any point in game.'' This suggests a desire to queer time itself in pursuit of identity, where \textdi{character} creation does not necessary happen first~\cite{ruberg2025queer,Hantsbarger_2022}. For \textbf{PP37}, euphoria was specifically linked to avoiding dysphoria, as well as maintaining freedom of expression: ``being able to change body when I felt dysmorphic, or maybe wear particular outfits.''

Players devised ways to \textsc{subvert} or undermine the authority of the game (\textdi{mechanics, world}) or \textdi{social} environment.  
Echoing the work of \citet{beischel2022little} on non-binary folk
, many participants described a particular kind of joy from being able to create \textdi{characters} that generated a kind of illegibility or confusion for other players attempting to gender them (\textdi{social}). In \textbf{P102}'s experience, ``an NPC addresses the player character, clearly unable to determine if the player character is a boy or a girl just by looking. The idea of someone being confused by my gender presentation made me feel euphoric, although I hadn't at the time realised that I was agender yet.'' As a child, \textbf{P38} deliberately sought euphoric moments in fooling others with plurality, as well as themselves: ``It made me incredibly happy to play a mind game with other players in this way, and allowing me to switch between the multiple personalities in my mind, mentally euphoric.'' The cross-dressing and evasion of sexism in the \textdi{world} was euphoric for \textbf{P58}: ``I am someone that is physically strong and not seen as an easy target.'' Players were ``queergaming''~\cite{Chang_2015queergaming} wizards in service of \emph{disidentification}~\cite{munoz2013disidentifications}, transgressing and reimagining game and social logics to attain agender euphoric states.

\begin{figure*}[t]
  \centering
  \includegraphics[width=1\linewidth]{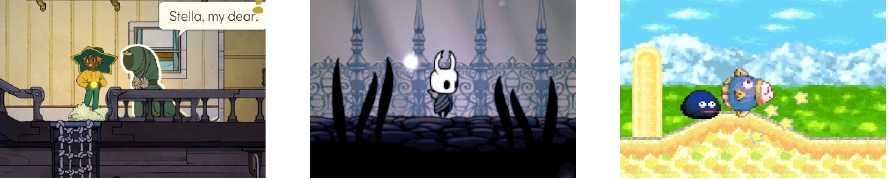}
  \caption{Three screenshots showing player representations and protagonists that participants identified as agender.
  Left, \textit{Spiritfarer}, \textbf{P144}: Stella---the smiling, human character wearing a green hat, yellow blouse, and green slacks---has no clear gender. 
  While many of the NPC spirits---e.g., the cloaked figure with a snake head 
  to the right of Stella---have a gender in life, their spirit forms are anthropomorphized animals. 
  Middle, \textit{Hollow Knight}, \textbf{P126}: The Knight---the player representation in \textit{Hollow Knight} who has an exoskeleton or armour that shows only two large, black eyes and horns---is canonically genderless. 
  Right, the Kirby series, \textbf{P97}: Kirby---a pink, amorphous blob 
  who can shapeshift by eating enemies---has a canonically unknown gender in the original Japanese games~\cite{kirby-flipbook}, despite often localized with ``he/him'' pronouns in English.
  Screenshots taken by the authors.}
  \Description{Three game screenshots. Left: A smiling, dark-skinned person wearing a large hat, blouse, and slacks standing on a boat next to a snake in a cloak. A white cat lounges on the deck. The snake-person has a speech bubble that reads ``Stella, my dear.''
  Middle: A small figure with a helmet-like head that has two large, black eyes and horns. Behind is a filigreed fence on bumpy ground. Right: A dark-coloured blob enemy next to a pink, smiling blob 
  in a fish suit.}
  \label{fig:agender-protag}
\end{figure*}

Agender euphoric accounts were also linked to \textsc{reflective} states. 
Experiencing \textdi{characters} led to nuanced reflections on desire: ``
Seeing them on screen gives me gender envy and euphoria'' (\textbf{P2}). 
Similarly, \textbf{P126} expressed a desire for genderless \textdi{narratives} and \textdi{worlds}: ``the knight and the hollow knight are both canonically genderless children of the pale king, the game treats this as something that is neither abnormal or strange''---what \citet{liang-2025-euphoria} note as destigmatization---``the citizens respect both characters for who they are as people and their lack of gender is never a subject to be dissected or treated as `progressive'. they [sic] just are, they simply exist. and I liked that, I wanted that for myself too.'' 
\autoref{fig:agender-protag}, middle, shows the Knight from \textit{Hollow Knight}. 
\textbf{P58} notes: ``misogyny exists in the world but I am not a victim of it.''
\textdi{Social} play was an important mode of gender exploration for those unable to come out in real life or just coming into a sense of an agender self. 
As in \citet{skelton2024itjust}, euphoria was a product of close 
relationships, like friends and family; affirmation of agender identities when playing games with each other was a source of euphoria. 
\textbf{P144}, as an agender parent with an agender child, shared this experience: ``my kid recently came out (...) [I] introduced them to Spiritfarer (...) Upon seeing the main character Stella my kid mentioned how gender neutral they appear and that they're not used to seeing representation like that in games. (...) My kid felt seen and I was thrilled to share this with them.
'' 
\autoref{fig:agender-protag}, left, shows the characters of \textit{Spiritfarer}. 
Having community support, a source of euphoria~\cite{skelton2024itjust}, was critical for many. \textbf{P148} explained: ``The socialization with the community, was a definite plus. Once my `online friends' realized that I wasn't going to tell them my birth gender, they just treated me like everyone else. Not male, not female, just me.'' Such game \textdi{worlds} and \textdi{social} contexts left many wishing to play forever.

\subsubsection{The Avatar Dimension}
We created a list of bare minimum customizations for avatars in \autoref{tab:avatar}. While we focused on the avatar prompt, we also drew from the main accounts because, despite instructions, participants could not help but refer to if not discuss matters of representation and identity---their own and others. We stress that people need and expect freedom across the range of possible customizations. \textbf{P61}, for instance, expressed feeling ``so grateful to the devs'' of \emph{Baldur's Gate 3} for providing a sheer diversity of options, underscoring the importance of player choice.

\begin{table}[b]
\caption{Bare minimum customizations for autonomy in gender euphoric game experiences.}
\label{tab:avatar}
\begin{tabular}{lll}
\toprule
\textbf{Account} & \textbf{Avatar Appearance} & \textbf{Avatar Referents} \\
\midrule
Handle & Clothing & No referents \\
Flags & Hair & Pronouns \\
\multirow{3}{*}{Profile image} & Facial features & Names \\
 & Body, inc. genitals & Titles \\
 & Non-human & Roles \\
 & Invisible/none & \\
\bottomrule
\end{tabular}
\end{table}

\textdi{Account} customizations included \textsc{Handle} or nick/name, including not being forced to use one, \textsc{flags} to indicate identity (or not), and \textsc{profile images} that could be set or remain absent. \textbf{P76}, for instance, stressed the need to display a pride flag at all times. \textbf{P147} felt immense joy at being able to express their agender identity in their \emph{Overwatch 2} profile.

\textdi{Avatar Appearance} and level of customizability emerged as a feature across accounts. Players broadly pointed to \textsc{clothing}, \textsc{hair}, \textsc{facial features}, and the \textsc{body}. On the last, \textsc{genital} customization, or being able to ``change my genitals at will'' (\textbf{P27}) in games like \emph{Cyberpunk 2077} was ``a joyous moment and very freeing from a weight I didn't know existed before that'' (\textbf{P33}). \textbf{P10} noted that was ``not meant to be sexual in nature [but] purely to allow for representation and choice in the game.'' Sliders, even binary ones, allowed for agender customizations: ``I chose a jawline that was soft yet defined, eyes that carried an air of mystery, and brows that struck a balance between bold and gentle'' (\textbf{P32}). Clothes and accessories allowed for gender evasion, as articulated by \textbf{P95}: ``my avatar has their face covered to symbolise my gender fluidness, I don't have to explain to anyone or try to be someone I'm not.'' For \textbf{P141}, this was truly gender-evasive: ``The androgynous of the armor and facemask hiding her female characteristics.''

Player representation went beyond human embodiments. As in \citet{Whitehouse_2023agender}, \textsc{non-human} avatars, from animals to aliens to amorphous beings, were perceived as agender. \textbf{P36} explained that the player ``can acquire animal or inanimate body parts, progressively transforming into a wolf, bird, lightning monster, etc. (...) I felt very reflected (...) in this form, and really enjoyed looking at it.'' For \textbf{P136,} the non-binary character in \emph{Warframe} was ``so far off from anything human. I decided right there that was the one I would main, and 2.5 years later, they're still my most used character.'' 
\textbf{P97} pointed out that even games without customizable or selectable characters elicited agender euphoria: ``I appreciate adventure games that have a main character that can't be gendered (...) Kirby's formlessness as opposed to other games that have characters with distinct gender characteristics and plots that revolve around their gender identity.'' 
Kirby is shown in \autoref{fig:agender-protag}, right. 
Sheer customizability also gave rise to unexpected but pleasurable embodiments: ``messing around in the character generator to make some horrible genderless homumculus creature, that I ended up becoming really attached to, like I too would have liked to be some near featureless unidentifiable gunk'' (\textbf{P86}).

\begin{figure*}[t]
  \centering
  \includegraphics[width=1\linewidth]{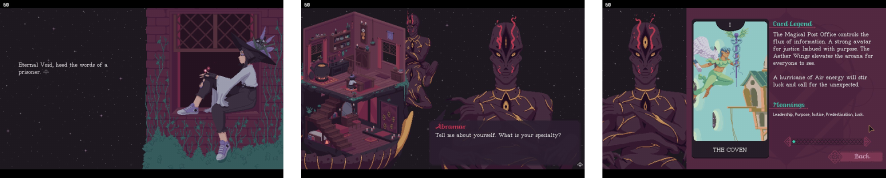}
  \caption{Three screenshots showing the introduction to \textit{Cosmic Wheel Sisterhood}, which \textbf{P131} appreciated for a \textit{lack of} character creation options.
  The player representation, the witch Fortuna (left, seated and hunched over a desk) canonically uses she/her pronouns. 
  Characterization happens through answers to questions by an otherworldly being (middle) and constructing cards (right), without addressing gender or gender expression.
  Screenshots taken by the authors.}
  \Description{Three screenshots from a game. Left: A person with a large witch hat, white blouse, black jeans, and purple sneakers perched in the window of a house floating in space. The text overlay reads ``Eternal Void, heed the words of a prisoner.''
  Middle: The interior of a house, from an isometric view, that is filled with witch paraphernalia (e.g., a cauldron, potions). The same character is seated at a desk. Outside, a giant being that is a cross between a human and a snake with eyes all over the body is wrapped around the floating house. An inset to the right shows the being and a text box reading ``Abramar: Tell me about yourself. What is your specialty?''
  Right: Abramar is on the left and to the right is a panel displaying a card, called The Coven, created by the player. The card features a fairy carrying a caduceus and provides the description of: ``Card Legend: The Magical Post Office controls the flux of information. A strong avatar for justice. Imbued with purpose. The Aether Wings elevates the arcana for everyone to see. A hurricane of Air energy will stir luck and call for the unexpected. Meanings: Leadership, Purpose, Justice, Predestination, Luck.''
  }
  \label{fig:cosmic-intro}
\end{figure*}

An agender-specific pattern was found in \textsc{invisible} embodiments. For instance, \textbf{P131} revelled in the lack of character customization procedures in \emph{Cosmic Wheel Sisterhood} (\autoref{fig:cosmic-intro}). 
Others were relieved to be free of the necessity of avatars, such as \textbf{P57}, who described the game as ``more anti experience,'' previously having difficulty relating to characters and avatar options across most games, and even choosing options different from themselves to maintain psychological distance. \textbf{P94} enjoyed that ``You could only pick the name [and] it didn't really change anything in-game.'' 

Key to the agender euphoric experience was, unlike other gender identities~\cite{Whitehouse_2023agender,austin2022gender,galupo2017like}, the absence of \textdi{avatar referents}. As \textbf{P16} explains: ``No one addresses you directly or refers to you to other characters. There is no use of pronouns or addressing you directly or character customisation (so far).'' \textbf{P136} stresses the point: ``NPCs also don't refer to you by any gendered pronouns.'' Text-based game aesthetics were particularly freeing in this regard: ``the game had no avatar, it was strictly text-based, so creation was choosing descriptions and setting pronouns/terms that would be used'' (\textbf{P101}). At the same time, being able to set pronouns or choose ``they/them'' was also a pattern. Notably, the social aspect of pronoun use was important: ``you will be called `they/them' for the entire game'' (\textbf{P3}). Also clear was the importance of being able to set names. \textbf{P67} asserted their identity by recrafting their gendered name as the name of the game character: ``Evgenius.'' Names need not be formal names; for \textbf{P128}, ``The Alchemist'' felt ``just felt so right,'' while \textbf{102} liked ``the idea of my gender just being `Pokémon Trainer.''  The use of titles in place of all else was important: ``Going solely by titles and not pronouns/names'' (\textbf{P109}).
The \textsc{role} assigned to the character also helped with evading gender, as in \emph{Satisfactory}: ``Your avatar is called `the pioneer' and is never gendered'' (\textbf{P130}).


\subsection{Non-Euphoric Experiences (RQ2)}
\label{sec:noneuph}

Non-euphoria was due to: (i) forced gendered actions, e.g., gendered dress-up in \emph{Pok\'{e}mon} (\textbf{P123}), gender binary saunas in \emph{Stardew Valley} (\textbf{P80}); and (ii) assumptions and misgendering, e.g., binary gender options and being referred to as ``boy/girl'' in \emph{Pok\'{e}mon (\textbf{P99})}. 
Specific elements included quests (\textdi{mechanics, narrative}), e.g., required gendered dress-up; space (\textdi{mechanics, world}), e.g., gendered saunas; dialogue (\textdi{social, narrative}), e.g., referent use by NPCs; character creation restrictions (\textdi{character}); and translation issues (\textdi{narrative, social}), e.g., stereotyped portrayals of ``xianxia/wuxia tropes and normative gender prescriptions for the imagined jianghu'' in \emph{Immortal Taoist} (\textbf{P123}).


\subsection{Barriers to Experiencing Agender Euphoria (RQ2)}
\label{sec:barriers}

Those who could not provide an account of an agender euphoric experience provided three key reasons why not. Two reasons speak to the \textdi{Agender Baseline} theme. Some had euphoric experiences that were unrelated to gender. \textbf{P1} explained that ``the euphoria I get from games is very uncommonly gender-related,'' perhaps representing a gender-apathetic perspective. Others never thought about gender as a function of their agender orientation. As \textbf{P140} explained: ``I have never experienced agender euphoria because my feeling about gender is an absence.'' Reflecting the \textdi{Agender Play:} \textsc{Reflection} theme, some were
questioning or struggling with their identity. \textbf{P127} shared: ``I am really struggling to answer (...) because I've just repressed my feelings towards what I truly feel about my gender expression.'' Finally, as in the \textdi{Agender Play: }\textsc{Projection} theme, respondents instead experienced identity transference or parasocial engagement. \textbf{P150} described a typical experience: ``When I'm playing RPGs, for example, I tend to create characters \emph{unlike} myself, rather than characters I feel represent me'' (our emphasis).

\subsection{Ideas to Enable Agender Euphoria (RQ3)}
\label{sec:ideas}

The ideas suggested by participants who could not provide an account of an agender euphoric experience mapped onto many of the experiences reported by those who could. We summarize the ideas as follows:

\begin{itemize}
    \item Complete avoidance of gender in general and specifically gender-reductive representations, e.g., ``I didn't want to deal with trans stuff like dysphoria and transphobia in my safe fun fantasy space, nor did I want to create an unrealistic hugboxing wish fulfillment world that would only make me feel worse about real life'' (\textbf{P90}); 
    
    \item Nonprescriptive agender avatar options, e.g., androgynous (\textbf{P6, P21, P71, P71, P90}), non-binary (\textbf{P78, P83, P103, P150}), genderless sex but gendered appearance (\textbf{P6}), non-reductive avatars (as \textbf{P90} explained: ``they used the male avatar as the `base' for the new avatar system,'' which was dysphoric); 
    
    \item Gender fluidity during play (\textbf{P28, P54, P90, P119, P138}), e.g.., ``My character flows between aesthetics- masculine, feminine, and everything in between- without limits (...) how the world embraces my fluidity'' (\textbf{P54}); 
    
    \item Non-human options, e.g., blob, creature (\textbf{P64}), robot (\textbf{P83}); 
    
    \item Nonprescriptive referents, e.g., pronoun or referent choice (\textbf{P6, P69, P90, P103, P114, P137, P138, P150}), choice of ``they/them'' (\textbf{P14, P54, P64, P90}), no pronouns (\textbf{P69, P90, P137}, \textbf{137}), use of roles or titles like ``friend'' (\textbf{P98}), no misgendering (\textbf{P133}); 
    
    \item No gendered play or mechanics, e.g., unrelated to the gendered self (\textbf{P11, P85, P90, P119, P133}), absent or ``functionally genderless'' (\textbf{P100, P114, P140, P150}); 
    
    \item Ungendered relations, e.g., treated by NPCs in a ``gender-neutral way'' (\textbf{P21}), ``a world unrestricted with specific gender roles and expectations'' (\textbf{P119}), non-cisgendered relationships (\textbf{P149}); 
    
    \item Agender representation, e.g., an agender NPC (\textbf{P52}) or playable character (\textbf{P96}, \textbf{150}); and
    
    \item Identity celebration, e.g., identity labels (\textbf{P77}), an agender background (\textbf{P103}).
\end{itemize}

\subsection{Quantitative Results (RQ1)}
\label{sec:results}

The following results represent broad patterns of emotion and meaning in the short and long terms related to the agender euphoric experience reported, helping to characterize ``euphoria'' in a consistent way across players.

\subsubsection{Affect (I-PANAS-SF and Quantified Feelings)}

\begin{figure*}[!t]
  \centering
  \includegraphics[width=.7\linewidth]{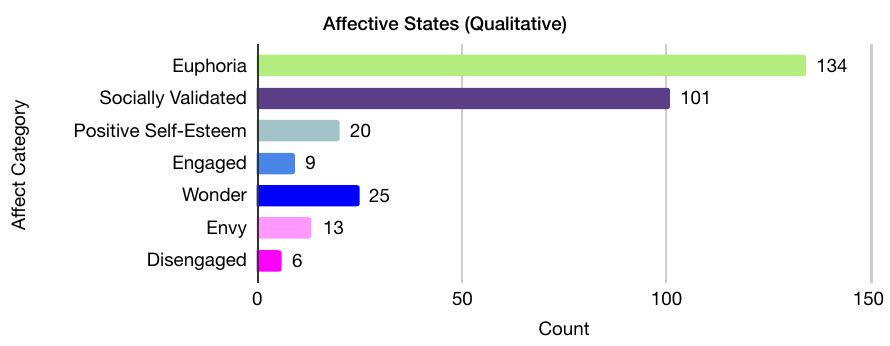}
  \caption{Affective states from the qualitative prompts.}
  \Description{A graph indicating a range feelings, emotions, moods, and affective states representing agender euphoric experiences: euphoria, socially validated, positive self-esteem, engaged, wonder, envy, and disengaged. The vast majority counted as euphoria and socially validated.}
  \label{fig:affect}
\end{figure*}

Participants reported an array of feelings connected to their agender euphoric experience (\autoref{fig:affect}). We broadly captured these descriptors as follows: euphoria (134, 44\%), including feeling emotional, excited, happy, and content; socially validated (101, 33\%), like feeling accepted, safe, comfortable, free, ``right,'' and relaxed; having positive self esteem (20, 6\%), like feeling confident, cool, empowered, proud, accomplished, and motivated; feeling engaged (9, 3\%), including focused, immersed, entertained, and amused; wonder (25, 8\%), including awe, surprise, creativity, curiosity, enlightenment, and appreciation; envy (13, 4\%), with feelings of longing, hope, and jealousy; and feeling disengaged (6, 2\%), including being confused, embarrassed, detached, and disconnected. These affective states map onto the existing euphoria literature beyond agender identities, notably feelings of comfort, confidence, and joy~\cite{austin2022gender}, feelings of ``distinct enjoyment or satisfaction'' \cite[p.~6]{ashley2018favor}, feelings of relief or freedom~\cite{beischel2022little}, feelings of being socially recognized and accepted~\cite{skelton2024itjust}, and achievement, alignment, authenticity, congruence, and validation, etc.~\cite{alutalica2021transgender,kai2022euphor}.
This suggests that agender euphoria is very similar to other experiences of euphoria.

\begin{figure*}[t]
  \centering
  \includegraphics[width=.7\linewidth]{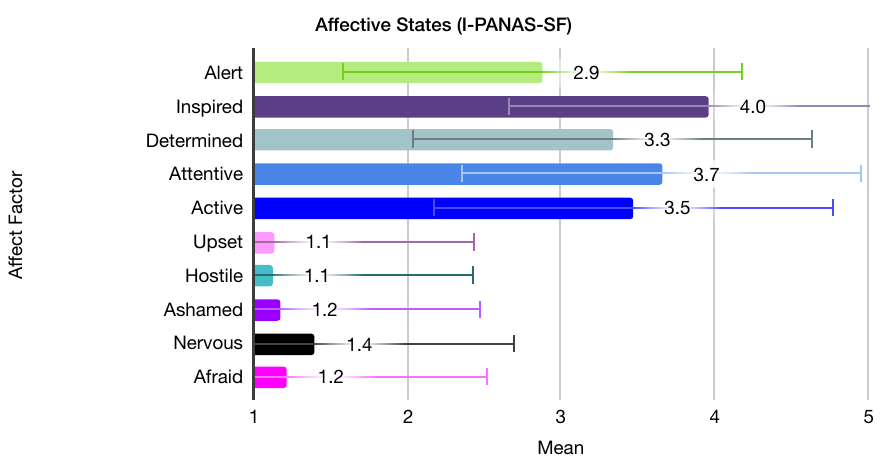}
  \caption{Affective states via the quantitative measure (I-PANAS-SF). Details are in \autoref{tab:affect}.}
  \Description{A graph indicating high positive affect, especially inspiration, and low negative affect.}
  \label{fig:panas}
\end{figure*}

The quantitative measures (\autoref{fig:panas}, \autoref{tab:affect}) showed higher positive affect ($M = 3.5, SD = 0.8$) than negative affect ($M = 1.2 ,SD = 0.3$), according to a paired t-test $t(108) = 30.9, p < .001$, 95\% CI [-1.98, 1.98]. Still, positive affect was not very high. This could be explained by the design of the instrument, which did not include most of the keywords and sentiments expressed in the qualitative descriptions of agender euphoric feelings. The exception was ``inspired'' ($M=4.0, SD=1.0$), which directly links to the reported feelings of ``creativity'' above and expressiveness to euphoria in general~\cite{skelton2024itjust,Whitehouse_2023agender,reisner2023exploring}.

\begin{table}[b]
\caption{Descriptive statistics for the affect measure. $\alpha$: Cronbach's alpha.}
\label{tab:affect}
\begin{tabular}{lrrrrrrr}
\toprule
\textbf{Measure}    & \multicolumn{1}{l}{\textbf{Statistic}} & \multicolumn{1}{l}{}   & \multicolumn{1}{l}{}   & \multicolumn{1}{l}{}    & \multicolumn{1}{l}{}    & \multicolumn{1}{l}{}  &  \\
I-PANAS-SF & \multicolumn{1}{r}{M}    & \multicolumn{1}{l}{SD} & \multicolumn{1}{l}{MD} & \multicolumn{1}{l}{IQR} & \multicolumn{1}{l}{Min} & \multicolumn{1}{l}{Max} & $\alpha$\\
\midrule
Alert & 2.9 & 1.3 & 3.0 & 2.0  & 1 & 5 & \\
Inspired & 4.0 & 1.0 & 4.0 & 2.0  & 1 & 5&  \\
Determined & 3.3 & 1.3 & 4.0 & 2.0  & 1 & 5 & \\
Attentive  & 3.7 & 1.1 & 4.0 & 1.0  & 1 & 5 & \\
Active   & 3.5 & 1.2 & 4.0 & 1.0  & 1 & 5 & \\
\textbf{Positive}   & \textbf{3.5}     & \textbf{0.8}  & \textbf{3.4}  & \textbf{1.0} & \textbf{1.2} & \textbf{5.0} & \textbf{0.71}\\
Upset & 1.1 & 0.4 & 1.0 & 0.0  & 1 & 3 &\\
Hostile  & 1.1 & 0.6 & 1.0 & 0.0  & 1 & 5 & \\
Ashamed  & 1.2 & 0.4 & 1.0 & 0.0  & 1 & 3 & \\
Nervous  & 1.4 & 0.6 & 1.0 & 1.0  & 1 & 3 & \\
Afraid   & 1.2 & 0.6 & 1.0 & 0.0  & 1 & 5 & \\
\textbf{Negative}   & \textbf{1.2}     & \textbf{0.3}  & \textbf{1.0}  & \textbf{0.4} & \textbf{1.0} & \textbf{2.4} & \textbf{0.61}  \\
\bottomrule
\end{tabular}
\end{table}

\subsubsection{Hedonia and Eudaimonia (HEMA-R)}

\begin{figure*}[t]
  \centering
  \includegraphics[width=.7\linewidth]{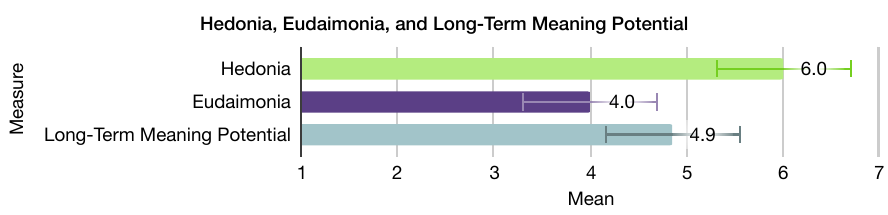}
  \caption{Well-being orientations via hedonia and eudaimonia, as well as long-term meaning potential.}
  \Description{A graph indicating high levels of hedonia, medium eudaimonia, and somewhat high levels of long-term meaning potential.}
  \label{fig:wellbeing}
\end{figure*}

Hedonia ($M=6.0, SD=0.7, MD=6.2, IQR=0.8, Min=3.0, Max=7.0, \alpha=0.676$) and eudaimonia ($M=4.0, SD=1.5, MD=4.0, IQR=2.2, Min=1.0, Max=7.0, \alpha=0.903$) ratings are visualized in \autoref{fig:wellbeing}. Both were strongly, positively correlated, $r(107) = .245, p = .010$, 95\% CI [0.06, 0.41], as expected~\cite{huta_pursuing_2010, Huta_2013,Huta_2016}. Statistically significant relationships were not found for hedonia and positive affect, $p=.995$, and negative affect, $p=992$. This again suggests that the positive items do not reflect euphoria. Statistically significant correlations were found between eudaimonia and positive affect, $r(107) = .316, p < .001$, 95\% CI [-0.19, 0.19], and negative affect, $r(107) = .253, p = .008$, 95\% CI [0.07, 0.42]. Both were positive, suggesting a nuanced and complex relationship between meaning as uplifting \emph{and} difficult.

\subsubsection{Player Experience (PXI)}

\begin{figure*}[t]
  \centering
  \includegraphics[width=.7\linewidth]{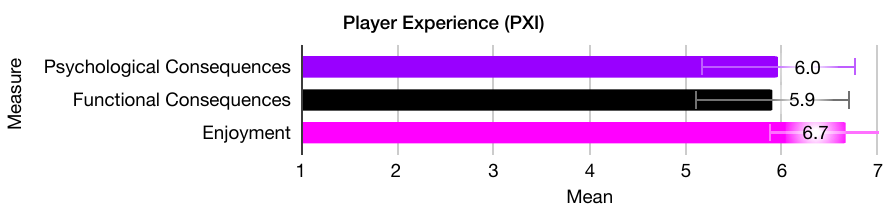}
  \caption{Player experience via the PXI. Details are provided in \autoref{tab:pxi}.} 
  \Description{A graph indicating high levels of psychosocial consequences, functional consequences, and enjoyment.}
  \label{fig:pxi}
\end{figure*}

PXI scores are visualized in \autoref{fig:pxi}, with descriptive statistics presented in \autoref{tab:pxi}. All scores were very high. Several statistically significant correlations were found between the subscales and well-being measures. Psychosocial consequences was correlated with hedonia, $r(107) = .431, p < .001$, 95\% CI [0.26, 0.57], and eudaimonia, $r(107) = .367, p < .001$, 95\% CI [0.19, 0.52]. Functional consequences was correlated with hedonia, $r(107) = .304, p = .001$, 95\% CI [0.12, 0.47], but not eudaimonia, $p=.099$. Enjoyment was correlated with hedonia, $(r(107) = .39, p < .001$, 95\% CI [0.22, 0.54], and eudaimonia, $r(107) = .199, p = .038$, 95\% CI [0.01, 0.37]. These relationships support the pleasurable and meaningful recollection of agender euphoric accounts. The lack of a statistically significant relationship between eudaimonia and functional consequences may be explained by the focus of that subscale on usability matters (Progress Feedback, Ease of Control, Goals and Rules) and media-related (rather than purely experiential) pleasure, i.e., Audiovisual Appeal.

\begin{table*}[ht]
\caption{Descriptive statistics for the player experience (PXI) measure. $\alpha$: Cronbach's alpha.}
\label{tab:pxi}
\begin{tabular}{lrrrrrrr}
\toprule
\textbf{Measure} & \multicolumn{1}{l}{\textbf{Statistic}} & \multicolumn{1}{l}{}   & \multicolumn{1}{l}{} & \multicolumn{1}{l}{} & \multicolumn{1}{l}{} & \multicolumn{1}{l}{} &  \\
PXI & \multicolumn{1}{r}{M}  & \multicolumn{1}{l}{SD} & \multicolumn{1}{l}{MD} & \multicolumn{1}{l}{IQR} & \multicolumn{1}{l}{Min} & \multicolumn{1}{l}{Max} & $\alpha$ \\
\midrule
Psychosocial Consequences & 6.0 & 0.8 & 6.1 & 0.8  & 2.4  & 7.0 & 0.902 \\
Functional Consequences   & 5.9 & 0.7 & 6.0 & 0.9  & 2.1  & 7.0 & 0.866 \\
Enjoyment & 6.7 & 0.5 & 7.0 & 0.3  & 4.0  & 7.0 & 0.837 \\
\bottomrule
\end{tabular}
\end{table*}

\subsubsection{Long-Term Meaning Potential (LTMP)}

LTMP (\autoref{fig:wellbeing}) was high ($M=4.9, SD=1.5, MD=5.0, IQR=2.0, Min=1, Max=7$), lending validity to the CIT and importance of the incidents~\cite{mekler_momentary_2016,Seaborn_2024maldai}.
While no statistically significant relationship was found between LTMP and hedonia, $p = .103$, a positive relationship was found between LTMP and eudaimonia, $r(107) = .416, p < .001$,
95\% CI [0.25, 0.56], indicating long-lasting meaning in the experiences.


\section{Discussion}
\label{sec:discussion}


Games \textbf{do} enable agender euphoric experiences (\textbf{overarching RQ}). Their nature varies widely, intersecting with each player's specific agender identity, preferences for expression, and the game medium in a ``prism'' (\textbf{RQ1}). 
Still, the accounts clearly reflect elements of gender euphoric experiences~\cite{kai2022euphor,Grant_2024euph,liang-2025-euphoria, reisner2023exploring,tacit2020joyful}, distinguished by a need for various forms of removing or avoiding rather than embracing gender. 
Our findings show coverage of known agender expressions: genderless, not defined by the binary, and gender-neutral~\cite{ketola2022identity}. 
Experiences were connected to great pleasure (hedonia) and meaningful self-expression (eudaimonia) with long-term meaning potential. 
They were emotionally complex but largely positive, meeting most players' psychosocial, functional, and enjoyment needs. 
Still, participants made us aware of barriers to agender euphoric play experiences, including barriers to imagining such experiences (\textbf{RQ2}). 
By critically engaging with the diversity of accounts on game and avatar experiences, as well as eliciting ideas from those who wished to experience agender euphoria, we were able to surface bare minimum requirements, guidelines, and opportunities for game designers and developers (\textbf{RQ3}). 
We highlight findings of agender euphoria as its own concept next, then provide design guidelines in digest form (\autoref{sec:guidelines}).

\subsection{Agender Euphoria as Independent from Gender Euphoria}

While the greater research community---within and beyond games and play---has begun to identify how gender euphoria affects positive experiences and wellbeing---and has all indications of being an established concept~\cite{Grant_2024euph,Blacklock-gender-euphoria-scale-2025}---\textbf{\underline{a}gender euphoria} was not yet studied. 
Our findings demarcate agender euphoria as its own concept, yet situated within the euphoria spectrum; markedly different from experiences of gender euphoria, yet occurring along similar lines of personal and social experiences. 
Designing for agender and gender euphoria may take vastly different forms. 

Agender identities and experiences are characterized by forms of gender \textit{absence}, unique from other gendered modes of being. 
Constructs that aim at understanding gender euphoria may apply, but the phrasing---which expressly calls on participants to consider their gender identity and that of those around them---may miss agender euphoria entirely. The GES, for instance, excites us greatly, but we must also note that it does not list ``agender'' in any of its example identities~\cite{Blacklock-gender-euphoria-scale-2025,blacklock2024gender}.
While our findings are limited to game experiences, they suggest that further research into the concept of agender euphoria is warranted, if not necessary for full inclusion.

\subsection{Agender Euphoria as a Design  Goal in Games and Interactive Media}

Games are powerful experiential engines that can enable complex engagements beyond the binary of winning or losing~\cite{liang-2025-euphoria,banks2015object,Alharthi_Alsaedi_Toups,collecting-toups-dugas,Livingston2014how}. Queer perspectives on play (e.g., \citet{ruberg2015nofun,ruberg2025queer,haimson2025trans}) and joy in digital spaces (e.g., \citet{Steeds2025joy,ketola2022identity}) have highlighted the potential of technology to transcend conventional notions of ``fun'' by fostering experiences of ambivalence, pain, confusion, sublimity, elation, and so on. The agender euphoric experiences we surfaced validate and underscore this potential for games.
The importance of experiential complexity found for literature and films~\cite{bartsch2012emotional,bartsch2014moved} is echoed in the concept of agender euphoria, expanding the horizons of game design and player experience.
This also reflects calls in queer game studies~\cite{ruberg2015nofun,ruberg2017queergamestudies} 
to go beyond critical examinations of identity representation in digital realms. Instead, we should creatively engage with what kinds of knowledge can be generated 
to design complex experiences and tap into the unique strengths 
that marginalized groups bring to design~\cite{to2023flourishing}. For instance, no participatory design~\cite{bodker2022participatory} work has involved agender+ gamers throughout the design cycle, even though opinion-gathering has been done on specific elements, like avatar design~\cite{Morgan2020avatar}. Comparatively, \citet{haimson2025trans} found that 19\% of trans technologies involved trans people throughout the design process, with 39\% involving trans people in a more limited capacity. The willingness of our agender+ cohort to engage in design ideation suggests this trajectory as a strong candidate for future work in research and game design practice.

Our reification of agender euphoria as an experiential element that crosses multiple agender and adjacent identities contributes to not only game design knowledge but also general design knowledge in HCI. The euphoric experience itself is a valuable and crucial design goal~\cite{liang-2025-euphoria}, grounded in a positive psychology framing~\cite{rao2015expanding} that elevates language and experiences connected to minority identities. 
While it might seem a bit counterintuitive, game designers could draw inspiration from work on queer-centred offerings~\cite{Zabala2024} to create games that centre and directly reference agender identities and agender euphoria, potentially uplifting agender+ players through radical saliency of agenderness.
Game designers could also leverage successful work on diversity-centred participatory methods~\cite{Gardner2024}, working side-by-side agender+ players to craft avatars and other game elements directed at agender euphoria. This could ensure baseline requirements are met and raise awareness among game designers about the logics that they employ even when attempting to enable player autonomy through features like avatar customization~\cite{Gardner2024}.
Agender euphoria, as a subject of study and design space, also has potential to be found in or translated to the broader landscape of interactive media. For instance, several browser plugins are available that ``neutralize'' gender by replacing or removing gendered terms on web pages. Our study provides a foundation for future qualitative and quantitative HCI research on positive gender-relevant experiences within agender+ communities.


\subsection{Guidelines for Agender Euphoric Experiences in Games and Play (RQ3)}
\label{sec:guidelines}

Our findings point to game design requirements, recommendations, and provocations for game designers. Many guidelines map onto the extant gender and queer games literatures, providing a strong cue that multiple communities can be served by following similar advice. We signpost the agender-relevant points explicitly.

\begin{itemize}
    \item \textbf{One-size-fits-none}: We verified the plurality of agender identities (\autoref{sec:participants}). Thus, specifying and designing for an ``agender identity'' will not elicit agender euphoria in an inclusive way. Instead, creators should adopt an ``ethos of multiplicity''~\cite[Location No. 3405]{ruberg2025queer} that accepts the multitude of agender identity expressions and their linkages with other identities. This includes recognizing, as a minimum, genderless, binary-irrelevant, and gender-neutral identities~\cite{ketola2022identity} that may coincide with non-binary, trans*, and cisgender identities. 
    Creators might consider how essential a gendered experience truly. 
    Cis/heteronormativity profoundly infuses the ``defaults'' of design, erasing agender folk and agender experiences regardless of gender identity, among others. 
    Our findings should provoke creators to reconsider the basis on which they design games. 

    \item \textbf{Gender-free play}: Many agender euphoric accounts spoke to an \textdi{agender baseline}. Whether the \textdi{characters, narrative, mechanics,} or \textdi{world}, ``gender-free'' was the ideal. This would include no mention of gender or gendering by others, real people, or NPCs (\textdi{social}). Character-less worlds, first-person POVs, non-human embodiments, ambiguous avatars, and pronoun avoidance would be characteristic of this baseline. 

    \item \textbf{Gender on/off switch}: Not all games may be able to provide an \textdi{agender baseline}. 
    There is also a risk of preventing \emph{gender} euphoric experiences, which lean on gender cues in visual appearance and pronoun use~\cite{Whitehouse_2023agender,ketola2022identity,liang-2025-euphoria}. 
    One option is a gender on/off switch presented once and then accessible by the player. 
    While perhaps not internally consistent or ``realistic,'' 
    allowing players to easily 
    transform their player representation is inclusive. 
    Further, such approaches are reflective of how people engage with games in long-term ways that may transcend their senses of identity---supporting such experiences enables enjoyment. 

    \item \textbf{Bare minimum representation}: Creators could go the \textdi{agender by design} route, taking note of \autoref{tab:avatar} in particular. At a high-level, this means not making assumptions about or prescribing gender, including not conflating non-binary with agender identities.
    Agender would be the default option on \textdi{character} creation screens. Agender \textdi{character} would exist, playable or not. This \textdi{validates} agender identities. Even if gendering was needed or \textdi{chosen} by the player, it would not reflect on the \textdi{mechanics, narrative,} or \textdi{world} (\textdi{irrelevance}). 
    Other players and NPCs would honour the handles and (lack of) pronoun use.

    \item \textbf{Sandbox-style freedom}: Players are clever and crafty with what the game provides, finding such \textdi{agender play} euphoric. 
    Players can \textdi{project} genderlessness onto characters, regardless of in-game assertions. 
    Players can \textdi{evade}, \textdi{explore}, and \textdi{subvert} gender-boundedness through freedom of expression in the \textdi{characters, narrative,} and \textdi{social} context. 
    Players will transgress gendered \textdi{mechanics} and \textdi{worlds} when the choices offered do not preclude taking action. 
    Fluidity and a range of options would be the baseline.

    \item \textbf{Enabling meaningful experiences}: Agender euphoria is tied up with meaningful personal expression and \textdi{reflection}: eudaimonic orientations with long-term meaning potential. 
    Designers and developers should be aware that games offer a safe and inspirational magic circle~\cite{huizinga_homo_2000,rules-of-play} for identity play, gender relief, and even agender envy. 
    Games with a \textdi{social} component may empower friends, family members, teams, and communities to explore and affirm identity. 
    Creators should be conscious of these opportunities and especially of unintentional restrictions, like binary options and forced disclosures of handles and pronouns.
    
\end{itemize}


\subsection{Limitations}
\label{sec:limit}
Like much participant research~\cite{henrich2010most,henrich2010weirdest}, notably in HCI~\cite{linxen2021weird,seaborn2024subtle} and adjacent fields~\cite{seaborn2023not}, our sample was drawn from white and English-speaking populations~\cite{Trammell_2023}. 
Future work should sample outside of Anglocentric and WEIRD contexts. 
The CIT method relies on accounts retrieved from memory, i.e., post-hoc recall, a constructive process prone to decay~\cite{huta_pursuing_2010,kim_pleasure_2014,mekler_momentary_2016,Seaborn_2024maldai}, especially if the incident happened a long time before. 
Follow-up work could include real-time observations of play, similar to a stimulated-recall technique \cite{tracing-change-chi-25}.
Our use of the I-PANAS-SF~\cite{Karim_2011panas,Thompson_2007panas} was guided by future cross-cultural work. 
However, the scale items did not fully match euphoric experiences. 
We recommend using the upcoming Gender Euphoria Scale~\cite{blacklock2024gender} instead. Finally, we included agender+ populations offering agender (non)euphoric experiences. Differences among agender variants~\cite{Whitehouse_2023agender} may have been obscured by our approach. Moreover, those who identified as cisgender but had agender euphoric experiences may have been filtered out. Future work may involve a fine-grained analysis by agender identity. Also, the recruitment materials referenced ``agender,'' acting as an identity prime with unknown effects. Future work could broadly capture ``euphoria'' in games and use either platform demographics data or post-task items to gather identity characteristics without priming.


\section{Conclusion}
\label{sec:conclude}

We have gathered a diverse array of game-based agender euphoric experiences across a diverse agender+ player sample. Our participants have offered a glimpse at a queer posthuman world~\cite{ruberg2025queer,Ruberg_2022post}---beyond gender and gender dynamics---that we call \textbf{radical gender neutrality.} The rich accounts offered by our agender experience-aligned player cohort invite us to reconsider not only ``the centrality of \emph{human} agency''~\cite[p.~413]{Ruberg_2022post} (our emphasis) but also the centrality of \emph{gender} in player experiences of agency and identity work. The next steps are clear:

\begin{itemize}
    \item For scholars, explore agender euphoria in other cultural and geographic contexts. 
    \item For game designers, design and playtest agender euphoric experiences with players of all genders. 
    \item Ultimately, offer these games to the community for play and critique.
\end{itemize}

Games are a powerful medium of expression and euphoria in many ways, and (a)gender alignment is one of them. Games can offer a radical gender neutral space and play experiences for anyone on or outside of the gender spectrum, but especially those who identify as agender. We encourage game designers, scholars, and the community of practice to engage with the full prism of player experience within and outside of the magic circle and beyond the ties of identity. We offer agender euphoria as a novel but needed shard in the prism of player experience across (a)gendered worlds.

\begin{acks}
\subsubsection*{General Acknowledgements}
We thank Rose Bohrer for allowing us liberal use of her Discord server. Liang thanks Rose Bohrer and Lane Harrison for general research support during the time this work was carried out.

\subsubsection*{Entirely Human-Made}

No form of generative artificial intelligence (AI), large language models (LLMs), or similar software was used in the production of research or writing of this work.

\subsubsection*{Acknowledgement of Country (Toups Dugas)}

I acknowledge the Traditional Owners and Custodians of Country throughout Australia and the lands on which I carry out my research: the Bunurong people of the Kulin Nations. 
I recognize their continuing connection to the land, waters, and community since time immemorial; I also recognize that they have never ceded sovereignty. 
I pay my respects to their Elders, past and present, and emerging leaders.

\end{acks}

\bibliographystyle{ACM-Reference-Format}
\balance
\bibliography{★refs}


\begin{thebibliography}{169}


\ifx \showCODEN    \undefined \def \showCODEN     #1{\unskip}     \fi
\ifx \showDOI      \undefined \def \showDOI       #1{#1}\fi
\ifx \showISBNx    \undefined \def \showISBNx     #1{\unskip}     \fi
\ifx \showISBNxiii \undefined \def \showISBNxiii  #1{\unskip}     \fi
\ifx \showISSN     \undefined \def \showISSN      #1{\unskip}     \fi
\ifx \showLCCN     \undefined \def \showLCCN      #1{\unskip}     \fi
\ifx \shownote     \undefined \def \shownote      #1{#1}          \fi
\ifx \showarticletitle \undefined \def \showarticletitle #1{#1}   \fi
\ifx \showURL      \undefined \def \showURL       {\relax}        \fi
\providecommand\bibfield[2]{#2}
\providecommand\bibinfo[2]{#2}
\providecommand\natexlab[1]{#1}
\providecommand\showeprint[2][]{arXiv:#2}

\bibitem[Abeele et~al\mbox{.}(2020)]%
        {abeele_development_2020}
\bibfield{author}{\bibinfo{person}{Vero~Vanden Abeele}, \bibinfo{person}{Katta Spiel}, \bibinfo{person}{Lennart Nacke}, \bibinfo{person}{Daniel Johnson}, {and} \bibinfo{person}{Kathrin Gerling}.} \bibinfo{year}{2020}\natexlab{}.
\newblock \showarticletitle{Development and validation of the player experience inventory: {A} scale to measure player experiences at the level of functional and psychosocial consequences}.
\newblock \bibinfo{journal}{\emph{International Journal of Human-Computer Studies}}  \bibinfo{volume}{135} (\bibinfo{date}{March} \bibinfo{year}{2020}), \bibinfo{pages}{102370}.
\newblock
\showISSN{1071-5819}
\urldef\tempurl%
\url{https://doi.org/10.1016/j.ijhcs.2019.102370}
\showDOI{\tempurl}


\bibitem[Adams and Dormans(2012)]%
        {Adams:2012:GMA:2385822}
\bibfield{author}{\bibinfo{person}{Ernest Adams} {and} \bibinfo{person}{Joris Dormans}.} \bibinfo{year}{2012}\natexlab{}.
\newblock \bibinfo{booktitle}{\emph{Game Mechanics: Advanced Game Design} (\bibinfo{edition}{1st} ed.)}.
\newblock \bibinfo{publisher}{New Riders Publishing}, \bibinfo{address}{Thousand Oaks, CA, USA}.
\newblock
\showISBNx{0321820274, 9780321820273}


\bibitem[Ahmed(2014)]%
        {ahmed2013cultural}
\bibfield{author}{\bibinfo{person}{Sara Ahmed}.} \bibinfo{year}{2014}\natexlab{}.
\newblock \showarticletitle{Queer Feelings}.
\newblock In \bibinfo{booktitle}{\emph{The Cultural Politics of Emotion} (\bibinfo{edition}{second} ed.)}. \bibinfo{publisher}{Edinburgh University Press}, \bibinfo{address}{Edinburgh, Scotland, UK}, \bibinfo{pages}{144--167}.
\newblock
\showISBNx{978-0-415-97254-3}


\bibitem[Ahmed(2020)]%
        {ahmed2020orientations}
\bibfield{author}{\bibinfo{person}{Sara Ahmed}.} \bibinfo{year}{2020}\natexlab{}.
\newblock \bibinfo{booktitle}{\emph{Orientations Matter}}.
\newblock \bibinfo{publisher}{Duke University Press}, \bibinfo{address}{New York, NY, USA}, \bibinfo{pages}{234–257}.
\newblock
\showISBNx{9780822392996}
\urldef\tempurl%
\url{https://doi.org/10.1515/9780822392996-013}
\showDOI{\tempurl}


\bibitem[Alharthi et~al\mbox{.}(2018)]%
        {Alharthi_Alsaedi_Toups}
\bibfield{author}{\bibinfo{person}{Sultan~A. Alharthi}, \bibinfo{person}{Olaa Alsaedi}, \bibinfo{person}{Phoebe~O. Toups~Dugas}, \bibinfo{person}{Theresa~Jean Tanenbaum}, {and} \bibinfo{person}{Jessica Hammer}.} \bibinfo{year}{2018}\natexlab{}.
\newblock \showarticletitle{Playing to Wait: A Taxonomy of Idle Games}. In \bibinfo{booktitle}{\emph{Proceedings of the 2018 CHI Conference on Human Factors in Computing Systems}} \emph{(\bibinfo{series}{CHI '18})}. \bibinfo{publisher}{Association for Computing Machinery}, \bibinfo{address}{New York, NY, USA}, \bibinfo{pages}{1--15}.
\newblock
\showISBNx{978-1-4503-5620-6}
\urldef\tempurl%
\url{https://doi.org/10.1145/3173574.3174195}
\showDOI{\tempurl}


\bibitem[Alutalica(2021)]%
        {alutalica2021transgender}
\bibfield{author}{\bibinfo{person}{Jaxx Alutalica}.} \bibinfo{year}{2021}\natexlab{}.
\newblock \showarticletitle{Transgender bliss: From gender dysphoria to gender euphoria through decolonizing gender}.
\newblock In \bibinfo{booktitle}{\emph{An intersectional approach to sex therapy}}. \bibinfo{publisher}{Routledge}, \bibinfo{address}{New York, USA}, \bibinfo{pages}{109--123}.
\newblock
\showISBNx{9781003034063}


\bibitem[Ashley and Ells(2018)]%
        {ashley2018favor}
\bibfield{author}{\bibinfo{person}{Florence Ashley} {and} \bibinfo{person}{Carolyn Ells}.} \bibinfo{year}{2018}\natexlab{}.
\newblock \showarticletitle{In Favor of Covering Ethically Important Cosmetic Surgeries: Facial Feminization Surgery for Transgender People}.
\newblock \bibinfo{journal}{\emph{The American Journal of Bioethics}} \bibinfo{volume}{18}, \bibinfo{number}{12} (\bibinfo{year}{2018}), \bibinfo{pages}{23--25}.
\newblock
\urldef\tempurl%
\url{https://doi.org/10.1080/15265161.2018.1531162}
\showDOI{\tempurl}


\bibitem[Austin et~al\mbox{.}(2022)]%
        {austin2022gender}
\bibfield{author}{\bibinfo{person}{Ashley Austin}, \bibinfo{person}{Ryan Papciak}, {and} \bibinfo{person}{Lindsay Lovins}.} \bibinfo{year}{2022}\natexlab{}.
\newblock \showarticletitle{Gender euphoria: A grounded theory exploration of experiencing gender affirmation}.
\newblock \bibinfo{journal}{\emph{Psychology \& Sexuality}} \bibinfo{volume}{13}, \bibinfo{number}{5} (\bibinfo{year}{2022}), \bibinfo{pages}{1406--1426}.
\newblock
\urldef\tempurl%
\url{https://doi.org/10.1080/19419899.2022.2049632}
\showDOI{\tempurl}


\bibitem[Bachert(2023)]%
        {bachert2023apathetic}
\bibfield{author}{\bibinfo{person}{Alex Bachert}.} \bibinfo{year}{2023}\natexlab{}.
\newblock \bibinfo{title}{What Does It Mean to Be Gender Apathetic?}
\newblock
\newblock
\urldef\tempurl%
\url{https://www.charliehealth.com/post/what-does-it-mean-to-be-gender-apathetic}
\showURL{%
\tempurl}


\bibitem[Badgley et~al\mbox{.}(2023)]%
        {dypshoriabible}
\bibfield{author}{\bibinfo{person}{Jocelyn Badgley} {et~al\mbox{.}}} \bibinfo{year}{2023}\natexlab{}.
\newblock \bibinfo{title}{The Gender Dysphoria Bible}.
\newblock
\newblock
\urldef\tempurl%
\url{https://genderdysphoria.fyi/en/}
\showURL{%
\tempurl}


\bibitem[Banks(2015)]%
        {banks2015object}
\bibfield{author}{\bibinfo{person}{Jaime Banks}.} \bibinfo{year}{2015}\natexlab{}.
\newblock \showarticletitle{Object, Me, Symbiote, Other: A social typology of player-avatar relationships}.
\newblock \bibinfo{journal}{\emph{First Monday}} \bibinfo{volume}{20}, \bibinfo{number}{2} (\bibinfo{date}{Feb} \bibinfo{year}{2015}).
\newblock
\urldef\tempurl%
\url{https://doi.org/10.5210/fm.v20i2.5433}
\showDOI{\tempurl}


\bibitem[Bardzell(2010)]%
        {Bardzell_2010}
\bibfield{author}{\bibinfo{person}{Shaowen Bardzell}.} \bibinfo{year}{2010}\natexlab{}.
\newblock \showarticletitle{Feminist HCI: taking stock and outlining an agenda for design}. In \bibinfo{booktitle}{\emph{Proceedings of the SIGCHI Conference on Human Factors in Computing Systems}} \emph{(\bibinfo{series}{CHI ’10})}. \bibinfo{publisher}{ACM}, \bibinfo{address}{New York, NY, USA}, \bibinfo{pages}{1301–1310}.
\newblock
\urldef\tempurl%
\url{https://doi.org/10.1145/1753326.1753521}
\showDOI{\tempurl}


\bibitem[Bartsch(2012)]%
        {bartsch2012emotional}
\bibfield{author}{\bibinfo{person}{Anne Bartsch}.} \bibinfo{year}{2012}\natexlab{}.
\newblock \showarticletitle{Emotional gratification in entertainment experience. Why viewers of movies and television series find it rewarding to experience emotions}.
\newblock \bibinfo{journal}{\emph{Media Psychology}} \bibinfo{volume}{15}, \bibinfo{number}{3} (\bibinfo{year}{2012}), \bibinfo{pages}{267--302}.
\newblock


\bibitem[Bartsch et~al\mbox{.}(2014)]%
        {bartsch2014moved}
\bibfield{author}{\bibinfo{person}{Anne Bartsch}, \bibinfo{person}{Anja Kalch}, {and} \bibinfo{person}{Mary~Beth Oliver}.} \bibinfo{year}{2014}\natexlab{}.
\newblock \showarticletitle{Moved to think}.
\newblock \bibinfo{journal}{\emph{Journal of Media Psychology}}  \bibinfo{volume}{26} (\bibinfo{year}{2014}), \bibinfo{pages}{125--140}.
\newblock
Issue 3.


\bibitem[Bayliss(2010)]%
        {Bayliss2010Videogames-Inte}
\bibfield{author}{\bibinfo{person}{Peter Bayliss}.} \bibinfo{year}{2010}\natexlab{}.
\newblock \emph{\bibinfo{title}{Videogames, Interfaces, and the Body: The importance of embodied phenomena to the experience of videogame play}}.
\newblock \bibinfo{thesistype}{Ph.\,D. Dissertation}. \bibinfo{school}{RMIT University}.
\newblock


\bibitem[Beischel(2022)]%
        {beischel2022gender}
\bibfield{author}{\bibinfo{person}{Will Beischel}.} \bibinfo{year}{2022}\natexlab{}.
\newblock \emph{\bibinfo{title}{Gender Pleasure: The Positive Affective Component of Gender/Sex}}.
\newblock \bibinfo{thesistype}{Ph.\,D. Dissertation}. \bibinfo{school}{University of Michigan}.
\newblock
\urldef\tempurl%
\url{https://deepblue.lib.umich.edu/handle/2027.42/174393}
\showURL{%
\tempurl}


\bibitem[Beischel et~al\mbox{.}(2022)]%
        {beischel2022little}
\bibfield{author}{\bibinfo{person}{Will~J. Beischel}, \bibinfo{person}{St{\'e}phanie E.~M. Gauvin}, {and} \bibinfo{person}{Sari~M. van Anders}.} \bibinfo{year}{2022}\natexlab{}.
\newblock \showarticletitle{``A little shiny gender breakthrough'': Community understandings of gender euphoria}.
\newblock \bibinfo{journal}{\emph{International Journal of Transgender Health}} \bibinfo{volume}{23}, \bibinfo{number}{3} (\bibinfo{year}{2022}), \bibinfo{pages}{274--294}.
\newblock
\urldef\tempurl%
\url{https://doi.org/10.1080/26895269.2021.1915223}
\showDOI{\tempurl}


\bibitem[Beres et~al\mbox{.}(2021)]%
        {Beres_2021toxic}
\bibfield{author}{\bibinfo{person}{Nicole~A Beres}, \bibinfo{person}{Julian Frommel}, \bibinfo{person}{Elizabeth Reid}, \bibinfo{person}{Regan~L Mandryk}, {and} \bibinfo{person}{Madison Klarkowski}.} \bibinfo{year}{2021}\natexlab{}.
\newblock \showarticletitle{Don’t You Know That You’re Toxic: Normalization of Toxicity in Online Gaming}. In \bibinfo{booktitle}{\emph{Proceedings of the 2021 CHI Conference on Human Factors in Computing Systems}} \emph{(\bibinfo{series}{CHI ’21})}. \bibinfo{publisher}{ACM}, \bibinfo{address}{New York, NY, USA}, \bibinfo{pages}{1–15}.
\newblock
\urldef\tempurl%
\url{https://doi.org/10.1145/3411764.3445157}
\showDOI{\tempurl}


\bibitem[Blackless et~al\mbox{.}(2000)]%
        {blackless2000}
\bibfield{author}{\bibinfo{person}{Melanie Blackless}, \bibinfo{person}{Anthony Charuvastra}, \bibinfo{person}{Amanda Derryck}, \bibinfo{person}{Anne Fausto-Sterling}, \bibinfo{person}{Karl Lauzanne}, {and} \bibinfo{person}{Ellen Lee}.} \bibinfo{year}{2000}\natexlab{}.
\newblock \showarticletitle{How sexually dimorphic are we? Review and synthesis}.
\newblock \bibinfo{journal}{\emph{American Journal of Human Biology}}  \bibinfo{volume}{12} (\bibinfo{year}{2000}), \bibinfo{pages}{151--166}.
\newblock


\bibitem[Blacklock et~al\mbox{.}(2024)]%
        {blacklock2024gender}
\bibfield{author}{\bibinfo{person}{C.~A. Blacklock}, \bibinfo{person}{M.~A. Tollit}, \bibinfo{person}{C.~C. Pace}, \bibinfo{person}{B. Elphinstone}, \bibinfo{person}{K.~C. Pang}, {and} \bibinfo{person}{S. Buzwell}.} \bibinfo{year}{2024}\natexlab{}.
\newblock \showarticletitle{The Gender Euphoria Scale (GES): a protocol for developing and validating a tool to measure gender euphoria in transgender and gender diverse individuals}.
\newblock \bibinfo{journal}{\emph{Frontiers in Psychology}}  \bibinfo{volume}{14} (\bibinfo{year}{2024}), \bibinfo{pages}{1284991}.
\newblock
\showISSN{1664-1078}
\urldef\tempurl%
\url{https://doi.org/10.3389/fpsyg.2023.1284991}
\showDOI{\tempurl}


\bibitem[Blacklock et~al\mbox{.}(2025)]%
        {Blacklock-gender-euphoria-scale-2025}
\bibfield{author}{\bibinfo{person}{Charlotte~A. Blacklock}, \bibinfo{person}{Michelle~A. Tollit}, \bibinfo{person}{Carmen~C. Pace}, \bibinfo{person}{Bradley Elphinstone}, \bibinfo{person}{Sav Zwickl}, \bibinfo{person}{Ada~S. Cheung}, \bibinfo{person}{Kyra Citron}, \bibinfo{person}{Samuel Marsan}, \bibinfo{person}{Nancy Zucker}, \bibinfo{person}{Simone Buzwell}, {and} \bibinfo{person}{Ken~C. Pang}.} \bibinfo{year}{2025}\natexlab{}.
\newblock \showarticletitle{The Gender Euphoria Scale (GES): Development of a tool to measure gender euphoria}.
\newblock \bibinfo{journal}{\emph{International Journal of Transgender Health}} \bibinfo{volume}{0}, \bibinfo{number}{0} (\bibinfo{year}{2025}), \bibinfo{pages}{1--12}.
\newblock
\urldef\tempurl%
\url{https://doi.org/10.1080/26895269.2024.2447768}
\showDOI{\tempurl}


\bibitem[Blanco-Fern{\'a}ndez and Moreno(2023)]%
        {blanco2023video}
\bibfield{author}{\bibinfo{person}{V{\'\i}tor Blanco-Fern{\'a}ndez} {and} \bibinfo{person}{Jose~A Moreno}.} \bibinfo{year}{2023}\natexlab{}.
\newblock \showarticletitle{“Video Games Were My First Safe Space”: Queer Gaming in the Animal Crossing New Horizons LGBTIQA+ Community}.
\newblock \bibinfo{journal}{\emph{Games and Culture}} \bibinfo{volume}{20}, \bibinfo{number}{4} (\bibinfo{year}{2023}), \bibinfo{pages}{438--458}.
\newblock
\urldef\tempurl%
\url{https://doi.org/10.1177/15554120231205638}
\showDOI{\tempurl}


\bibitem[B{\o}dker et~al\mbox{.}(2022)]%
        {bodker2022participatory}
\bibfield{author}{\bibinfo{person}{Susanne B{\o}dker}, \bibinfo{person}{Christian Dindler}, \bibinfo{person}{Ole~S. Iversen}, {and} \bibinfo{person}{Rachel~C. Smith}.} \bibinfo{year}{2022}\natexlab{}.
\newblock \showarticletitle{What is participatory design?}
\newblock In \bibinfo{booktitle}{\emph{Participatory Design}}. \bibinfo{publisher}{Springer}, \bibinfo{address}{London, UK}, \bibinfo{pages}{5--13}.
\newblock


\bibitem[Bradford et~al\mbox{.}(2021)]%
        {bradford2021hair}
\bibfield{author}{\bibinfo{person}{Nova~J. Bradford}, \bibinfo{person}{G.~Nic Rider}, {and} \bibinfo{person}{Katherine~G. Spencer}.} \bibinfo{year}{2021}\natexlab{}.
\newblock \showarticletitle{Hair removal and psychological well-being in transfeminine adults: associations with gender dysphoria and gender euphoria}.
\newblock \bibinfo{journal}{\emph{Journal of Dermatological Treatment}} \bibinfo{volume}{32}, \bibinfo{number}{6} (\bibinfo{year}{2021}), \bibinfo{pages}{635--642}.
\newblock
\urldef\tempurl%
\url{https://doi.org/10.1080/09546634.2019.1687823}
\showDOI{\tempurl}


\bibitem[Bragan{\c{c}}a et~al\mbox{.}(2016)]%
        {braganca2016twinegame}
\bibfield{author}{\bibinfo{person}{L. Bragan{\c{c}}a}, \bibinfo{person}{R. Mota}, {and} \bibinfo{person}{E. Fantini}.} \bibinfo{year}{2016}\natexlab{}.
\newblock \showarticletitle{Twine Game Narrative and discussion about LGBTQ representation}. In \bibinfo{booktitle}{\emph{Proceedings of SBGames}}. \bibinfo{publisher}{ACM}, \bibinfo{address}{New York, NY, USA}, \bibinfo{pages}{937--946}.
\newblock


\bibitem[Braun and Clarke(2019)]%
        {Braun_2019}
\bibfield{author}{\bibinfo{person}{Virginia Braun} {and} \bibinfo{person}{Victoria Clarke}.} \bibinfo{year}{2019}\natexlab{}.
\newblock \showarticletitle{Reflecting on reflexive thematic analysis}.
\newblock \bibinfo{journal}{\emph{Qualitative Research in Sport, Exercise and Health}} \bibinfo{volume}{11}, \bibinfo{number}{4} (\bibinfo{date}{June} \bibinfo{year}{2019}), \bibinfo{pages}{589–597}.
\newblock
\showISSN{2159-6778}
\urldef\tempurl%
\url{https://doi.org/10.1080/2159676x.2019.1628806}
\showDOI{\tempurl}


\bibitem[Braun and Clarke(2022)]%
        {Braun_2022}
\bibfield{author}{\bibinfo{person}{Virginia Braun} {and} \bibinfo{person}{Victoria Clarke}.} \bibinfo{year}{2022}\natexlab{}.
\newblock \showarticletitle{Toward good practice in thematic analysis: Avoiding common problems and be(com)ing aknowingresearcher}.
\newblock \bibinfo{journal}{\emph{International Journal of Transgender Health}} \bibinfo{volume}{24}, \bibinfo{number}{1} (\bibinfo{date}{Oct.} \bibinfo{year}{2022}), \bibinfo{pages}{1–6}.
\newblock
\showISSN{2689-5269}
\urldef\tempurl%
\url{https://doi.org/10.1080/26895269.2022.2129597}
\showDOI{\tempurl}


\bibitem[Brock(2011)]%
        {Brock_2011racism}
\bibfield{author}{\bibinfo{person}{André Brock}.} \bibinfo{year}{2011}\natexlab{}.
\newblock \showarticletitle{‘“When Keeping it Real Goes Wrong”’: Resident Evil 5, Racial Representation, and Gamers}.
\newblock \bibinfo{journal}{\emph{Games and Culture}} \bibinfo{volume}{6}, \bibinfo{number}{5} (\bibinfo{date}{July} \bibinfo{year}{2011}), \bibinfo{pages}{429–452}.
\newblock
\showISSN{1555-4139}
\urldef\tempurl%
\url{https://doi.org/10.1177/1555412011402676}
\showDOI{\tempurl}


\bibitem[Carpenter(2024)]%
        {carpenter-intersex-2024}
\bibfield{author}{\bibinfo{person}{Morgan Carpenter}.} \bibinfo{year}{2024}\natexlab{}.
\newblock \showarticletitle{Fixing bodies and shaping narratives: Epistemic injustice and the responses of medicine and bioethics to intersex human rights demands}.
\newblock \bibinfo{journal}{\emph{Clinical Ethics}} \bibinfo{volume}{19}, \bibinfo{number}{1} (\bibinfo{year}{2024}), \bibinfo{pages}{3--17}.
\newblock
\urldef\tempurl%
\url{https://doi.org/10.1177/14777509231180412}
\showDOI{\tempurl}
\showeprint{https://doi.org/10.1177/14777509231180412}


\bibitem[Chang(2015)]%
        {Chang_2015queergaming}
\bibfield{author}{\bibinfo{person}{Edmond~Y. Chang}.} \bibinfo{year}{2015}\natexlab{}.
\newblock \showarticletitle{Love Is in the Air: Queer (Im)Possibility and Straightwashing in FrontierVille and World of Warcraft}.
\newblock \bibinfo{journal}{\emph{QED: A Journal in GLBTQ Worldmaking}} \bibinfo{volume}{2}, \bibinfo{number}{2} (\bibinfo{date}{June} \bibinfo{year}{2015}), \bibinfo{pages}{6–31}.
\newblock
\showISSN{2327-1590}
\urldef\tempurl%
\url{https://doi.org/10.14321/qed.2.2.0006}
\showDOI{\tempurl}


\bibitem[Chimakonam(2020)]%
        {chimakonam2020othering}
\bibfield{author}{\bibinfo{person}{Jonathan~O. Chimakonam}.} \bibinfo{year}{2020}\natexlab{}.
\newblock \showarticletitle{Othering, Re-othering, and De-othering: Interrogating the Skolombo's Fight-Back Strategy}.
\newblock In \bibinfo{booktitle}{\emph{Handbook of African Philosophy of Difference}}, \bibfield{editor}{\bibinfo{person}{Elvis Imafidon}} (Ed.). \bibinfo{publisher}{Springer International Publishing}, \bibinfo{address}{Cham, Germany}, \bibinfo{pages}{433--448}.
\newblock
\showISBNx{978-3-030-14835-5}
\urldef\tempurl%
\url{https://doi.org/10.1007/978-3-030-14835-5_20}
\showDOI{\tempurl}


\bibitem[Christoph et~al\mbox{.}(2009)]%
        {Christoph_2009self}
\bibfield{author}{\bibinfo{person}{Klimmt Christoph}, \bibinfo{person}{Hefner Dorothée}, {and} \bibinfo{person}{Vorderer Peter}.} \bibinfo{year}{2009}\natexlab{}.
\newblock \showarticletitle{The Video Game Experience as “True” Identification: A Theory of Enjoyable Alterations of Players’ Self-Perception}.
\newblock \bibinfo{journal}{\emph{Communication Theory}} \bibinfo{volume}{19}, \bibinfo{number}{4} (\bibinfo{date}{Nov.} \bibinfo{year}{2009}), \bibinfo{pages}{351–373}.
\newblock
\showISSN{1468-2885}
\urldef\tempurl%
\url{https://doi.org/10.1111/j.1468-2885.2009.01347.x}
\showDOI{\tempurl}


\bibitem[Cook(1995)]%
        {dnd2ephb}
\bibfield{author}{\bibinfo{person}{David~Zeb Cook}.} \bibinfo{year}{1995}\natexlab{}.
\newblock \bibinfo{booktitle}{\emph{Advanced Dungeons and Dragons Player's Handbook} (\bibinfo{edition}{2nd} ed.)}.
\newblock \bibinfo{publisher}{TSR Inc.}, \bibinfo{address}{Wisconsin}.
\newblock


\bibitem[Cooley(1983)]%
        {cooley1983human}
\bibfield{author}{\bibinfo{person}{Charles~Horton Cooley}.} \bibinfo{year}{1983}\natexlab{}.
\newblock \bibinfo{booktitle}{\emph{Human nature and the social order} (\bibinfo{edition}{1st} ed.)}.
\newblock \bibinfo{publisher}{Routledge}, \bibinfo{address}{Oxfordshire, England, UK}.
\newblock
\urldef\tempurl%
\url{https://doi.org/10.4324/9780203789513}
\showDOI{\tempurl}


\bibitem[Cosgrove et~al\mbox{.}(2021)]%
        {cosgrove2021service}
\bibfield{author}{\bibinfo{person}{Darren Cosgrove}, \bibinfo{person}{Christine Bozlak}, {and} \bibinfo{person}{Parker Reid}.} \bibinfo{year}{2021}\natexlab{}.
\newblock \showarticletitle{Service barriers for gender nonbinary young adults: Using photovoice to understand support and stigma}.
\newblock \bibinfo{journal}{\emph{Affilia}} \bibinfo{volume}{36}, \bibinfo{number}{2} (\bibinfo{year}{2021}), \bibinfo{pages}{220--239}.
\newblock
\urldef\tempurl%
\url{https://doi.org/10.1177/0886109920944535}
\showDOI{\tempurl}


\bibitem[Creswell and Plano~Clark(2007)]%
        {creswell2007mixed}
\bibfield{author}{\bibinfo{person}{John~W. Creswell} {and} \bibinfo{person}{Vicki~L. Plano~Clark}.} \bibinfo{year}{2007}\natexlab{}.
\newblock \bibinfo{booktitle}{\emph{Designing and conducting mixed methods research}}.
\newblock \bibinfo{publisher}{Sage}, \bibinfo{address}{Thousand Oaks, CA, USA}.
\newblock


\bibitem[D'Aloia(2009)]%
        {DAloia:2009}
\bibfield{author}{\bibinfo{person}{Adriano D'Aloia}.} \bibinfo{year}{2009}\natexlab{}.
\newblock \showarticletitle{Adamant bodies. {T}he avatar-body and the problem of autoempathy}.
\newblock \bibinfo{journal}{\emph{E$|$C}} \bibinfo{volume}{III}, \bibinfo{number}{5} (\bibinfo{year}{2009}), \bibinfo{pages}{51--58}.
\newblock


\bibitem[Dalton(2013)]%
        {Dalton_2013}
\bibfield{author}{\bibinfo{person}{Nicholas~Sheep Dalton}.} \bibinfo{year}{2013}\natexlab{}.
\newblock \showarticletitle{Neurodiversity \& HCI}. In \bibinfo{booktitle}{\emph{CHI ’13 Extended Abstracts on Human Factors in Computing Systems}} \emph{(\bibinfo{series}{CHI ’13})}. \bibinfo{publisher}{ACM}, \bibinfo{address}{New York, NY, USA}, \bibinfo{pages}{2295–2304}.
\newblock
\urldef\tempurl%
\url{https://doi.org/10.1145/2468356.2468752}
\showDOI{\tempurl}


\bibitem[Deci and Ryan(2006)]%
        {Deci_2006}
\bibfield{author}{\bibinfo{person}{Edward~L. Deci} {and} \bibinfo{person}{Richard~M. Ryan}.} \bibinfo{year}{2006}\natexlab{}.
\newblock \showarticletitle{Hedonia, eudaimonia, and well-being: An introduction}.
\newblock \bibinfo{journal}{\emph{Journal of Happiness Studies}} \bibinfo{volume}{9}, \bibinfo{number}{1} (\bibinfo{date}{Nov.} \bibinfo{year}{2006}), \bibinfo{pages}{1–11}.
\newblock
\showISSN{1573-7780}
\urldef\tempurl%
\url{https://doi.org/10.1007/s10902-006-9018-1}
\showDOI{\tempurl}


\bibitem[Denisova and Cairns(2015)]%
        {denisova2015first}
\bibfield{author}{\bibinfo{person}{Alena Denisova} {and} \bibinfo{person}{Paul Cairns}.} \bibinfo{year}{2015}\natexlab{}.
\newblock \showarticletitle{First person vs. third person perspective in digital games: do player preferences affect immersion?}. In \bibinfo{booktitle}{\emph{Proceedings of the 33rd annual ACM conference on human factors in computing systems}} (Seoul, Republic of Korea) \emph{(\bibinfo{series}{CHI '15})}. \bibinfo{publisher}{Association for Computing Machinery}, \bibinfo{address}{New York, NY, USA}, \bibinfo{pages}{145--148}.
\newblock
\urldef\tempurl%
\url{https://doi.org/10.1145/2702123.2702256}
\showDOI{\tempurl}


\bibitem[Derrick et~al\mbox{.}(2008)]%
        {derrick2008parasocial}
\bibfield{author}{\bibinfo{person}{Jaye~L. Derrick}, \bibinfo{person}{Shira Gabriel}, {and} \bibinfo{person}{Brooke Tippin}.} \bibinfo{year}{2008}\natexlab{}.
\newblock \showarticletitle{Parasocial relationships and self-discrepancies: Faux relationships have benefits for low self-esteem individuals}.
\newblock \bibinfo{journal}{\emph{Personal Relationships}} \bibinfo{volume}{15}, \bibinfo{number}{2} (\bibinfo{year}{2008}), \bibinfo{pages}{261--280}.
\newblock
\urldef\tempurl%
\url{https://doi.org/10.1111/j.1475-6811.2008.00197.x}
\showDOI{\tempurl}


\bibitem[Diamond(2020)]%
        {diamond2020gender}
\bibfield{author}{\bibinfo{person}{Lisa~M. Diamond}.} \bibinfo{year}{2020}\natexlab{}.
\newblock \showarticletitle{Gender fluidity and nonbinary gender identities among children and adolescents}.
\newblock \bibinfo{journal}{\emph{Child Development Perspectives}} \bibinfo{volume}{14}, \bibinfo{number}{2} (\bibinfo{year}{2020}), \bibinfo{pages}{110--115}.
\newblock
\urldef\tempurl%
\url{https://doi.org/10.1111/cdep.12366}
\showDOI{\tempurl}


\bibitem[Drenten et~al\mbox{.}(2022)]%
        {Drenten_2022}
\bibfield{author}{\bibinfo{person}{Jenna Drenten}, \bibinfo{person}{Robert~L Harrison}, {and} \bibinfo{person}{Nicholas~J Pendarvis}.} \bibinfo{year}{2022}\natexlab{}.
\newblock \showarticletitle{More Gamer, Less Girl: Gendered Boundaries, Tokenism, and the Cultural Persistence of Masculine Dominance}.
\newblock \bibinfo{journal}{\emph{Journal of Consumer Research}} \bibinfo{volume}{50}, \bibinfo{number}{1} (\bibinfo{date}{Oct.} \bibinfo{year}{2022}), \bibinfo{pages}{2–24}.
\newblock
\showISSN{1537-5277}
\urldef\tempurl%
\url{https://doi.org/10.1093/jcr/ucac046}
\showDOI{\tempurl}


\bibitem[Erikson(1968)]%
        {erikson1968identity}
\bibfield{author}{\bibinfo{person}{Erik~H. Erikson}.} \bibinfo{year}{1968}\natexlab{}.
\newblock \bibinfo{booktitle}{\emph{Identity youth and crisis}}.
\newblock \bibinfo{publisher}{WW Norton \& company}, \bibinfo{address}{New York, NY, USA}.
\newblock
\showISBNx{9780393347340}


\bibitem[Fausto-Sterling(2000)]%
        {fausto2000sexing}
\bibfield{author}{\bibinfo{person}{Anne Fausto-Sterling}.} \bibinfo{year}{2000}\natexlab{}.
\newblock \bibinfo{booktitle}{\emph{Sexing the body: Gender politics and the construction of sexuality}}.
\newblock \bibinfo{publisher}{Basic Books}, \bibinfo{address}{New York, New York, USA}.
\newblock


\bibitem[Fern{\'a}ndez-Vara(2009)]%
        {fernandez2009play}
\bibfield{author}{\bibinfo{person}{Clara Fern{\'a}ndez-Vara}.} \bibinfo{year}{2009}\natexlab{}.
\newblock \showarticletitle{Play's the thing: A framework to study videogames as performance}. In \bibinfo{booktitle}{\emph{2009 DiGRA International Conference: Breaking New Ground: Innovation in Games, Play, Practice and Theory}}. \bibinfo{publisher}{Digital Games Research Association}, \bibinfo{address}{N/A}, \bibinfo{numpages}{9}~pages.
\newblock


\bibitem[Freeman et~al\mbox{.}(2022)]%
        {Freeman_2022}
\bibfield{author}{\bibinfo{person}{Guo Freeman}, \bibinfo{person}{Divine Maloney}, \bibinfo{person}{Dane Acena}, {and} \bibinfo{person}{Catherine Barwulor}.} \bibinfo{year}{2022}\natexlab{}.
\newblock \showarticletitle{(Re)discovering the Physical Body Online: Strategies and Challenges to Approach Non-Cisgender Identity in Social Virtual Reality}. In \bibinfo{booktitle}{\emph{CHI Conference on Human Factors in Computing Systems}} \emph{(\bibinfo{series}{CHI ’22})}. \bibinfo{publisher}{ACM}, \bibinfo{address}{New York, NY, USA}, \bibinfo{pages}{1–15}.
\newblock
\urldef\tempurl%
\url{https://doi.org/10.1145/3491102.3502082}
\showDOI{\tempurl}


\bibitem[Galupo et~al\mbox{.}(2017)]%
        {galupo2017like}
\bibfield{author}{\bibinfo{person}{M.~Paz Galupo}, \bibinfo{person}{Lex Pulice-Farrow}, {and} \bibinfo{person}{Johanna~L. Ramirez}.} \bibinfo{year}{2017}\natexlab{}.
\newblock \showarticletitle{“Like a constantly flowing river”: Gender identity flexibility among nonbinary transgender individuals}.
\newblock In \bibinfo{booktitle}{\emph{Identity Flexibility during Adulthood: Perspectives in Adult Development}}, \bibfield{editor}{\bibinfo{person}{Jan~D. Sinnott}} (Ed.). \bibinfo{publisher}{Springer International Publishing}, \bibinfo{address}{Cham}, \bibinfo{pages}{163--177}.
\newblock
\urldef\tempurl%
\url{https://doi.org/10.1007/978-3-319-55658-1_10}
\showDOI{\tempurl}


\bibitem[Gardner et~al\mbox{.}(2024)]%
        {Gardner2024}
\bibfield{author}{\bibinfo{person}{Daniel~L. Gardner}, \bibinfo{person}{LouAnne Boyd}, {and} \bibinfo{person}{Reginald~T. Gardner}.} \bibinfo{year}{2024}\natexlab{}.
\newblock \showarticletitle{Piecing Together Performance: Collaborative, Participatory Research-Through-Design for Better Diversity in Games}.
\newblock \bibinfo{journal}{\emph{IEEE Transactions on Games}} \bibinfo{volume}{16}, \bibinfo{number}{3} (\bibinfo{date}{Sept.} \bibinfo{year}{2024}), \bibinfo{pages}{683–696}.
\newblock
\showISSN{2475-1510}
\urldef\tempurl%
\url{https://doi.org/10.1109/tg.2023.3349369}
\showDOI{\tempurl}


\bibitem[Gardner and Tanenbaum(2018)]%
        {gardner-tanenbaum-census-2018}
\bibfield{author}{\bibinfo{person}{Daniel~L. Gardner} {and} \bibinfo{person}{Theresa~Jean Tanenbaum}.} \bibinfo{year}{2018}\natexlab{}.
\newblock \showarticletitle{Dynamic Demographics: Lessons from a Large-Scale Census of Performative Possibilities in Games}. In \bibinfo{booktitle}{\emph{Proceedings of the 2018 CHI Conference on Human Factors in Computing Systems}} (Montreal QC, Canada) \emph{(\bibinfo{series}{CHI '18})}. \bibinfo{publisher}{Association for Computing Machinery}, \bibinfo{address}{New York, NY, USA}, \bibinfo{pages}{1–12}.
\newblock
\showISBNx{9781450356206}
\urldef\tempurl%
\url{https://doi.org/10.1145/3173574.3173667}
\showDOI{\tempurl}


\bibitem[George(2021)]%
        {George_2021}
\bibfield{author}{\bibinfo{person}{Leya George}.} \bibinfo{year}{2021}\natexlab{}.
\newblock \showarticletitle{Investigating the Role of Technology in Supporting Exploration of Gender Identity Through Games and Play}. In \bibinfo{booktitle}{\emph{Extended Abstracts of the 2021 Annual Symposium on Computer-Human Interaction in Play}} \emph{(\bibinfo{series}{CHI PLAY ’21})}. \bibinfo{publisher}{ACM}, \bibinfo{address}{New York, NY, USA}, \bibinfo{pages}{399–400}.
\newblock
\urldef\tempurl%
\url{https://doi.org/10.1145/3450337.3483516}
\showDOI{\tempurl}


\bibitem[George et~al\mbox{.}(2023)]%
        {george2023}
\bibfield{author}{\bibinfo{person}{Leya George}, \bibinfo{person}{Aneesha Singh}, \bibinfo{person}{Nadia Berthouze}, \bibinfo{person}{Lorna Hobbs}, {and} \bibinfo{person}{Jo Gibbs}.} \bibinfo{year}{2023}\natexlab{}.
\newblock \showarticletitle{Jamming-as-Exploration: Creating and Playing Games to Explore Gender Identity}. In \bibinfo{booktitle}{\emph{Proceedings of the 2023 CHI Conference on Human Factors in Computing Systems}} (Hamburg, Germany) \emph{(\bibinfo{series}{CHI '23})}. \bibinfo{publisher}{Association for Computing Machinery}, \bibinfo{address}{New York, NY, USA}, Article \bibinfo{articleno}{208}, \bibinfo{numpages}{19}~pages.
\newblock
\showISBNx{9781450394215}
\urldef\tempurl%
\url{https://doi.org/10.1145/3544548.3580646}
\showDOI{\tempurl}


\bibitem[Gesler(2003)]%
        {gesler2003healing}
\bibfield{author}{\bibinfo{person}{Wilbert~M Gesler}.} \bibinfo{year}{2003}\natexlab{}.
\newblock \bibinfo{booktitle}{\emph{Healing places}}.
\newblock \bibinfo{publisher}{Rowman \& Littlefield}, \bibinfo{address}{Lanham, MD, US}.
\newblock
\showISBNx{0742519554}


\bibitem[Grant et~al\mbox{.}(2024)]%
        {Grant_2024euph}
\bibfield{author}{\bibinfo{person}{Ruby Grant}, \bibinfo{person}{Natalie Amos}, \bibinfo{person}{Teddy Cook}, \bibinfo{person}{Ashleigh Lin}, \bibinfo{person}{Adam Hill}, \bibinfo{person}{Marina Carman}, {and} \bibinfo{person}{Adam Bourne}.} \bibinfo{year}{2024}\natexlab{}.
\newblock \showarticletitle{From euphoria to wellbeing: Correlates of gender euphoria and its association with mental wellbeing among transgender adults}.
\newblock \bibinfo{journal}{\emph{International Journal of Transgender Health}} (\bibinfo{date}{March} \bibinfo{year}{2024}), \bibinfo{pages}{1–9}.
\newblock
\showISSN{2689-5269}
\urldef\tempurl%
\url{https://doi.org/10.1080/26895269.2024.2324100}
\showDOI{\tempurl}


\bibitem[{Grindr Editorial team}(2024)]%
        {grindr2024apathetic}
\bibfield{author}{\bibinfo{person}{{Grindr Editorial team}}.} \bibinfo{year}{2024}\natexlab{}.
\newblock \bibinfo{title}{Gender Apathetic: Exploring the Meaning Behind Apagender}.
\newblock
\newblock
\urldef\tempurl%
\url{https://www.grindr.com/blog/gender-apathetic}
\showURL{%
\tempurl}


\bibitem[Haimson(2025)]%
        {haimson2025trans}
\bibfield{author}{\bibinfo{person}{Oliver~L. Haimson}.} \bibinfo{year}{2025}\natexlab{}.
\newblock \bibinfo{booktitle}{\emph{Trans Technologies}}.
\newblock \bibinfo{publisher}{MIT Press}, \bibinfo{address}{Cambridge, MA, USA}.
\newblock


\bibitem[Halberstam(2000)]%
        {halberstam2000tellingtalesbrandonteena}
\bibfield{author}{\bibinfo{person}{Jack Halberstam}.} \bibinfo{year}{2000}\natexlab{}.
\newblock \showarticletitle{Telling Tales: Brandon Teena, Billy Tipton, and Transgender Biography}.
\newblock \bibinfo{journal}{\emph{a/b: Auto/Biography Studies}} \bibinfo{volume}{15}, \bibinfo{number}{1} (\bibinfo{year}{2000}), \bibinfo{pages}{62--81}.
\newblock
\showISSN{0898-9575}


\bibitem[Halberstam(2017)]%
        {halberstam2017queergaminggaming}
\bibfield{author}{\bibinfo{person}{Jack Halberstam}.} \bibinfo{year}{2017}\natexlab{}.
\newblock \bibinfo{booktitle}{\emph{Queer gaming: Gaming, hacking, and going turbo}}.
\newblock \bibinfo{publisher}{University of Minnesota Press Minneapolis}, \bibinfo{address}{London, UK}, \bibinfo{pages}{187--200}.
\newblock


\bibitem[Hantsbarger et~al\mbox{.}(2022)]%
        {Hantsbarger_2022}
\bibfield{author}{\bibinfo{person}{Matthew Hantsbarger}, \bibinfo{person}{Giovanni~Maria Troiano}, \bibinfo{person}{Alexandra To}, {and} \bibinfo{person}{Casper Harteveld}.} \bibinfo{year}{2022}\natexlab{}.
\newblock \showarticletitle{Alienated Serendipity and Reflective Failure: Exploring Queer Game Mechanics and Queerness in Games via Queer Temporality}.
\newblock \bibinfo{journal}{\emph{Proceedings of the ACM on Human-Computer Interaction}} \bibinfo{volume}{6}, \bibinfo{number}{CHI PLAY} (\bibinfo{year}{2022}), \bibinfo{pages}{1–27}.
\newblock
\showISSN{2573-0142}
\urldef\tempurl%
\url{https://doi.org/10.1145/3549484}
\showDOI{\tempurl}


\bibitem[Haraway(1988)]%
        {Haraway_1988}
\bibfield{author}{\bibinfo{person}{Donna Haraway}.} \bibinfo{year}{1988}\natexlab{}.
\newblock \showarticletitle{Situated knowledges: The science question in feminism and the privilege of partial perspective}.
\newblock \bibinfo{journal}{\emph{Feminist Studies}} \bibinfo{volume}{14}, \bibinfo{number}{3} (\bibinfo{year}{1988}), \bibinfo{pages}{575}.
\newblock
\showISSN{0046-3663}
\urldef\tempurl%
\url{https://doi.org/10.2307/3178066}
\showDOI{\tempurl}


\bibitem[Hassenzahl(2010)]%
        {hassenzahl_experience_2010}
\bibfield{author}{\bibinfo{person}{Marc Hassenzahl}.} \bibinfo{year}{2010}\natexlab{}.
\newblock \showarticletitle{Experience design: {Technology} for all the right reasons}.
\newblock \bibinfo{journal}{\emph{Synthesis Lectures on Human-Centered Informatics}} \bibinfo{volume}{3}, \bibinfo{number}{1} (\bibinfo{date}{Jan.} \bibinfo{year}{2010}), \bibinfo{pages}{1--95}.
\newblock
\showISSN{1946-7680}
\urldef\tempurl%
\url{https://doi.org/10.2200/S00261ED1V01Y201003HCI008}
\showDOI{\tempurl}


\bibitem[Henrich et~al\mbox{.}(2010a)]%
        {henrich2010most}
\bibfield{author}{\bibinfo{person}{Joseph Henrich}, \bibinfo{person}{Steven~J Heine}, {and} \bibinfo{person}{Ara Norenzayan}.} \bibinfo{year}{2010}\natexlab{a}.
\newblock \showarticletitle{Most people are not WEIRD}.
\newblock \bibinfo{journal}{\emph{Nature}} \bibinfo{volume}{466}, \bibinfo{number}{7302} (\bibinfo{year}{2010}), \bibinfo{pages}{29--29}.
\newblock


\bibitem[Henrich et~al\mbox{.}(2010b)]%
        {henrich2010weirdest}
\bibfield{author}{\bibinfo{person}{Joseph Henrich}, \bibinfo{person}{Steven~J. Heine}, {and} \bibinfo{person}{Ara Norenzayan}.} \bibinfo{year}{2010}\natexlab{b}.
\newblock \showarticletitle{The weirdest people in the world?}
\newblock \bibinfo{journal}{\emph{Behavioral and Brain Sciences}} \bibinfo{volume}{33}, \bibinfo{number}{2-3} (\bibinfo{year}{2010}), \bibinfo{pages}{61--83}.
\newblock


\bibitem[Horton and Wohl(1956)]%
        {horton1956mass}
\bibfield{author}{\bibinfo{person}{Donald Horton} {and} \bibinfo{person}{R.~Richard Wohl}.} \bibinfo{year}{1956}\natexlab{}.
\newblock \showarticletitle{Mass Communication and Para-Social Interaction}.
\newblock \bibinfo{journal}{\emph{Psychiatry}} \bibinfo{volume}{19}, \bibinfo{number}{3} (\bibinfo{year}{1956}), \bibinfo{pages}{215--229}.
\newblock
\urldef\tempurl%
\url{https://doi.org/10.1080/00332747.1956.11023049}
\showDOI{\tempurl}


\bibitem[Huizinga(2000)]%
        {huizinga_homo_2000}
\bibfield{author}{\bibinfo{person}{J. Huizinga}.} \bibinfo{year}{2000}\natexlab{}.
\newblock \bibinfo{booktitle}{\emph{Homo {Ludens}: {A} {Study} of the {Play}-{Element} in {Culture}}}.
\newblock \bibinfo{publisher}{Routledge}, \bibinfo{address}{London, England}.
\newblock


\bibitem[Huta(2016)]%
        {Huta_2016}
\bibfield{author}{\bibinfo{person}{Veronika Huta}.} \bibinfo{year}{2016}\natexlab{}.
\newblock \bibinfo{booktitle}{\emph{Eudaimonic and Hedonic Orientations: Theoretical Considerations and Research Findings}}.
\newblock \bibinfo{publisher}{Springer International Publishing}, \bibinfo{address}{Cham, Switzerland}, \bibinfo{pages}{215–231}.
\newblock
\showISBNx{9783319424453}
\showISSN{2468-7235}
\urldef\tempurl%
\url{https://doi.org/10.1007/978-3-319-42445-3_15}
\showDOI{\tempurl}


\bibitem[Huta and Ryan(2010)]%
        {huta_pursuing_2010}
\bibfield{author}{\bibinfo{person}{Veronika Huta} {and} \bibinfo{person}{Richard~M. Ryan}.} \bibinfo{year}{2010}\natexlab{}.
\newblock \showarticletitle{Pursuing pleasure or virtue: {The} differential and overlapping well-being benefits of hedonic and eudaimonic motives}.
\newblock \bibinfo{journal}{\emph{Journal of Happiness Studies}} \bibinfo{volume}{11}, \bibinfo{number}{6} (\bibinfo{date}{Dec.} \bibinfo{year}{2010}), \bibinfo{pages}{735--762}.
\newblock
\showISSN{1389-4978, 1573-7780}
\urldef\tempurl%
\url{https://doi.org/10.1007/s10902-009-9171-4}
\showDOI{\tempurl}


\bibitem[Huta and Waterman(2013)]%
        {Huta_2013}
\bibfield{author}{\bibinfo{person}{Veronika Huta} {and} \bibinfo{person}{Alan~S. Waterman}.} \bibinfo{year}{2013}\natexlab{}.
\newblock \showarticletitle{Eudaimonia and Its Distinction from Hedonia: Developing a Classification and Terminology for Understanding Conceptual and Operational Definitions}.
\newblock \bibinfo{journal}{\emph{Journal of Happiness Studies}} \bibinfo{volume}{15}, \bibinfo{number}{6} (\bibinfo{date}{Dec.} \bibinfo{year}{2013}), \bibinfo{pages}{1425–1456}.
\newblock
\showISSN{1573-7780}
\urldef\tempurl%
\url{https://doi.org/10.1007/s10902-013-9485-0}
\showDOI{\tempurl}


\bibitem[Hyde et~al\mbox{.}(2019)]%
        {Hyde_2019sexgender}
\bibfield{author}{\bibinfo{person}{Janet~Shibley Hyde}, \bibinfo{person}{Rebecca~S. Bigler}, \bibinfo{person}{Daphna Joel}, \bibinfo{person}{Charlotte~Chucky Tate}, {and} \bibinfo{person}{Sari~M. van Anders}.} \bibinfo{year}{2019}\natexlab{}.
\newblock \showarticletitle{The future of sex and gender in psychology: Five challenges to the gender binary}.
\newblock \bibinfo{journal}{\emph{American Psychologist}} \bibinfo{volume}{74}, \bibinfo{number}{2} (\bibinfo{date}{Feb.} \bibinfo{year}{2019}), \bibinfo{pages}{171–193}.
\newblock
\showISSN{0003-066X}
\urldef\tempurl%
\url{https://doi.org/10.1037/amp0000307}
\showDOI{\tempurl}


\bibitem[Jaroszewski et~al\mbox{.}(2018)]%
        {Jaroszewski_2018}
\bibfield{author}{\bibinfo{person}{Samantha Jaroszewski}, \bibinfo{person}{Danielle Lottridge}, \bibinfo{person}{Oliver~L. Haimson}, {and} \bibinfo{person}{Katie Quehl}.} \bibinfo{year}{2018}\natexlab{}.
\newblock \showarticletitle{“Genderfluid” or “Attack Helicopter”: Responsible HCI Research Practice with Non-binary Gender Variation in Online Communities}. In \bibinfo{booktitle}{\emph{Proceedings of the 2018 CHI Conference on Human Factors in Computing Systems}} \emph{(\bibinfo{series}{CHI ’18})}. \bibinfo{publisher}{ACM}, \bibinfo{address}{New York, NY, USA}, \bibinfo{pages}{1–15}.
\newblock
\urldef\tempurl%
\url{https://doi.org/10.1145/3173574.3173881}
\showDOI{\tempurl}


\bibitem[Jensen(2011)]%
        {jensen2011othering}
\bibfield{author}{\bibinfo{person}{Sune~Qvotrup Jensen}.} \bibinfo{year}{2011}\natexlab{}.
\newblock \showarticletitle{Othering, identity formation and agency}.
\newblock \bibinfo{journal}{\emph{Qualitative Studies}} \bibinfo{volume}{2}, \bibinfo{number}{2} (\bibinfo{date}{Oct.} \bibinfo{year}{2011}), \bibinfo{pages}{63--78}.
\newblock
\urldef\tempurl%
\url{https://doi.org/10.7146/qs.v2i2.5510}
\showDOI{\tempurl}


\bibitem[Jenson et~al\mbox{.}(2010)]%
        {jenson2010gender}
\bibfield{author}{\bibinfo{person}{Jennifer Jenson} {et~al\mbox{.}}} \bibinfo{year}{2010}\natexlab{}.
\newblock \showarticletitle{Gender and digital gameplay: Theories, oversights, accidents, and surprises}.
\newblock In \bibinfo{booktitle}{\emph{Educational gameplay and simulation environments: Case studies and lessons learned}}. \bibinfo{publisher}{IGI Global}, \bibinfo{address}{Hershey, Pennsylvania, US}, \bibinfo{pages}{96--105}.
\newblock


\bibitem[J{\o}rgensen(2013)]%
        {gameworld-jorgensen}
\bibfield{author}{\bibinfo{person}{Kristine J{\o}rgensen}.} \bibinfo{year}{2013}\natexlab{}.
\newblock \bibinfo{booktitle}{\emph{Gameworld Interfaces}}.
\newblock \bibinfo{publisher}{The MIT Press}, \bibinfo{address}{Cambridge, Massachusetts, USA}.
\newblock
\urldef\tempurl%
\url{https://doi.org/10.7551/mitpress/9780262026864.001.0001}
\showDOI{\tempurl}


\bibitem[Jungselius and Weilenmann(2025)]%
        {tracing-change-chi-25}
\bibfield{author}{\bibinfo{person}{Beata Jungselius} {and} \bibinfo{person}{Alexandra Weilenmann}.} \bibinfo{year}{2025}\natexlab{}.
\newblock \showarticletitle{Tracing Change in Social Media Use: A Qualitative Longitudinal Study}. In \bibinfo{booktitle}{\emph{Proceedings of the 2025 CHI Conference on Human Factors in Computing Systems}} \emph{(\bibinfo{series}{CHI '25})}. \bibinfo{publisher}{Association for Computing Machinery}, \bibinfo{address}{New York, NY, USA}, Article \bibinfo{articleno}{957}, \bibinfo{numpages}{14}~pages.
\newblock
\showISBNx{9798400713941}
\urldef\tempurl%
\url{https://doi.org/10.1145/3706598.3713813}
\showDOI{\tempurl}


\bibitem[Kai and Devor(2022)]%
        {kai2022euphor}
\bibfield{author}{\bibinfo{person}{Jacobsen Kai} {and} \bibinfo{person}{Aaron Devor}.} \bibinfo{year}{2022}\natexlab{}.
\newblock \showarticletitle{Moving From Gender Dysphoria to Gender Euphoria}.
\newblock \bibinfo{journal}{\emph{Bulletin of Applied Transgender Studies}}  \bibinfo{volume}{1} (\bibinfo{year}{2022}), \bibinfo{pages}{119--143}.
\newblock
\urldef\tempurl%
\url{https://doi.org/10.57814/GGFG-4J14}
\showDOI{\tempurl}


\bibitem[Karim et~al\mbox{.}(2011)]%
        {Karim_2011panas}
\bibfield{author}{\bibinfo{person}{Jahanvash Karim}, \bibinfo{person}{Robert Weisz}, {and} \bibinfo{person}{Shafiq~Ur Rehman}.} \bibinfo{year}{2011}\natexlab{}.
\newblock \showarticletitle{International positive and negative affect schedule short-form (I-PANAS-SF): Testing for factorial invariance across cultures}.
\newblock \bibinfo{journal}{\emph{Procedia - Social and Behavioral Sciences}}  \bibinfo{volume}{15} (\bibinfo{year}{2011}), \bibinfo{pages}{2016–2022}.
\newblock
\showISSN{1877-0428}
\urldef\tempurl%
\url{https://doi.org/10.1016/j.sbspro.2011.04.046}
\showDOI{\tempurl}


\bibitem[Ketola et~al\mbox{.}(2022)]%
        {ketola2022identity}
\bibfield{author}{\bibinfo{person}{Morgan Ketola}, \bibinfo{person}{Schyler Selander}, {and} \bibinfo{person}{Ayalla Ruvio}.} \bibinfo{year}{2022}\natexlab{}.
\newblock \showarticletitle{Identity expressions of agender individuals in a digital world}.
\newblock In \bibinfo{booktitle}{\emph{The Routledge Handbook of Digital Consumption}}. \bibinfo{publisher}{Routledge}, \bibinfo{address}{New York}, \bibinfo{pages}{425--435}.
\newblock
\showISBNx{9781032329604}


\bibitem[Keyes(2018)]%
        {Keyes_2018}
\bibfield{author}{\bibinfo{person}{Os Keyes}.} \bibinfo{year}{2018}\natexlab{}.
\newblock \showarticletitle{The Misgendering Machines: Trans/HCI Implications of Automatic Gender Recognition}.
\newblock \bibinfo{journal}{\emph{Proceedings of the ACM on Human-Computer Interaction}} \bibinfo{volume}{2}, \bibinfo{number}{CSCW} (\bibinfo{date}{Nov.} \bibinfo{year}{2018}), \bibinfo{pages}{1–22}.
\newblock
\showISSN{2573-0142}
\urldef\tempurl%
\url{https://doi.org/10.1145/3274357}
\showDOI{\tempurl}


\bibitem[Kim et~al\mbox{.}(2014)]%
        {kim_pleasure_2014}
\bibfield{author}{\bibinfo{person}{Jinhyung Kim}, \bibinfo{person}{Pyungwon Kang}, {and} \bibinfo{person}{Incheol Choi}.} \bibinfo{year}{2014}\natexlab{}.
\newblock \showarticletitle{Pleasure now, meaning later: {Temporal} dynamics between pleasure and meaning}.
\newblock \bibinfo{journal}{\emph{Journal of Experimental Social Psychology}}  \bibinfo{volume}{55} (\bibinfo{date}{Nov.} \bibinfo{year}{2014}), \bibinfo{pages}{262--270}.
\newblock
\showISSN{0022-1031}
\urldef\tempurl%
\url{https://doi.org/10.1016/j.jesp.2014.07.018}
\showDOI{\tempurl}


\bibitem[Kivijärvi and Katila(2021)]%
        {Kivij_rvi_2021}
\bibfield{author}{\bibinfo{person}{Marke Kivijärvi} {and} \bibinfo{person}{Saija Katila}.} \bibinfo{year}{2021}\natexlab{}.
\newblock \showarticletitle{Becoming a Gamer: Performative Construction of Gendered Gamer Identities}.
\newblock \bibinfo{journal}{\emph{Games and Culture}} \bibinfo{volume}{17}, \bibinfo{number}{3} (\bibinfo{date}{Sept.} \bibinfo{year}{2021}), \bibinfo{pages}{461–481}.
\newblock
\showISSN{1555-4139}
\urldef\tempurl%
\url{https://doi.org/10.1177/15554120211042260}
\showDOI{\tempurl}


\bibitem[Kosciesza(2025)]%
        {kosciesza2025doing}
\bibfield{author}{\bibinfo{person}{Aiden~James Kosciesza}.} \bibinfo{year}{2025}\natexlab{}.
\newblock \showarticletitle{Doing gender in game spaces: Transgender and non-binary players’ gender signaling strategies in online games}.
\newblock \bibinfo{journal}{\emph{new media \& society}} \bibinfo{volume}{27}, \bibinfo{number}{1} (\bibinfo{year}{2025}), \bibinfo{pages}{5--23}.
\newblock
\urldef\tempurl%
\url{https://doi.org/10.1177/14614448231168107}
\showDOI{\tempurl}


\bibitem[Krobov{\'a} et~al\mbox{.}(2015)]%
        {krobova2015dressing}
\bibfield{author}{\bibinfo{person}{Tereza Krobov{\'a}}, \bibinfo{person}{Ond{\v{r}}ej Moravec}, {and} \bibinfo{person}{Jaroslav {\v{S}}velch}.} \bibinfo{year}{2015}\natexlab{}.
\newblock \showarticletitle{Dressing Commander Shepard in pink: Queer playing in a heteronormative game culture}.
\newblock \bibinfo{journal}{\emph{Cyberpsychology: Journal of Psychosocial Research on Cyberspace}} \bibinfo{volume}{9}, \bibinfo{number}{3} (\bibinfo{year}{2015}), \bibinfo{numpages}{14}~pages.
\newblock
\urldef\tempurl%
\url{https://doi.org/10.5817/CP2015-3-3}
\showDOI{\tempurl}


\bibitem[Kuss et~al\mbox{.}(2022)]%
        {Kuss_2022femgamer}
\bibfield{author}{\bibinfo{person}{Daria~J. Kuss}, \bibinfo{person}{Anne~Marie Kristensen}, \bibinfo{person}{A.~Jess Williams}, {and} \bibinfo{person}{Olatz Lopez-Fernandez}.} \bibinfo{year}{2022}\natexlab{}.
\newblock \showarticletitle{To Be or Not to Be a Female Gamer: A Qualitative Exploration of Female Gamer Identity}.
\newblock \bibinfo{journal}{\emph{International Journal of Environmental Research and Public Health}} \bibinfo{volume}{19}, \bibinfo{number}{3} (\bibinfo{date}{Jan.} \bibinfo{year}{2022}), \bibinfo{pages}{1169}.
\newblock
\showISSN{1660-4601}
\urldef\tempurl%
\url{https://doi.org/10.3390/ijerph19031169}
\showDOI{\tempurl}


\bibitem[Lambrou et~al\mbox{.}(2020)]%
        {lambrou2020learning}
\bibfield{author}{\bibinfo{person}{Nickolas~H. Lambrou}, \bibinfo{person}{Katherine~M. Cochran}, \bibinfo{person}{Samantha Everhart}, \bibinfo{person}{Jason~D. Flatt}, \bibinfo{person}{Megan Zuelsdorff}, \bibinfo{person}{John~B. O'Hara}, \bibinfo{person}{Lance Weinhardt}, \bibinfo{person}{Susan Flowers~Benton}, {and} \bibinfo{person}{Carey~E. Gleason}.} \bibinfo{year}{2020}\natexlab{}.
\newblock \showarticletitle{Learning from transmasculine experiences with health care: Tangible inlets for reducing health disparities through patient--provider relationships}.
\newblock \bibinfo{journal}{\emph{Transgender Health}} \bibinfo{volume}{5}, \bibinfo{number}{1} (\bibinfo{year}{2020}), \bibinfo{pages}{18--32}.
\newblock
\urldef\tempurl%
\url{https://doi.org/10.1089/trgh.2019.0054}
\showDOI{\tempurl}
\newblock
\shownote{PMID: 32322685}.


\bibitem[Lawrence(2018)]%
        {lawrence2018if}
\bibfield{author}{\bibinfo{person}{Chris Lawrence}.} \bibinfo{year}{2018}\natexlab{}.
\newblock \showarticletitle{What if Zelda wasn't a girl? Problematizing \emph{Ocarina of Time's} great gender debate}.
\newblock In \bibinfo{booktitle}{\emph{Queerness in Play}}. \bibinfo{publisher}{Palgrave Macmillan, Cham}, \bibinfo{address}{Cham, Switzerland}, \bibinfo{pages}{97--113}.
\newblock
\urldef\tempurl%
\url{https://doi.org/10.1007/978-3-319-90542-6_6}
\showDOI{\tempurl}


\bibitem[Lev(2013)]%
        {lev2013gender}
\bibfield{author}{\bibinfo{person}{Arlene~Istar Lev}.} \bibinfo{year}{2013}\natexlab{}.
\newblock \showarticletitle{Gender dysphoria: Two steps forward, one step back}.
\newblock \bibinfo{journal}{\emph{Clinical social work journal}} \bibinfo{volume}{41}, \bibinfo{number}{3} (\bibinfo{year}{2013}), \bibinfo{pages}{288--296}.
\newblock
\showISSN{1573-3343}
\urldef\tempurl%
\url{https://doi.org/10.1007/s10615-013-0447-0}
\showDOI{\tempurl}


\bibitem[Liang et~al\mbox{.}(2025a)]%
        {liang-2025-euphoria}
\bibfield{author}{\bibinfo{person}{Shano Liang}, \bibinfo{person}{Michelle~V Cormier}, \bibinfo{person}{Rose Bohrer}, {and} \bibinfo{person}{Phoebe~O. Toups~Dugas}.} \bibinfo{year}{2025}\natexlab{a}.
\newblock \showarticletitle{Designed \& Discovered Euphoria: Insights from Trans-Femme Players' Experiences of Gender Euphoria in Video Games}. In \bibinfo{booktitle}{\emph{Proceedings of the 2025 CHI Conference on Human Factors in Computing Systems}} \emph{(\bibinfo{series}{CHI '25})}. \bibinfo{publisher}{Association for Computing Machinery}, \bibinfo{address}{New York, NY, USA}, Article \bibinfo{articleno}{104}, \bibinfo{numpages}{21}~pages.
\newblock
\showISBNx{9798400713941}
\urldef\tempurl%
\url{https://doi.org/10.1145/3706598.3714081}
\showDOI{\tempurl}


\bibitem[Liang et~al\mbox{.}(2023)]%
        {Liang2023misrepresentation}
\bibfield{author}{\bibinfo{person}{Shano Liang}, \bibinfo{person}{Michelle~V. Cormier}, \bibinfo{person}{Phoebe~O. Toups~Dugas}, {and} \bibinfo{person}{Rose Bohrer}.} \bibinfo{year}{2023}\natexlab{}.
\newblock \showarticletitle{Analyzing Trans (Mis)Representation in Video Games to Remediate Gender Dysphoria Triggers}.
\newblock \bibinfo{journal}{\emph{Proc. ACM Hum.-Comput. Interact.}} \bibinfo{volume}{7}, \bibinfo{number}{CHI PLAY}, Article \bibinfo{articleno}{388} (\bibinfo{date}{oct} \bibinfo{year}{2023}), \bibinfo{numpages}{33}~pages.
\newblock
\urldef\tempurl%
\url{https://doi.org/10.1145/3611034}
\showDOI{\tempurl}


\bibitem[Liang et~al\mbox{.}(2025b)]%
        {liang-2025-three-steps}
\bibfield{author}{\bibinfo{person}{Shano Liang}, \bibinfo{person}{Michelle~V Cormier}, \bibinfo{person}{Phoebe~O. Toups~Dugas}, {and} \bibinfo{person}{Rose Bohrer}.} \bibinfo{year}{2025}\natexlab{b}.
\newblock \showarticletitle{The Three Steps to Trans Death: Introducing Trans Cyber-Necropolitics in Digital Media}.
\newblock \bibinfo{journal}{\emph{ACM Trans. Comput.-Hum. Interact.}} (\bibinfo{date}{June} \bibinfo{year}{2025}), \bibinfo{numpages}{34}~pages.
\newblock
\showISSN{1073-0516}
\urldef\tempurl%
\url{https://doi.org/10.1145/3745767}
\showDOI{\tempurl}


\bibitem[Linxen et~al\mbox{.}(2021)]%
        {linxen2021weird}
\bibfield{author}{\bibinfo{person}{Sebastian Linxen}, \bibinfo{person}{Christian Sturm}, \bibinfo{person}{Florian Br\"{u}hlmann}, \bibinfo{person}{Vincent Cassau}, \bibinfo{person}{Klaus Opwis}, {and} \bibinfo{person}{Katharina Reinecke}.} \bibinfo{year}{2021}\natexlab{}.
\newblock \showarticletitle{How WEIRD is CHI?}. In \bibinfo{booktitle}{\emph{Proceedings of the 2021 CHI Conference on Human Factors in Computing Systems}} (Yokohama, Japan) \emph{(\bibinfo{series}{CHI '21})}. \bibinfo{publisher}{Association for Computing Machinery}, \bibinfo{address}{New York, NY, USA}, Article \bibinfo{articleno}{143}, \bibinfo{numpages}{14}~pages.
\newblock
\showISBNx{9781450380966}
\urldef\tempurl%
\url{https://doi.org/10.1145/3411764.3445488}
\showDOI{\tempurl}


\bibitem[Livingston et~al\mbox{.}(2014)]%
        {Livingston2014how}
\bibfield{author}{\bibinfo{person}{Ian~J. Livingston}, \bibinfo{person}{Carl Gutwin}, \bibinfo{person}{Regan~L. Mandryk}, {and} \bibinfo{person}{Max Birk}.} \bibinfo{year}{2014}\natexlab{}.
\newblock \showarticletitle{How Players Value Their Characters in World of Warcraft}. In \bibinfo{booktitle}{\emph{Proceedings of the 17th ACM Conference on Computer Supported Cooperative Work \& Social Computing}} \emph{(\bibinfo{series}{CSCW '14})}. \bibinfo{publisher}{ACM}, \bibinfo{address}{New York, NY, USA}, \bibinfo{pages}{1333--1343}.
\newblock


\bibitem[Maletska(2024)]%
        {maletska2024queer}
\bibfield{author}{\bibinfo{person}{Mark Maletska}.} \bibinfo{year}{2024}\natexlab{}.
\newblock \showarticletitle{Queer Players' Strategies for Queering Character Interactions}. In \bibinfo{booktitle}{\emph{Conference Proceedings of DiGRA 2024 Conference: Playgrounds}}. \bibinfo{publisher}{DiGRA}, \bibinfo{address}{Tampere}, \bibinfo{numpages}{22}~pages.
\newblock
\urldef\tempurl%
\url{https://dl.digra.org/index.php/dl/article/view/2239}
\showURL{%
\tempurl}


\bibitem[McArthur et~al\mbox{.}(2015)]%
        {McArthur-avatar-affordances-2015}
\bibfield{author}{\bibinfo{person}{Victoria McArthur}, \bibinfo{person}{Robert~John Teather}, {and} \bibinfo{person}{Jennifer Jenson}.} \bibinfo{year}{2015}\natexlab{}.
\newblock \showarticletitle{The Avatar Affordances Framework: Mapping Affordances and Design Trends in Character Creation Interfaces}. In \bibinfo{booktitle}{\emph{Proceedings of the 2015 Annual Symposium on Computer-Human Interaction in Play}} (London, United Kingdom) \emph{(\bibinfo{series}{CHI PLAY '15})}. \bibinfo{publisher}{Association for Computing Machinery}, \bibinfo{address}{New York, NY, USA}, \bibinfo{pages}{231–240}.
\newblock
\showISBNx{9781450334662}
\urldef\tempurl%
\url{https://doi.org/10.1145/2793107.2793121}
\showDOI{\tempurl}


\bibitem[Mekler and Hornb{\ae}k(2016)]%
        {mekler_momentary_2016}
\bibfield{author}{\bibinfo{person}{Elisa~D. Mekler} {and} \bibinfo{person}{Kasper Hornb{\ae}k}.} \bibinfo{year}{2016}\natexlab{}.
\newblock \showarticletitle{Momentary pleasure or lasting meaning? {Distinguishing} eudaimonic and hedonic user experiences}. In \bibinfo{booktitle}{\emph{Proceedings of the 2016 {CHI} {Conference} on {Human} {Factors} in {Computing} {Systems}}} \emph{(\bibinfo{series}{{CHI} '16})}. \bibinfo{publisher}{Association for Computing Machinery}, \bibinfo{address}{New York, NY, USA}, \bibinfo{pages}{4509--4520}.
\newblock
\showISBNx{978-1-4503-3362-7}
\urldef\tempurl%
\url{https://doi.org/10.1145/2858036.2858225}
\showDOI{\tempurl}


\bibitem[{@moistmogai}(2023)]%
        {moistmogai2023agenderandadjacent}
\bibfield{author}{\bibinfo{person}{{@moistmogai}}.} \bibinfo{year}{2023}\natexlab{}.
\newblock \bibinfo{title}{Agender-Adjacent Genders}.
\newblock
\newblock
\urldef\tempurl%
\url{https://archive.ph/Ay4bw}
\showURL{%
\tempurl}


\bibitem[Morgan et~al\mbox{.}(2020)]%
        {Morgan2020avatar}
\bibfield{author}{\bibinfo{person}{Helen Morgan}, \bibinfo{person}{Amanda O’Donovan}, \bibinfo{person}{Renita Almeida}, \bibinfo{person}{Ashleigh Lin}, {and} \bibinfo{person}{Yael Perry}.} \bibinfo{year}{2020}\natexlab{}.
\newblock \showarticletitle{The Role of the Avatar in Gaming for Trans and Gender Diverse Young People}.
\newblock \bibinfo{journal}{\emph{International Journal of Environmental Research and Public Health}} \bibinfo{volume}{17}, \bibinfo{number}{22} (\bibinfo{date}{Nov.} \bibinfo{year}{2020}), \bibinfo{pages}{8617}.
\newblock
\showISSN{1660-4601}
\urldef\tempurl%
\url{https://doi.org/10.3390/ijerph17228617}
\showDOI{\tempurl}


\bibitem[M{\"u}ller et~al\mbox{.}(2015)]%
        {muller_facets_2015}
\bibfield{author}{\bibinfo{person}{Livia~J. M{\"u}ller}, \bibinfo{person}{Elisa~D. Mekler}, {and} \bibinfo{person}{Klaus Opwis}.} \bibinfo{year}{2015}\natexlab{}.
\newblock \showarticletitle{Facets in {HCI}: {Towards} understanding eudaimonic {UX} -- preliminary findings}. In \bibinfo{booktitle}{\emph{Proceedings of the 33rd {Annual} {ACM} {Conference} {Extended} {Abstracts} on {Human} {Factors} in {Computing} {Systems}}} \emph{(\bibinfo{series}{{CHI} {EA} '15})}. \bibinfo{publisher}{Association for Computing Machinery}, \bibinfo{address}{New York, NY, USA}, \bibinfo{pages}{2283--2288}.
\newblock
\showISBNx{978-1-4503-3146-3}
\urldef\tempurl%
\url{https://doi.org/10.1145/2702613.2732836}
\showDOI{\tempurl}


\bibitem[Mu{\~n}oz(2013)]%
        {munoz2013disidentifications}
\bibfield{author}{\bibinfo{person}{Jos{\'e}~Esteban Mu{\~n}oz}.} \bibinfo{year}{2013}\natexlab{}.
\newblock \bibinfo{booktitle}{\emph{Disidentifications: Queers of color and the performance of politics}}. Vol.~\bibinfo{volume}{2}.
\newblock \bibinfo{publisher}{University of Minnesota Press}, \bibinfo{address}{Minneapolis, MN, USA}.
\newblock


\bibitem[Mu{\~n}oz(2020)]%
        {munoz2019cruising}
\bibfield{author}{\bibinfo{person}{Jos{\'e}~Esteban Mu{\~n}oz}.} \bibinfo{year}{2020}\natexlab{}.
\newblock \bibinfo{booktitle}{\emph{Cruising Utopia, 10th Anniversary Edition: The Then and There of Queer Futurity}}.
\newblock \bibinfo{publisher}{New York University Press}, \bibinfo{address}{New York, NY, USA}.
\newblock
\showISBNx{9781479868780}
\urldef\tempurl%
\url{https://doi.org/10.18574/nyu/9781479868780.001.0001}
\showDOI{\tempurl}


\bibitem[Murray(2021)]%
        {murray2021video}
\bibfield{author}{\bibinfo{person}{Soraya Murray}.} \bibinfo{year}{2021}\natexlab{}.
\newblock \bibinfo{booktitle}{\emph{On video games: The visual politics of race, gender and space}}. Vol.~\bibinfo{volume}{27}.
\newblock \bibinfo{publisher}{Bloomsbury Publishing}, \bibinfo{address}{New York, NY, USA}.
\newblock


\bibitem[Nakamura(2012)]%
        {nakamura2012queer}
\bibfield{author}{\bibinfo{person}{Lisa Nakamura}.} \bibinfo{year}{2012}\natexlab{}.
\newblock \showarticletitle{Queer female of color: The highest difficulty setting there is? {G}aming rhetoric as gender capital}.
\newblock \bibinfo{journal}{\emph{Ada: A Journal of Gender, New Media, and Technology}} (\bibinfo{year}{2012}), \bibinfo{numpages}{5}~pages.
\newblock
Issue 1.
\showISSN{2325-0496}
\urldef\tempurl%
\url{https://doi.org/10.7264/N37P8W9V}
\showDOI{\tempurl}


\bibitem[Nakashima et~al\mbox{.}(2012)]%
        {nakashima2012group}
\bibfield{author}{\bibinfo{person}{Ken'ichiro Nakashima}, \bibinfo{person}{Chikae Isobe}, {and} \bibinfo{person}{Mitsuhiro Ura}.} \bibinfo{year}{2012}\natexlab{}.
\newblock \showarticletitle{In-group representation and social value affect the use of in-group identification for maintaining and enhancing self-evaluation}.
\newblock \bibinfo{journal}{\emph{Asian Journal of Social Psychology}} \bibinfo{volume}{15}, \bibinfo{number}{1} (\bibinfo{year}{2012}), \bibinfo{pages}{49--59}.
\newblock
\urldef\tempurl%
\url{https://doi.org/10.1111/j.1467-839X.2011.01361.x}
\showDOI{\tempurl}


\bibitem[Pace et~al\mbox{.}(2009)]%
        {tyler-mmorpg}
\bibfield{author}{\bibinfo{person}{Tyler Pace}, \bibinfo{person}{Aaron Houssian}, {and} \bibinfo{person}{Victoria McArthur}.} \bibinfo{year}{2009}\natexlab{}.
\newblock \showarticletitle{Are socially exclusive values embedded in the avatar creation interfaces of MMORPGs?}
\newblock \bibinfo{journal}{\emph{Journal of Information, Communication and Ethics in Society}} \bibinfo{volume}{7}, \bibinfo{number}{2/3} (\bibinfo{date}{2025/02/13} \bibinfo{year}{2009}), \bibinfo{pages}{192--210}.
\newblock
\showISBNx{1477-996X}
\urldef\tempurl%
\url{https://doi.org/10.1108/14779960910955909}
\showDOI{\tempurl}


\bibitem[Papacharissi(2012)]%
        {papacharissi2012without}
\bibfield{author}{\bibinfo{person}{Zizi Papacharissi}.} \bibinfo{year}{2012}\natexlab{}.
\newblock \showarticletitle{Without you, I'm nothing: Performances of the self on {Twitter}}.
\newblock \bibinfo{journal}{\emph{International Journal of Communication}}  \bibinfo{volume}{6} (\bibinfo{year}{2012}), \bibinfo{pages}{18}.
\newblock


\bibitem[Papisova(2020)]%
        {papisova2016wahtitmeans}
\bibfield{author}{\bibinfo{person}{Vera Papisova}.} \bibinfo{year}{2020}\natexlab{}.
\newblock \bibinfo{title}{What It Means to Identify as Agender}.
\newblock
\newblock
\urldef\tempurl%
\url{https://www.teenvogue.com/story/what-is-agender}
\showURL{%
\tempurl}


\bibitem[{PBS Independent Lens}(2015)]%
        {gendermap}
\bibfield{author}{\bibinfo{person}{{PBS Independent Lens}}.} \bibinfo{year}{2015}\natexlab{}.
\newblock \bibinfo{title}{A map of gender-diverse cultures}.
\newblock \bibinfo{howpublished}{Annotated Map}.
\newblock
\urldef\tempurl%
\url{https://www.pbs.org/independentlens/content/two-spirits_map-html}
\showURL{%
\tempurl}


\bibitem[Phillips(2017)]%
        {phillips2017welcome}
\bibfield{author}{\bibinfo{person}{Amanda Phillips}.} \bibinfo{year}{2017}\natexlab{}.
\newblock \bibinfo{booktitle}{\emph{Welcome to my fantasy zone: Bayonetta and queer femme disturbance}}.
\newblock \bibinfo{publisher}{University of Minnesota Press}, \bibinfo{address}{London, UK}, Chapter~12, \bibinfo{pages}{109--123}.
\newblock


\bibitem[Pozo(2018)]%
        {pozo2018queer}
\bibfield{author}{\bibinfo{person}{Teddy Pozo}.} \bibinfo{year}{2018}\natexlab{}.
\newblock \showarticletitle{Queer games after empathy: Feminism and haptic game design aesthetics from consent to cuteness to the radically soft}.
\newblock \bibinfo{journal}{\emph{Game Studies}} \bibinfo{volume}{18}, \bibinfo{number}{3} (\bibinfo{year}{2018}).
\newblock
\showISSN{1604-7982}
\urldef\tempurl%
\url{https://gamestudies.org/1803/articles/pozo}
\showURL{%
\tempurl}


\bibitem[Proudfoot(2022)]%
        {Proudfoot_2022}
\bibfield{author}{\bibinfo{person}{Kevin Proudfoot}.} \bibinfo{year}{2022}\natexlab{}.
\newblock \showarticletitle{Inductive/Deductive Hybrid Thematic Analysis in Mixed Methods Research}.
\newblock \bibinfo{journal}{\emph{Journal of Mixed Methods Research}} \bibinfo{volume}{17}, \bibinfo{number}{3} (\bibinfo{date}{Sept.} \bibinfo{year}{2022}), \bibinfo{pages}{308–326}.
\newblock
\showISSN{1558-6901}
\urldef\tempurl%
\url{https://doi.org/10.1177/15586898221126816}
\showDOI{\tempurl}


\bibitem[Pulos(2013)]%
        {pulos2013confronting}
\bibfield{author}{\bibinfo{person}{Alexis Pulos}.} \bibinfo{year}{2013}\natexlab{}.
\newblock \showarticletitle{Confronting Heteronormativity in Online Games: A Critical Discourse Analysis of LGBTQ Sexuality in World of Warcraft}.
\newblock \bibinfo{journal}{\emph{Games and Culture}} \bibinfo{volume}{8}, \bibinfo{number}{2} (\bibinfo{year}{2013}), \bibinfo{pages}{77--97}.
\newblock
\urldef\tempurl%
\url{https://doi.org/10.1177/1555412013478688}
\showDOI{\tempurl}


\bibitem[Purdie-Vaughns et~al\mbox{.}(2008)]%
        {purdie2008social}
\bibfield{author}{\bibinfo{person}{Valerie Purdie-Vaughns}, \bibinfo{person}{Claude~M Steele}, \bibinfo{person}{Paul~G Davies}, \bibinfo{person}{Ruth Ditlmann}, {and} \bibinfo{person}{Jennifer~Randall Crosby}.} \bibinfo{year}{2008}\natexlab{}.
\newblock \showarticletitle{Social identity contingencies: how diversity cues signal threat or safety for African Americans in mainstream institutions.}
\newblock \bibinfo{journal}{\emph{Journal of Personality and Social Psychology}} \bibinfo{volume}{94}, \bibinfo{number}{4} (\bibinfo{year}{2008}), \bibinfo{pages}{615--630}.
\newblock
\showISSN{1939-1315(Electronic),0022-3514(Print)}
\urldef\tempurl%
\url{https://doi.org/10.1037/0022-3514.94.4.615}
\showDOI{\tempurl}


\bibitem[Rahilly(2022)]%
        {Rahilly_2022}
\bibfield{author}{\bibinfo{person}{Elizabeth Rahilly}.} \bibinfo{year}{2022}\natexlab{}.
\newblock \showarticletitle{“WellDuh, That’s How You Raise a Kid”: Gender-Open Parenting in a (Non)Binary World}.
\newblock \bibinfo{journal}{\emph{LGBTQ+ Family: An Interdisciplinary Journal}} \bibinfo{volume}{18}, \bibinfo{number}{3} (\bibinfo{date}{May} \bibinfo{year}{2022}), \bibinfo{pages}{262–280}.
\newblock
\showISSN{2770-338X}
\urldef\tempurl%
\url{https://doi.org/10.1080/27703371.2022.2089309}
\showDOI{\tempurl}


\bibitem[Rao and Donaldson(2015)]%
        {rao2015expanding}
\bibfield{author}{\bibinfo{person}{Meghana~A Rao} {and} \bibinfo{person}{Stewart~I Donaldson}.} \bibinfo{year}{2015}\natexlab{}.
\newblock \showarticletitle{Expanding opportunities for diversity in positive psychology: An examination of gender, race, and ethnicity.}
\newblock \bibinfo{journal}{\emph{Canadian Psychology/Psychologie Canadienne}} \bibinfo{volume}{56}, \bibinfo{number}{3} (\bibinfo{year}{2015}), \bibinfo{pages}{271--282}.
\newblock
\showISSN{1878-7304(Electronic),0708-5591(Print)}
\urldef\tempurl%
\url{https://doi.org/10.1037/cap0000036}
\showDOI{\tempurl}


\bibitem[Reisner et~al\mbox{.}(2023)]%
        {reisner2023exploring}
\bibfield{author}{\bibinfo{person}{Sari~L Reisner}, \bibinfo{person}{David~R Pletta}, \bibinfo{person}{Alexander Harris}, \bibinfo{person}{Juwan Campbell}, \bibinfo{person}{Andrew Asquith}, \bibinfo{person}{Dana~J Pardee}, \bibinfo{person}{Madeline~B Deutsch}, \bibinfo{person}{Rodrigo Aguayo-Romero}, \bibinfo{person}{Meg Quint}, \bibinfo{person}{Alex~S Keuroghlian}, {et~al\mbox{.}}} \bibinfo{year}{2023}\natexlab{}.
\newblock \showarticletitle{Exploring gender euphoria in a sample of transgender and gender diverse patients at two US urban community health centers}.
\newblock \bibinfo{journal}{\emph{Psychiatry Research}}  \bibinfo{volume}{329} (\bibinfo{year}{2023}), \bibinfo{pages}{115541}.
\newblock
\showISSN{0165-1781}
\urldef\tempurl%
\url{https://doi.org/10.1016/j.psychres.2023.115541}
\showDOI{\tempurl}


\bibitem[Reyes and Fisher(2022)]%
        {reyes2022theimpacts}
\bibfield{author}{\bibinfo{person}{Zoey Reyes} {and} \bibinfo{person}{Joshua Fisher}.} \bibinfo{year}{2022}\natexlab{}.
\newblock \showarticletitle{The Impacts of Virtual Reality Avatar Creation and Embodiment on Transgender and Genderqueer Individuals in Games: A Grounded Theory Analysis of Survey and Interview Data from Transgender and Genderqueer Individuals about Their Experiences with Avatar Creation Interfaces in Virtual Reality}. In \bibinfo{booktitle}{\emph{Proceedings of the 17th International Conference on the Foundations of Digital Games}} (Athens, Greece) \emph{(\bibinfo{series}{FDG '22})}. \bibinfo{publisher}{Association for Computing Machinery}, \bibinfo{address}{New York, NY, USA}, Article \bibinfo{articleno}{25}, \bibinfo{numpages}{9}~pages.
\newblock
\showISBNx{9781450397957}
\urldef\tempurl%
\url{https://doi.org/10.1145/3555858.3555882}
\showDOI{\tempurl}


\bibitem[Richard and Gray(2018)]%
        {Gabriela_T_Richard_2018}
\bibfield{author}{\bibinfo{person}{Gabriela~T. Richard} {and} \bibinfo{person}{Kishonna~L. Gray}.} \bibinfo{year}{2018}\natexlab{}.
\newblock \showarticletitle{Gendered Play, Racialized Reality: Black Cyberfeminism, Inclusive Communities of Practice, and the Intersections of Learning, Socialization, and Resilience in Online Gaming}.
\newblock \bibinfo{journal}{\emph{Frontiers: A Journal of Women Studies}} \bibinfo{volume}{39}, \bibinfo{number}{1} (\bibinfo{year}{2018}), \bibinfo{pages}{112}.
\newblock
\showISSN{0160-9009}
\urldef\tempurl%
\url{https://doi.org/10.5250/fronjwomestud.39.1.0112}
\showDOI{\tempurl}


\bibitem[Richards et~al\mbox{.}(2016)]%
        {richards2016non}
\bibfield{author}{\bibinfo{person}{Christina Richards}, \bibinfo{person}{Walter~Pierre Bouman}, \bibinfo{person}{Leighton Seal}, \bibinfo{person}{Meg~John Barker}, \bibinfo{person}{Timo~O. Nieder}, {and} \bibinfo{person}{Guy T’Sjoen}.} \bibinfo{year}{2016}\natexlab{}.
\newblock \showarticletitle{Non-binary or genderqueer genders}.
\newblock \bibinfo{journal}{\emph{International Review of Psychiatry}} \bibinfo{volume}{28}, \bibinfo{number}{1} (\bibinfo{year}{2016}), \bibinfo{pages}{95--102}.
\newblock
\urldef\tempurl%
\url{https://doi.org/10.3109/09540261.2015.1106446}
\showDOI{\tempurl}


\bibitem[Rode et~al\mbox{.}(2025)]%
        {rode-2025-reframing-gender}
\bibfield{author}{\bibinfo{person}{Jennifer~A. Rode}, \bibinfo{person}{Phoebe~O. Toups~Dugas}, {and} \bibinfo{person}{Andruid Kerne}.} \bibinfo{year}{2025}\natexlab{}.
\newblock \showarticletitle{Reframing Diversity in Computing on the Basis of Genders}.
\newblock \bibinfo{journal}{\emph{Computer Supported Cooperative Work (CSCW)}}  \bibinfo{volume}{34} (\bibinfo{year}{2025}), \bibinfo{pages}{785--833}.
\newblock
\showISBNx{1573-7551}
\urldef\tempurl%
\url{https://doi.org/10.1007/s10606-025-09518-0}
\showDOI{\tempurl}


\bibitem[Ruberg(2015)]%
        {ruberg2015nofun}
\bibfield{author}{\bibinfo{person}{Bo Ruberg}.} \bibinfo{year}{2015}\natexlab{}.
\newblock \showarticletitle{{No Fun: The Queer Potential of Video Games that Annoy, Anger, Disappoint, Sadden, and Hurt}}.
\newblock \bibinfo{journal}{\emph{QED: A Journal in GLBTQ Worldmaking}} \bibinfo{volume}{2}, \bibinfo{number}{2} (\bibinfo{date}{06} \bibinfo{year}{2015}), \bibinfo{pages}{108--124}.
\newblock
\showISSN{2327-1574}
\urldef\tempurl%
\url{https://doi.org/10.14321/qed.2.2.0108}
\showDOI{\tempurl}


\bibitem[Ruberg(2019)]%
        {ruberg2019videogameshavealwaysbeenqueer}
\bibfield{author}{\bibinfo{person}{Bonnie Ruberg}.} \bibinfo{year}{2019}\natexlab{}.
\newblock \bibinfo{booktitle}{\emph{Video games have always been queer}}.
\newblock \bibinfo{publisher}{nyu Press}, \bibinfo{address}{New York, US}.
\newblock
\showISBNx{1479843741}


\bibitem[Ruberg(2020)]%
        {ruberg2020thequeergamesavant}
\bibfield{author}{\bibinfo{person}{Bo Ruberg}.} \bibinfo{year}{2020}\natexlab{}.
\newblock \bibinfo{booktitle}{\emph{The queer games avant-garde: How LGBTQ game makers are reimagining the medium of video games}}.
\newblock \bibinfo{publisher}{Duke University Press}, \bibinfo{address}{Duharm, NC, USA}.
\newblock
\showISBNx{1478007303}


\bibitem[Ruberg(2022)]%
        {Ruberg_2022post}
\bibfield{author}{\bibinfo{person}{Bo Ruberg}.} \bibinfo{year}{2022}\natexlab{}.
\newblock \showarticletitle{After agency: The queer posthumanism of video games that cannot be played}.
\newblock \bibinfo{journal}{\emph{Convergence: The International Journal of Research into New Media Technologies}} \bibinfo{volume}{28}, \bibinfo{number}{2} (\bibinfo{date}{April} \bibinfo{year}{2022}), \bibinfo{pages}{413–430}.
\newblock
\showISSN{1748-7382}
\urldef\tempurl%
\url{https://doi.org/10.1177/13548565221094257}
\showDOI{\tempurl}


\bibitem[Ruberg(2025)]%
        {ruberg2025queer}
\bibfield{author}{\bibinfo{person}{Bo Ruberg}.} \bibinfo{year}{2025}\natexlab{}.
\newblock \bibinfo{booktitle}{\emph{How to Queer the World: Radical Worldbuilding through Video Games}}.
\newblock \bibinfo{publisher}{NYU Press}, \bibinfo{address}{New York, NY, USA}.
\newblock


\bibitem[Ruberg and Shaw(2017)]%
        {ruberg2017queergamestudies}
\bibfield{author}{\bibinfo{person}{Bonnie Ruberg} {and} \bibinfo{person}{Adrienne Shaw}.} \bibinfo{year}{2017}\natexlab{}.
\newblock \bibinfo{booktitle}{\emph{Queer game studies}}.
\newblock \bibinfo{publisher}{University of Minnesota Press}, \bibinfo{address}{London, UK}.
\newblock
\showISBNx{1452954631}


\bibitem[Salen~Tekinba{\c s} and Zimmerman(2003)]%
        {rules-of-play}
\bibfield{author}{\bibinfo{person}{Katie Salen~Tekinba{\c s}} {and} \bibinfo{person}{Eric Zimmerman}.} \bibinfo{year}{2003}\natexlab{}.
\newblock \bibinfo{booktitle}{\emph{Rules of Play: Game Design Fundamentals}}.
\newblock \bibinfo{publisher}{MIT Press}, \bibinfo{address}{Cambridge, MA, USA}.
\newblock


\bibitem[Scheuerman et~al\mbox{.}(2020)]%
        {hcigender}
\bibfield{author}{\bibinfo{person}{Morgan~Klaus Scheuerman}, \bibinfo{person}{Katta Spiel}, \bibinfo{person}{Oliver~L. Haimson}, \bibinfo{person}{Foad Hamidi}, {and} \bibinfo{person}{Stacy~M. Branham}.} \bibinfo{year}{2020}\natexlab{}.
\newblock \bibinfo{title}{HCI Guidelines for Gender Equity and Inclusivity}.
\newblock
\newblock
\urldef\tempurl%
\url{https://www.morgan-klaus.com/gender-guidelines.html}
\showURL{%
\tempurl}


\bibitem[Seaborn(2023)]%
        {seaborn2023link}
\bibfield{author}{\bibinfo{person}{Katie Seaborn}.} \bibinfo{year}{2023}\natexlab{}.
\newblock \showarticletitle{Link, user-centred designer: Game characters as transcendent models}. In \bibinfo{booktitle}{\emph{Companion Proceedings of the Annual Symposium on Computer-Human Interaction in Play}} (Stratford, ON, Canada) \emph{(\bibinfo{series}{CHI PLAY Companion '23})}. \bibinfo{publisher}{Association for Computing Machinery}, \bibinfo{address}{New York, NY, USA}, \bibinfo{pages}{238--240}.
\newblock
\urldef\tempurl%
\url{https://doi.org/10.1145/3573382.3616052}
\showDOI{\tempurl}


\bibitem[Seaborn et~al\mbox{.}(2023)]%
        {seaborn2023not}
\bibfield{author}{\bibinfo{person}{Katie Seaborn}, \bibinfo{person}{Giulia Barbareschi}, {and} \bibinfo{person}{Shruti Chandra}.} \bibinfo{year}{2023}\natexlab{}.
\newblock \showarticletitle{Not Only WEIRD but ``Uncanny''? A Systematic Review of Diversity in Human--Robot Interaction Research}.
\newblock \bibinfo{journal}{\emph{International Journal of Social Robotics}}  \bibinfo{volume}{15} (\bibinfo{year}{2023}), \bibinfo{pages}{1841--1870}.
\newblock


\bibitem[Seaborn and Chang(2024)]%
        {seaborn2024subtle}
\bibfield{author}{\bibinfo{person}{Katie Seaborn} {and} \bibinfo{person}{Weichen~Joe Chang}.} \bibinfo{year}{2024}\natexlab{}.
\newblock \showarticletitle{Another subtle pattern: {Examining} demographic biases in dark patterns and deceptive design research}. In \bibinfo{booktitle}{\emph{Mobilizing {Research} and {Regulatory} {Action} on {Dark} {Patterns} and {Deceptive} {Design} {Practices} {Workshop} at {CHI} {Conference} on {Human} {Factors} in {Computing} {Systems}}} \emph{(\bibinfo{series}{{DDP} {CHI} `24})}. \bibinfo{publisher}{CEUR-WS.org}, \bibinfo{address}{online}, \bibinfo{numpages}{10}~pages.
\newblock
\urldef\tempurl%
\url{https://chi2024.darkpatternsresearchandimpact.com/}
\showURL{%
\tempurl}


\bibitem[Seaborn and Iseya(2023)]%
        {Seaborn_2023maldai}
\bibfield{author}{\bibinfo{person}{Katie Seaborn} {and} \bibinfo{person}{Satoru Iseya}.} \bibinfo{year}{2023}\natexlab{}.
\newblock \showarticletitle{Meaningful Play and Malicious Delight: Exploring Maldaimonic Game UX}. In \bibinfo{booktitle}{\emph{Companion Proceedings of the Annual Symposium on Computer-Human Interaction in Play}} \emph{(\bibinfo{series}{CHI PLAY ’23})}. \bibinfo{publisher}{ACM}, \bibinfo{address}{New York, NY, USA}, \bibinfo{pages}{174–180}.
\newblock
\urldef\tempurl%
\url{https://doi.org/10.1145/3573382.3616095}
\showDOI{\tempurl}


\bibitem[Seaborn et~al\mbox{.}(2024)]%
        {Seaborn_2024maldai}
\bibfield{author}{\bibinfo{person}{Katie Seaborn}, \bibinfo{person}{Satoru Iseya}, \bibinfo{person}{Shun Hidaka}, \bibinfo{person}{Sota Kobuki}, {and} \bibinfo{person}{Shruti Chandra}.} \bibinfo{year}{2024}\natexlab{}.
\newblock \showarticletitle{Play Across Boundaries: Exploring Cross-Cultural Maldaimonic Game Experiences}. In \bibinfo{booktitle}{\emph{Proceedings of the CHI Conference on Human Factors in Computing Systems}} \emph{(\bibinfo{series}{CHI ’24})}. \bibinfo{publisher}{ACM}, \bibinfo{address}{New York, NY, USA}, \bibinfo{pages}{1–15}.
\newblock
\urldef\tempurl%
\url{https://doi.org/10.1145/3613904.3642273}
\showDOI{\tempurl}


\bibitem[Seaborn and Pennefather(2022)]%
        {Seaborn_2022neut}
\bibfield{author}{\bibinfo{person}{Katie Seaborn} {and} \bibinfo{person}{Peter Pennefather}.} \bibinfo{year}{2022}\natexlab{}.
\newblock \showarticletitle{Neither “Hear” Nor “Their”: Interrogating Gender Neutrality in Robots}. In \bibinfo{booktitle}{\emph{2022 17th ACM/IEEE International Conference on Human-Robot Interaction (HRI)}}. \bibinfo{publisher}{IEEE}, \bibinfo{address}{New York, NY, USA}, \bibinfo{pages}{1030–1034}.
\newblock
\urldef\tempurl%
\url{https://doi.org/10.1109/hri53351.2022.9889350}
\showDOI{\tempurl}


\bibitem[Shaw(2011)]%
        {Shaw_2011identity}
\bibfield{author}{\bibinfo{person}{Adrienne Shaw}.} \bibinfo{year}{2011}\natexlab{}.
\newblock \showarticletitle{Do you identify as a gamer? Gender, race, sexuality, and gamer identity}.
\newblock \bibinfo{journal}{\emph{New Media \& Society}} \bibinfo{volume}{14}, \bibinfo{number}{1} (\bibinfo{date}{June} \bibinfo{year}{2011}), \bibinfo{pages}{28–44}.
\newblock
\showISSN{1461-7315}
\urldef\tempurl%
\url{https://doi.org/10.1177/1461444811410394}
\showDOI{\tempurl}


\bibitem[Shaw(2015)]%
        {shaw2015gaming}
\bibfield{author}{\bibinfo{person}{Adrienne Shaw}.} \bibinfo{year}{2015}\natexlab{}.
\newblock \bibinfo{booktitle}{\emph{Gaming at the edge: Sexuality and gender at the margins of gamer culture}}.
\newblock \bibinfo{publisher}{University of Minnesota Press}, \bibinfo{address}{Minneapolis, MN, USA}.
\newblock


\bibitem[Shaw and Friesem(2016)]%
        {shaw2016whereisthequeerness}
\bibfield{author}{\bibinfo{person}{Adrienne Shaw} {and} \bibinfo{person}{Elizaveta Friesem}.} \bibinfo{year}{2016}\natexlab{}.
\newblock \showarticletitle{Where is the queerness in games?: Types of lesbian, gay, bisexual, transgender, and queer content in digital games}.
\newblock \bibinfo{journal}{\emph{International Journal of Communication}}  \bibinfo{volume}{10} (\bibinfo{year}{2016}), \bibinfo{pages}{13}.
\newblock
\showISSN{1932-8036}


\bibitem[{Shogakukan} and {Nintendo}(2012)]%
        {kirby-flipbook}
\bibfield{author}{\bibinfo{person}{{Shogakukan}} {and} \bibinfo{person}{{Nintendo}}.} \bibinfo{year}{2012}\natexlab{}.
\newblock \bibinfo{booktitle}{\emph{Hoshi No Kirby 20th Anniversary Book Japan Game Art and Guide Book}}.
\newblock \bibinfo{publisher}{Shogakukan}, \bibinfo{address}{Tokyo, Japan}.
\newblock


\bibitem[Skelton et~al\mbox{.}(2024)]%
        {skelton2024itjust}
\bibfield{author}{\bibinfo{person}{Salem Skelton}, \bibinfo{person}{Damien~W. Riggs}, \bibinfo{person}{Annie Pullen~Sansfacon}, \bibinfo{person}{Sabra~L. Katz-Wise}, \bibinfo{person}{Manvi Arora}, {and} \bibinfo{person}{Charles-Antoine Thibeault}.} \bibinfo{year}{2024}\natexlab{}.
\newblock \showarticletitle{`It just feels really nice when people call me by my name': Accounts of gender euphoria among Australian trans young people and their parents}.
\newblock \bibinfo{journal}{\emph{Journal of Gender Studies}} \bibinfo{volume}{33}, \bibinfo{number}{4} (\bibinfo{year}{2024}), \bibinfo{pages}{470--482}.
\newblock
\urldef\tempurl%
\url{https://doi.org/10.1080/09589236.2023.2285984}
\showDOI{\tempurl}


\bibitem[Smith and Decker(2016)]%
        {smith2016understanding}
\bibfield{author}{\bibinfo{person}{Roger Smith} {and} \bibinfo{person}{Adrienne Decker}.} \bibinfo{year}{2016}\natexlab{}.
\newblock \showarticletitle{Understanding the impact of QPOC representation in video games}. In \bibinfo{booktitle}{\emph{2016 Research on Equity and Sustained Participation in Engineering, Computing, and Technology (RESPECT)}}. \bibinfo{publisher}{IEEE}, \bibinfo{address}{New York, New York, USA}, \bibinfo{pages}{1--8}.
\newblock
\urldef\tempurl%
\url{https://doi.org/10.1109/RESPECT.2016.7836164}
\showDOI{\tempurl}


\bibitem[Smith(2025)]%
        {Smith2025binbarriers}
\bibfield{author}{\bibinfo{person}{Xan Smith}.} \bibinfo{year}{2025}\natexlab{}.
\newblock \showarticletitle{Binary Barriers: Avatar Creation and Play for Nonbinary Video Game Players}.
\newblock \bibinfo{journal}{\emph{Games and Culture}} (\bibinfo{date}{Feb.} \bibinfo{year}{2025}), \bibinfo{numpages}{22}~pages.
\newblock
\showISSN{1555-4139}
\urldef\tempurl%
\url{https://doi.org/10.1177/15554120251322209}
\showDOI{\tempurl}


\bibitem[Spencer-Hall et~al\mbox{.}(2021)]%
        {Spencer_Hall_2021terms}
\bibfield{author}{\bibinfo{person}{Alicia Spencer-Hall}, \bibinfo{person}{Blake Gutt}, \bibinfo{person}{Martha~G. Newman}, \bibinfo{person}{Caitlyn McLoughlin}, \bibinfo{person}{Kevin C.~A. Elphick}, \bibinfo{person}{Felix Szabo}, \bibinfo{person}{Sophie Sexon}, \bibinfo{person}{Vanessa Wright}, \bibinfo{person}{Lee Colwill}, \bibinfo{person}{Amy~V. Ogden}, \bibinfo{person}{Gabrielle M.~W. Bychowski}, {and} \bibinfo{person}{Mathilde van Dijk}.} \bibinfo{year}{2021}\natexlab{}.
\newblock \bibinfo{booktitle}{\emph{Trans and Genderqueer Subjects in Medieval Hagiography}}.
\newblock \bibinfo{publisher}{Amsterdam University Press}, \bibinfo{address}{Amsterdam, Netherlands}.
\newblock
\showISBNx{9789462988248}
\urldef\tempurl%
\url{https://doi.org/10.5117/9789462988248}
\showDOI{\tempurl}


\bibitem[{Statistics Kingdom}(2022)]%
        {statskingdom}
\bibfield{author}{\bibinfo{person}{{Statistics Kingdom}}.} \bibinfo{year}{2022}\natexlab{}.
\newblock \bibinfo{title}{Statistics Kingdom}.
\newblock \bibinfo{howpublished}{\url{https://www.statskingdom.com/index.html}}.
\newblock


\bibitem[Steeds et~al\mbox{.}(2025)]%
        {Steeds2025joy}
\bibfield{author}{\bibinfo{person}{Madeleine Steeds}, \bibinfo{person}{Sarah Clinch}, \bibinfo{person}{Carolina Are}, \bibinfo{person}{Genavee Brown}, \bibinfo{person}{Ben Dalton}, \bibinfo{person}{Lexi Webster}, \bibinfo{person}{Alice Wilson}, {and} \bibinfo{person}{Dawn Woolley}.} \bibinfo{year}{2025}\natexlab{}.
\newblock \showarticletitle{Queer Joy on Social Media: Exploring the Expression and Facilitation of Queer Joy in Online Platforms}. In \bibinfo{booktitle}{\emph{Proceedings of the 2025 CHI Conference on Human Factors in Computing Systems}} \emph{(\bibinfo{series}{CHI ’25})}. \bibinfo{publisher}{ACM}, \bibinfo{address}{New York, NY, USA}, \bibinfo{pages}{1–19}.
\newblock
\urldef\tempurl%
\url{https://doi.org/10.1145/3706598.3713592}
\showDOI{\tempurl}


\bibitem[Storr et~al\mbox{.}(2021)]%
        {Storr_2021}
\bibfield{author}{\bibinfo{person}{R. Storr}, \bibinfo{person}{L. Nicholas}, \bibinfo{person}{K. Robinson}, {and} \bibinfo{person}{C. Davies}.} \bibinfo{year}{2021}\natexlab{}.
\newblock \showarticletitle{‘Game to play?’: barriers and facilitators to sexuality and gender diverse young people’s participation in sport and physical activity}.
\newblock \bibinfo{journal}{\emph{Sport, Education and Society}} \bibinfo{volume}{27}, \bibinfo{number}{5} (\bibinfo{date}{March} \bibinfo{year}{2021}), \bibinfo{pages}{604–617}.
\newblock
\showISSN{1470-1243}
\urldef\tempurl%
\url{https://doi.org/10.1080/13573322.2021.1897561}
\showDOI{\tempurl}


\bibitem[Stryker(2017)]%
        {transhistory}
\bibfield{author}{\bibinfo{person}{Susan Stryker}.} \bibinfo{year}{2017}\natexlab{}.
\newblock \bibinfo{booktitle}{\emph{Transgender History: The Roots of Today's Revolution} (\bibinfo{edition}{2nd} ed.)}.
\newblock \bibinfo{publisher}{Seal Press}, \bibinfo{address}{New York}.
\newblock


\bibitem[Sund{\'e}n(2012)]%
        {sunden2012queer}
\bibfield{author}{\bibinfo{person}{Jenny Sund{\'e}n}.} \bibinfo{year}{2012}\natexlab{}.
\newblock \showarticletitle{A queer eye on transgressive play}.
\newblock In \bibinfo{booktitle}{\emph{Gender and sexuality in online game cultures}}. \bibinfo{publisher}{Routledge}, \bibinfo{address}{New York, NY}, \bibinfo{pages}{171--190}.
\newblock
\showISBNx{9780203143148}


\bibitem[Sutton(2020)]%
        {Sutton_2020ambig}
\bibfield{author}{\bibinfo{person}{Selina~Jeanne Sutton}.} \bibinfo{year}{2020}\natexlab{}.
\newblock \showarticletitle{Gender Ambiguous, not Genderless: Designing Gender in Voice User Interfaces (VUIs) with Sensitivity}. In \bibinfo{booktitle}{\emph{Proceedings of the 2nd Conference on Conversational User Interfaces}} \emph{(\bibinfo{series}{CUI ’20})}. \bibinfo{publisher}{ACM}, \bibinfo{address}{New York, NY, USA}, \bibinfo{numpages}{8}~pages.
\newblock
\urldef\tempurl%
\url{https://doi.org/10.1145/3405755.3406123}
\showDOI{\tempurl}


\bibitem[Tacit(2020)]%
        {tacit2020joyful}
\bibfield{author}{\bibinfo{person}{Sam Tacit}.} \bibinfo{year}{2020}\natexlab{}.
\newblock \emph{\bibinfo{title}{Joyful bodies, joyful minds: Gender euphoria among transgender adults living in Canada}}.
\newblock \bibinfo{thesistype}{Ph.\,D. Dissertation}. \bibinfo{school}{Memorial University of Newfoundland}.
\newblock
\urldef\tempurl%
\url{https://research.library.mun.ca/14905/1/thesis.pdf}
\showURL{%
\tempurl}


\bibitem[Tannenbaum et~al\mbox{.}(2019)]%
        {Tannenbaum_2019}
\bibfield{author}{\bibinfo{person}{Cara Tannenbaum}, \bibinfo{person}{Robert~P. Ellis}, \bibinfo{person}{Friederike Eyssel}, \bibinfo{person}{James Zou}, {and} \bibinfo{person}{Londa Schiebinger}.} \bibinfo{year}{2019}\natexlab{}.
\newblock \showarticletitle{Sex and gender analysis improves science and engineering}.
\newblock \bibinfo{journal}{\emph{Nature}} \bibinfo{volume}{575}, \bibinfo{number}{7781} (\bibinfo{date}{Nov.} \bibinfo{year}{2019}), \bibinfo{pages}{137–146}.
\newblock
\showISSN{1476-4687}
\urldef\tempurl%
\url{https://doi.org/10.1038/s41586-019-1657-6}
\showDOI{\tempurl}


\bibitem[Tatum(2017)]%
        {tatum2017all}
\bibfield{author}{\bibinfo{person}{Beverly~Daniel Tatum}.} \bibinfo{year}{2017}\natexlab{}.
\newblock \showarticletitle{``Why Are All the Black Kids Still Sitting Together in the Cafeteria?'': and Other Conversations about Race in the Twenty-First Century}.
\newblock \bibinfo{journal}{\emph{Liberal Education}} \bibinfo{volume}{103}, \bibinfo{number}{3-4} (\bibinfo{year}{2017}), \bibinfo{pages}{46--56}.
\newblock


\bibitem[Tekinbas and Zimmerman(2003)]%
        {tekinbas2003rules}
\bibfield{author}{\bibinfo{person}{Katie~Salen Tekinbas} {and} \bibinfo{person}{Eric Zimmerman}.} \bibinfo{year}{2003}\natexlab{}.
\newblock \bibinfo{booktitle}{\emph{Rules of play: Game design fundamentals}}.
\newblock \bibinfo{publisher}{MIT Press}, \bibinfo{address}{Cambridge, MA, USA}.
\newblock


\bibitem[{Them}(2018)]%
        {them2018inqueery}
\bibfield{author}{\bibinfo{person}{{Them}}.} \bibinfo{year}{2018}\natexlab{}.
\newblock \bibinfo{title}{Inqueery: What Does It Mean to Be Agender?}
\newblock
\newblock
\urldef\tempurl%
\url{https://www.them.us/story/inqueery-agender}
\showURL{%
\tempurl}


\bibitem[Thompson(2007)]%
        {Thompson_2007panas}
\bibfield{author}{\bibinfo{person}{Edmund~R. Thompson}.} \bibinfo{year}{2007}\natexlab{}.
\newblock \showarticletitle{Development and Validation of an Internationally Reliable Short-Form of the Positive and Negative Affect Schedule (PANAS)}.
\newblock \bibinfo{journal}{\emph{Journal of Cross-Cultural Psychology}} \bibinfo{volume}{38}, \bibinfo{number}{2} (\bibinfo{date}{March} \bibinfo{year}{2007}), \bibinfo{pages}{227–242}.
\newblock
\showISSN{1552-5422}
\urldef\tempurl%
\url{https://doi.org/10.1177/0022022106297301}
\showDOI{\tempurl}


\bibitem[To et~al\mbox{.}(2023)]%
        {to2023flourishing}
\bibfield{author}{\bibinfo{person}{Alexandra To}, \bibinfo{person}{Angela~D.R. Smith}, \bibinfo{person}{Dilruba Showkat}, \bibinfo{person}{Adinawa Adjagbodjou}, {and} \bibinfo{person}{Christina Harrington}.} \bibinfo{year}{2023}\natexlab{}.
\newblock \showarticletitle{Flourishing in the everyday: Moving beyond damage-centered design in HCI for BIPOC communities}. In \bibinfo{booktitle}{\emph{Proceedings of the 2023 ACM Designing Interactive Systems Conference}} \emph{(\bibinfo{series}{DIS '23})}. \bibinfo{publisher}{Association for Computing Machinery}, \bibinfo{address}{New York, NY, USA}, \bibinfo{pages}{917--933}.
\newblock
\urldef\tempurl%
\url{https://doi.org/10.1145/3563657.3596057}
\showDOI{\tempurl}


\bibitem[Toups~Dugas et~al\mbox{.}(2016)]%
        {collecting-toups-dugas}
\bibfield{author}{\bibinfo{person}{Phoebe~O. Toups~Dugas}, \bibinfo{person}{Nicole~K. Crenshaw}, \bibinfo{person}{Rina~R. Wehbe}, \bibinfo{person}{Gustavo~F. Tondello}, {and} \bibinfo{person}{Lennart~E. Nacke}.} \bibinfo{year}{2016}\natexlab{}.
\newblock \showarticletitle{``The Collecting Itself Feels Good'': Towards Collection Interfaces for Digital Game Objects}. In \bibinfo{booktitle}{\emph{Proceedings of the 2016 Annual Symposium on Computer-Human Interaction in Play}} (Austin, Texas, USA) \emph{(\bibinfo{series}{CHI PLAY '16})}. \bibinfo{publisher}{Association for Computing Machinery}, \bibinfo{address}{New York, NY, USA}, \bibinfo{pages}{276--290}.
\newblock
\showISBNx{9781450344562}
\urldef\tempurl%
\url{https://doi.org/10.1145/2967934.2968088}
\showDOI{\tempurl}


\bibitem[Toups~Dugas and Tanenbaum(2024)]%
        {toups-dugas-pole-2024}
\bibfield{author}{\bibinfo{person}{Phoebe~O. Toups~Dugas} {and} \bibinfo{person}{Theresa Tanenbaum}.} \bibinfo{year}{2024}\natexlab{}.
\newblock \showarticletitle{Learning to Fail Beautifully: Pole Dancing as a Case Study of Gender Euphoria \& Dysphoria for Embodied Interaction Design}. In \bibinfo{booktitle}{\emph{Proceedings of the Halfway to the Future Symposium}} (Santa Cruz, CA, USA) \emph{(\bibinfo{series}{HttF '24})}. \bibinfo{publisher}{Association for Computing Machinery}, \bibinfo{address}{New York, NY, USA}, Article \bibinfo{articleno}{23}, \bibinfo{numpages}{8}~pages.
\newblock
\showISBNx{9798400710421}
\urldef\tempurl%
\url{https://doi.org/10.1145/3686169.3686180}
\showDOI{\tempurl}


\bibitem[Trammell(2023)]%
        {Trammell_2023}
\bibfield{author}{\bibinfo{person}{Aaron Trammell}.} \bibinfo{year}{2023}\natexlab{}.
\newblock \bibinfo{booktitle}{\emph{The Privilege of Play: A History of Hobby Games, Race, and Geek Culture}}.
\newblock \bibinfo{publisher}{New York University Press}, \bibinfo{address}{New York, NY, USA}.
\newblock
\showISBNx{9781479818419}
\urldef\tempurl%
\url{https://doi.org/10.18574/nyu/9781479818419.001.0001}
\showDOI{\tempurl}


\bibitem[Tucker(2010)]%
        {tucker2010mental}
\bibfield{author}{\bibinfo{person}{Ian Tucker}.} \bibinfo{year}{2010}\natexlab{}.
\newblock \showarticletitle{Mental health service user territories: Enacting `safe spaces' in the community}.
\newblock \bibinfo{journal}{\emph{Health: An Interdisciplinary Journal for the Social Study of Health, Illness and Medicine}} \bibinfo{volume}{14}, \bibinfo{number}{4} (\bibinfo{year}{2010}), \bibinfo{pages}{434--448}.
\newblock
\urldef\tempurl%
\url{https://doi.org/10.1177/1363459309357485}
\showDOI{\tempurl}


\bibitem[Turner(1999)]%
        {Turner1999}
\bibfield{author}{\bibinfo{person}{Stephanie~S. Turner}.} \bibinfo{year}{1999}\natexlab{}.
\newblock \showarticletitle{Intersex Identities: Locating New Intersections of Sex and Gender}.
\newblock \bibinfo{journal}{\emph{Gender and Society}} \bibinfo{volume}{13}, \bibinfo{number}{4} (\bibinfo{year}{1999}), \bibinfo{pages}{457--479}.
\newblock


\bibitem[Utsch et~al\mbox{.}(2017)]%
        {utsch2017queer}
\bibfield{author}{\bibinfo{person}{Sofia Utsch}, \bibinfo{person}{Luiza~C Bragan{\c{c}}a}, \bibinfo{person}{Pedro Ramos}, \bibinfo{person}{Pedro Caldeira}, {and} \bibinfo{person}{Joao Tenorio}.} \bibinfo{year}{2017}\natexlab{}.
\newblock \showarticletitle{Queer Identities in Video Games: Data visualization for a quantitative analysis of representation}.
\newblock \bibinfo{journal}{\emph{Proceedings of SBGames}} (\bibinfo{year}{2017}), \bibinfo{pages}{850--851}.
\newblock


\bibitem[Van~Wert and Howansky(2024)]%
        {VanWert2024}
\bibfield{author}{\bibinfo{person}{Sonder Van~Wert} {and} \bibinfo{person}{Kristina Howansky}.} \bibinfo{year}{2024}\natexlab{}.
\newblock \showarticletitle{Fantasy Worlds, Real-Life Impact: The Benefits of RPGs for Transgender Identity Exploration}.
\newblock \bibinfo{journal}{\emph{Journal of Homosexuality}} \bibinfo{volume}{72}, \bibinfo{number}{2} (\bibinfo{date}{Feb.} \bibinfo{year}{2024}), \bibinfo{pages}{319–345}.
\newblock
\showISSN{1540-3602}
\urldef\tempurl%
\url{https://doi.org/10.1080/00918369.2024.2320242}
\showDOI{\tempurl}


\bibitem[Vorvoreanu et~al\mbox{.}(2019)]%
        {Vorvoreanu_2019neut}
\bibfield{author}{\bibinfo{person}{Mihaela Vorvoreanu}, \bibinfo{person}{Lingyi Zhang}, \bibinfo{person}{Yun-Han Huang}, \bibinfo{person}{Claudia Hilderbrand}, \bibinfo{person}{Zoe Steine-Hanson}, {and} \bibinfo{person}{Margaret Burnett}.} \bibinfo{year}{2019}\natexlab{}.
\newblock \showarticletitle{From Gender Biases to Gender-Inclusive Design: An Empirical Investigation}. In \bibinfo{booktitle}{\emph{Proceedings of the 2019 CHI Conference on Human Factors in Computing Systems}} \emph{(\bibinfo{series}{CHI ’19})}. \bibinfo{publisher}{ACM}, \bibinfo{address}{New York, NY, USA}, \bibinfo{pages}{1–14}.
\newblock
\urldef\tempurl%
\url{https://doi.org/10.1145/3290605.3300283}
\showDOI{\tempurl}


\bibitem[Weitekamp(2013)]%
        {weitekamp2013more}
\bibfield{author}{\bibinfo{person}{Margaret~A. Weitekamp}.} \bibinfo{year}{2013}\natexlab{}.
\newblock \showarticletitle{More than ``just Uhura'': Understanding Star Trek's Lt. Uhura, civil rights, and space history}.
\newblock In \bibinfo{booktitle}{\emph{Star Trek and History (Wiley Pop Culture and History Series)}}. \bibinfo{publisher}{Wiley}, \bibinfo{address}{Hoboken, New Jersey, U.S.}, \bibinfo{pages}{22--37}.
\newblock


\bibitem[Werft and S{\'a}nchez(2016)]%
        {third-gender-werft}
\bibfield{author}{\bibinfo{person}{Meghan Werft} {and} \bibinfo{person}{Erica S{\'a}nchez}.} \bibinfo{year}{2016}\natexlab{}.
\newblock \bibinfo{title}{Male, Female, and Muxes: Places Where a Third Gender is Accepted}.
\newblock
\newblock
\urldef\tempurl%
\url{https://www.globalcitizen.org/en/content/third-gender-gay-rights-equality/}
\showURL{%
\tempurl}


\bibitem[Whitehouse et~al\mbox{.}(2023)]%
        {Whitehouse_2023agender}
\bibfield{author}{\bibinfo{person}{Kayson Whitehouse}, \bibinfo{person}{Michael Hitchens}, {and} \bibinfo{person}{Nicole Matthews}.} \bibinfo{year}{2023}\natexlab{}.
\newblock \showarticletitle{Trans* and gender diverse players: Avatars and gender-alignment}.
\newblock \bibinfo{journal}{\emph{Entertainment Computing}}  \bibinfo{volume}{47} (\bibinfo{date}{Aug.} \bibinfo{year}{2023}), \bibinfo{pages}{100584}.
\newblock
\showISSN{1875-9521}
\urldef\tempurl%
\url{https://doi.org/10.1016/j.entcom.2023.100584}
\showDOI{\tempurl}


\bibitem[Williams et~al\mbox{.}(2009)]%
        {Williams_2009rep}
\bibfield{author}{\bibinfo{person}{Dmitri Williams}, \bibinfo{person}{Nicole Martins}, \bibinfo{person}{Mia Consalvo}, {and} \bibinfo{person}{James~D. Ivory}.} \bibinfo{year}{2009}\natexlab{}.
\newblock \showarticletitle{The virtual census: representations of gender, race and age in video games}.
\newblock \bibinfo{journal}{\emph{New Media \& Society}} \bibinfo{volume}{11}, \bibinfo{number}{5} (\bibinfo{date}{July} \bibinfo{year}{2009}), \bibinfo{pages}{815–834}.
\newblock
\showISSN{1461-7315}
\urldef\tempurl%
\url{https://doi.org/10.1177/1461444809105354}
\showDOI{\tempurl}


\bibitem[Wolf(2011)]%
        {wolf2011theorizing}
\bibfield{author}{\bibinfo{person}{Mark~J.P. Wolf}.} \bibinfo{year}{2011}\natexlab{}.
\newblock \showarticletitle{Theorizing navigable space in video games}.
\newblock \bibinfo{journal}{\emph{Digarec Series}} \bibinfo{number}{6} (\bibinfo{year}{2011}), \bibinfo{pages}{18--49}.
\newblock


\bibitem[Woolsey(1986)]%
        {woolsey_critical_1986}
\bibfield{author}{\bibinfo{person}{Lorette~K. Woolsey}.} \bibinfo{year}{1986}\natexlab{}.
\newblock \showarticletitle{The critical incident technique: {An} innovative qualitative method of research}.
\newblock \bibinfo{journal}{\emph{Canadian Journal of Counselling and Psychotherapy}} \bibinfo{volume}{20}, \bibinfo{number}{4} (\bibinfo{year}{1986}), \bibinfo{pages}{242--254}.
\newblock
\showISSN{1923-6182}
\urldef\tempurl%
\url{https://cjc-rcc.ucalgary.ca/article/view/59733}
\showURL{%
\tempurl}


\bibitem[Zabala et~al\mbox{.}(2024)]%
        {Zabala2024}
\bibfield{author}{\bibinfo{person}{Jailyn Zabala}, \bibinfo{person}{Josie Zvelebilova}, {and} \bibinfo{person}{Alexandra To}.} \bibinfo{year}{2024}\natexlab{}.
\newblock \showarticletitle{Queer TTRPGs' Visibility, Safety, and Allegory as Resistance}. In \bibinfo{booktitle}{\emph{Proceedings of the 19th International Conference on the Foundations of Digital Games (FDG)}} \emph{(\bibinfo{series}{FDG 2024})}. \bibinfo{publisher}{ACM}, \bibinfo{address}{New York, NY, USA}, \bibinfo{pages}{1–10}.
\newblock
\urldef\tempurl%
\url{https://doi.org/10.1145/3649921.3650022}
\showDOI{\tempurl}


\bibitem[Šisler(2008)]%
        {_isler_2008}
\bibfield{author}{\bibinfo{person}{Vít Šisler}.} \bibinfo{year}{2008}\natexlab{}.
\newblock \showarticletitle{Digital {Arabs}: Representation in video games}.
\newblock \bibinfo{journal}{\emph{European Journal of Cultural Studies}} \bibinfo{volume}{11}, \bibinfo{number}{2} (\bibinfo{date}{May} \bibinfo{year}{2008}), \bibinfo{pages}{203–220}.
\newblock
\showISSN{1460-3551}
\urldef\tempurl%
\url{https://doi.org/10.1177/1367549407088333}
\showDOI{\tempurl}


\end{thebibliography}





\end{document}